\documentclass[1p,fleqn,times]{elsarticle}

\usepackage{hyperref}

\usepackage{geometry} % Required for adjusting page dimensions

\newcommand{\Vbar}{\overline{\mathbf{V}}^e}%{\overline{V}^e}
\newcommand{\Jnorm}[1]{||{#1}||}
\newcommand{\dev}[1]{\mbox{dev}\left({#1}\right)}

\usepackage{natbib}
\usepackage{amsmath,amssymb}
\usepackage{mathrsfs}
\usepackage{mathtools}
\usepackage{graphicx}
\usepackage{rotating}
\usepackage{blindtext}
\usepackage{physics}
\usepackage{xcolor}
\usepackage{enumitem}
\usepackage{bm}
\usepackage{amsmath}
\usepackage{multirow}
\usepackage{array}
\newcolumntype{P}[1]{>{\centering\arraybackslash}p{#1}}
\newcolumntype{M}[1]{>{\centering\arraybackslash}m{#1}}
\usepackage{pdflscape}
\usepackage{ifthen}
\usepackage{verbatim}
\usepackage{tikz}
\usepackage{pgfplots}
\usepackage{subcaption}
\usepackage{rotating}
\usepackage{pifont}

\usetikzlibrary{plotmarks}
\usetikzlibrary{arrows}
\usetikzlibrary{decorations.pathreplacing}
\usepgfplotslibrary{groupplots} 
\pgfplotsset{width=\textwidth, compat=1.13}

\newcommand{\hstar}{\text{\ding{73}}}

\graphicspath{ {Images_new/}}

\makeatletter
\def\ps@pprintTitle{%
  \let\@oddhead\@empty
  \let\@evenhead\@empty
  \def\@oddfoot{}%
  \let\@evenfoot\@oddfoot}
\makeatletter
\def\ps@pprintTitle{%
  \let\@oddhead\@empty
  \let\@evenhead\@empty
  \def\@oddfoot{}%
  \let\@evenfoot\@oddfoot}
\makeatother
\usepackage{fancyhdr}

\fancypagestyle{pprintTitle}{
\lhead{}\chead{}\rhead{}\lfoot{\textcopyright \small ~British Crown Owned Copyright 2020/AWE}\cfoot{\thepage}\rfoot{}
}

\begin{document}

\begin{frontmatter}

\title{A Flux-enriched Godunov Method for Multi-material Problems with Interface Slide and Void Opening}


\author[mymainaddress]{Tim Wallis \corref{mycorrespondingauthor}}
\cortext[mycorrespondingauthor]{Corresponding author}
\ead{tnmw2@cam.ac.uk}

\author[mysecondaryaddress]{Philip T. Barton}
\author[mymainaddress]{Nikolaos Nikiforakis}

\address[mymainaddress]{Department of Physics, University of Cambridge, Cavendish Laboratory, JJ Thomson Avenue, CB3 0HE, UK}
\address[mysecondaryaddress]{AWE Aldermaston, Reading, Berkshire, RG7 4PR, UK}

\begin{abstract}

This work outlines a new three-dimensional diffuse interface finite volume method for the simulation of multiple solid and fluid components featuring large deformations, sliding and void opening. This is achieved by extending an existing reduced-equation diffuse interface method by means of a number of novel flux-modifiers and interface seeding routines that enable the application of different material boundary conditions. The method allows for slip boundary conditions across solid interfaces, material-void interaction, and interface separation. 
The method is designed to be straightforward to implement, inexpensive and highly parallelisable. This makes it suitable for use in large, multi-dimensional simulations that feature many complex materials and physical processes interacting over multiple levels of adaptive mesh refinement.
Furthermore, the method allows for the generation of new interfaces in a conservative fashion and therefore naturally facilitates the simulation of high-strain rate fracture. Hence, the model is augmented to include ductile damage to allow for validation of the method against demanding physical experiments. 
The method is shown to give excellent agreement with both experiment and existing Eulerian interface tracking algorithms that employ sharp interface methods.

\textcopyright ~British Crown Owned Copyright 2020/AWE
\end{abstract}


\end{frontmatter}

\pagestyle{pprintTitle}
\thispagestyle{pprintTitle}

\section{Introduction}

%multi-physics is important
Multi-material, multi-physics simulation techniques are vital for the accurate modelling of many real-world physical systems. These problems are dynamic and often highly non-linear. Such simulations often feature multiple complex materials with different equations of state, large deformations and strain rates, and additional physical effects such as material fracture or reactivity. The combination of all of these factors makes a multi-physics approach crucial; modelling the system in a single framework ensures full physical interaction between all the components. To this end, realistic physical models, suitable mathematical solution techniques, and efficient computer algorithms must be brought together in a single framework to enable a complete system to be modelled. 

Importantly, real-life problems of interest often involve large, three dimensional systems evolving over long time-scales. This makes the development of a practical numerical scheme crucial, and provides the motivation for the work at hand. This work will be focused on providing a straightforward algorithm that can form a base for large, 3D, multi-material, multi-physics simulations. As such, particular care will be taken to ensure the resulting algorithm is direct, prioritising the ease with which additional physics can be incorporated.

A variety of mature methods for solving multi-material shock-capturing compressible flow problems already exist. These are divided into a number of approaches, the broadest of which being the division between Lagrangian and Eulerian methods. In essence, this choice comes down to choosing the frame of reference in which to solve the equations of motion. Lagrangian methods (typically finite element methods) consider a frame that is fixed with the material which deforms as the simulation progresses. Eulerian methods (typically finite volume methods) consider a frame fixed in space, through which material flows.
These methods both have their advantages and disadvantages, but in the high-deformation, high-strain-rate problems considered in this work, large topological changes can cause Lagrangian meshing methods to fail due to the severe mesh distortion. As Eulerian meshes do not deform with the material, they can smoothly handle large material deformations and complex topological changes, and so are well suited to the task at hand.

%Explain existing techniques

However, Eulerian methods are not without drawbacks and have challenges associated with their implementation in higher dimensions and with AMR.  Most existing Eulerian models capable of simulating fluids and elastoplastic solids are based on tracking material interfaces and fall into one of three broad categories. The first is methods based on homogenised mixed cells and volume-of-fluid reconstruction, such as \citet{benson:1992}. These methods underpin many well-established and legacy multi-material codes. The next is Ghost Fluid methods, such as \citet{SchochCoupled,MiNi18,4StatesOfMatter,BartonSliding} and \citet{BartonLevelSetDamage}. These are a fairly recent development, based on the methods of \citet{FedkiwGFM,RiemannGFM} and \citet{rGFM}. Ghost Fluid methods capture material boundaries with the use of a level set and apply dynamic boundary conditions to allow for material interaction. These methods have the downside of being non-conservative at material interfaces. Lastly, Cut-cell methods such as \citet{NandanCutCell,miller:2002} and \citet{barton:2011} resolve the geometry of cells intersected by interfaces and apply a strict finite volume discretisation to capture the flux across material boundaries. All of these approaches have the advantage that they maintain arbitrarily sharp interfaces, but to do so they involve complex interface reconstructions and mixed-cell algorithms. 
Furthermore, applying load-balancing to many of these schemes when using AMR is difficult due to the non-uniform numerical methods. Robust implementations can therefore be challenging to construct.
Some of these difficulties have led to the development of hybrid approaches such as Arbitrary Lagrangian Eulerian methods such as \citet{BarlowALE} and co-simulation methods, based on embedding finite element grids in fluid domains, see for example \citet{deiterding2006}. These methods are popular for fluid-structure interaction problems, but introduce the natural complexities of mesh management associated with Lagrangian methods.

%Explain Diffuse interface models
Diffuse interface methods represent a different way to handle material interfaces in Eulerian methods. These methods track materials through the volume fraction that they occupy in a given computational cell. This volume fraction is continuous and multiple materials can be present in a single cell. This numerical mixture is then evolved using a single system of partial differential equations that encompasses the physics of all the components in the mixture.
A particular benefit of diffuse interface models is that they provide conservation equations for mass, momentum and energy across interfaces. Previous work has shown that these schemes have the considerable advantage over interface tracking methods in that they can support genuine fluid mixtures, allowing for the study of phenomena such as cavitation and chemical reaction which rely on physical mixtures \cite{PetitpasCaviation,WallisMultiPhysics}.

Within diffuse interface methods, there are a range of different approaches. These different approaches are often due to equilibrium assumptions made about the physical model in question. The most general approach is to assume full non-equilibrium between mixture components. This results in each material in the mixture having its own density, velocity, pressure and temperature, with relaxation terms augmenting these equations to allow for materials to interact and exchange conserved quantities. An example of such a model is the work of \citet{BaerNunziato}.
However, experiments reveal that the scales over which both pressure and velocity equilibrate are very small, and evolving a velocity and pressure for every material is an encumbrance. It is therefore common \cite{Kapila, PetitpasCaviation} to perform an asymptotic reduction of non-equilibrium models, applying equilibrium conditions to variables such as velocity, to obtain a reduced-equation model. This process leaves scope for how many equilibrium assumptions are to be applied, with authors such as \citet{Allaire} assuming both pressure and velocity equilibrium, whereas authors such as \citet{FavrieElasticDiffuse} retain pressure non-equilibrium. These approaches lead to components of a numerical mixture sharing a single equation for quantities such as momentum or energy. This in turn results in having to solve fewer equations overall, and is considered beneficial in terms of both efficiency and stability.
However, these equilibrium assumptions often come at a price. In some scenarios, such as when considering porous materials like \citet{BaerNunziato}, it is wholly inappropriate to assume velocity equilibrium between phases. At an even more basic level, velocity equilibrium prevents the use of slip boundary conditions at material interfaces and prevents material separation. This can be seen straightforwardly, as both of these problems require a velocity discontinuity at the interface: tangentially for slip and normally for separation. Such discontinuities are not permitted by a velocity equilibrium model, which will average the two velocities at the interface, effectively applying a stick boundary condition. 

Diffuse interface methods are well established for multi-fluid problems, see for example \citet{Allaire, SaurelAbgrall, Massoni}. 
By contrast, relatively few authors have applied diffuse interface methods to coupled solid-fluid dynamics, including \citet{Barton2019, FavrieElasticDiffuse, FavriePlasticDiffuse, HankExperimental} and \citet{DumbserDamage}. This is likely due to a number of factors. Firstly, Eulerian diffuse interface methods require qualitatively different numerical methods compared to those traditionally applied to solid dynamics, especially in Lagrangian schemes. Secondly, the diffuse nature of the material interfaces has, in the past, proved to be a limiting factor on the accuracy of such methods. This issue has, however, recently been addressed by the development of various interface sharpening methods, such as the THINC method designed by \citet{XiaoTHINC}, implemented by \citet{Barton2019}. Lastly, diffuse interface methods have not been generally applied to solid dynamics due the limiting nature of the enforced stick boundary condition when using reduced-equation models. Unlike with fluids, it is natural for materials with strength to be able to slide along each other and separate from each other. This problem is addressed by the work at hand.

This paper details a new diffuse interface method for the simulation of coupled elastoplastic solids and fluids. Namely, this work introduces methods to allow for slip boundary conditions, material-void interaction, and material fracture in a diffuse interface context. The work extends the method of \citet{Barton2019}, who developed a model for coupled solid-fluid dynamics based on the Allaire five-equation multi-fluid model \cite{Allaire}. The new method introduces a novel set of flux-modifiers and interface seeding routines that will enable the desired boundary conditions to be applied on top of the base model. Alongside this, a damage model is incorporated to enable material damage and fracture to be studied.

As has been alluded to, many other methods exist that can model multi-material systems, and these will be mentioned by way of contrast and comparison in the validation of the new method. However, the new method presents a number of specific advantages -- it is designed to be straightforward to implement, inexpensive and highly parallelisable, making it suitable for use in large, multi-dimensional simulations that feature multiple complex materials and physical processes over several levels of adaptive mesh refinement (AMR). 
As such, the method can be considered an `engineering solution' to the problems presented by multi-material interactions.

The highly-parallelisable nature of the method stems from the locality of the algorithm. There is no need to use the expensive level set reinitialisation and extrapolation routines used by Ghost Fluid methods, which require multi-patch and multi-level communication. The method is also highly numerically homogeneous. One single set of partial differential equations that encompasses the physics of all materials present is solved across the entire domain and material boundary conditions are facilitated by a number of simple flux-modifiers. These flux-modifiers only change how the flux between cells is calculated, rather than requiring any drastically different numerical machinery.

How well a given algorithm can scale with the number of materials present is also important as the number of materials in a simulation gets increasingly large. As has been mentioned, non-equilibrium equation systems allow for material shear by providing every material with its own complete set of variables. By way of example, \citet{NdanouDiffuseFracture} facilitate a form of fracture by giving each material its own deformation and stress tensor. However, this method gets increasingly expensive as the number of materials grows. In a scenario where compaction dynamics and velocity relaxation times are not of specific interest, the new method has the advantage of allowing material shear while only using a single velocity, momentum, deformation and stress. This greatly reduces how the algorithmic complexity scales with the number of materials. It should be stressed, however, that the void and fracture components of the method could still be used in a non-equilibrium type equation system, if so desired.

%AMR is important
The problems of interest often occur over disparate scales, with both regions of fine detail with strong interaction and regions with smoothly varying features. It is therefore desirable to use AMR to better focus computational resources where they are needed. With computational efficiency in mind, the algorithm outlined in this work is designed to work straightforwardly with hierarchical AMR algorithms, saving computational resources by allowing high resolution around regions of interest and lower resolution in smooth regions. 

Finally, the method facilitates a more conservative fracture algorithm than can be achieved with many other sharp interface methods, due to the continuous, diffuse representation of interfaces. It is commonplace in many existing methods for simulating fracture to introduce cracks by eroding material from critically damaged cells \cite[e.g.][]{ParticleLevelSetDamage,UdaykumarDamage,BartonLevelSetDamage}. This approach is highly non-conservative and leads to issues if the conserved state of eroded material is simply redistributed to its neighbours. The new method avoids this problem, instead allowing a material interface to open continuously.

\section{Governing Theory}

\subsection{Evolution Equations}
In the interests of focusing on multi-physics, only a summary of the core evolution equations is provided here. Full details, including the derivation, are outlined by \citet{Barton2019}.\\
Materials are allowed to mix at their interfaces. A material's contribution to a spatially averaged physical quantity is weighted by its volume fraction, $\phi$, in that region. This mixing is referred to as numerical mixing, to distinguish it from the physical mixtures such as those produced by reactive fluids. The state of any material $l$ is characterised by the phasic density $\rho_{(l)}$, volume fraction $\phi_{(l)}$, symmetric left unimodular stretch tensor $\Vbar$, velocity vector $\mathbf{u}$, and  internal energy $\mathscr{E}$.

The left stretch tensor $\mathbf{V}^e$ is related to the deformation tensor $\mathbf{F}$ by the polar decomposition:
\begin{equation}
 \mathbf{F} = \mathbf{V}^e\mathbf{R},
\end{equation}
after which it is normalised to obtain $\Vbar$:
\begin{equation}
 \Vbar = \det\left(\mathbf{V}^e\right)^{-1/3}\mathbf{V}^e.
\end{equation}

The base model of \citet{Barton2019} assumes mechanical equilibrium; materials in a mixture region share a single velocity, pressure and deviatoric strain. This assumption will be maintained, in that only a single equation for each of these quantities will be evolved, but flux-modifiers in the numerical method will allow for the correct material boundary conditions to be imposed on top.

For $l=1,\ldots, N$ materials:
\begin{eqnarray}
\frac{\partial \phi_{(l)}}{\partial t} + \frac{\partial \phi_{(l)} u_k}{\partial x_k} &=& \phi_{(l)}\frac{\partial u_k}{\partial x_k}\\
\frac{\partial \rho_{(l)}\phi_{(l)}}{\partial t} + \frac{\partial \rho_{(l)} \phi_{(l)} u_k}{\partial x_k} &=& 0 \\
\frac{\partial \rho u_i }{\partial t} + \frac{\partial (\rho u_i u_k -\sigma_{ik}) }{\partial x_k} &=& 0\\
\frac{\partial \rho E }{\partial t} + \frac{\partial (\rho E u_k - u_i \sigma_{ik}) }{\partial x_k} &=& 0\\
\frac{\partial \Vbar_{ij} }{\partial t} + \frac{\partial \left( \Vbar_{ij}  u_k - \Vbar_{kj} u_i \right) }{\partial x_k} &=& \frac{2}{3}\Vbar_{ij}\frac{\partial u_k}{\partial x_k} - u_i\beta_j - \boldsymbol\Phi_{ij} \ , %\\
\end{eqnarray}
where $E=\mathscr{E}+|\mathbf{u}|/2$ denotes the specific total energy, $\boldsymbol\sigma$ denotes the Cauchy stress tensor, $\beta_j=\partial\Vbar_{kj}/\partial x_k$, and $\boldsymbol\Phi$ represents the contribution from plastic effects.
Some multi-physics closure models introduce a dependence on material history variables such as the equivalent plastic strain, $\varepsilon_{p(l)}$, or the damage parameter $D_{(l)}$. 
For these materials, additional evolution equations are required:
\begin{eqnarray}
\frac{\partial \rho_{(l)}\phi_{(l)} \alpha_{(l)}}{\partial t} + \frac{\partial \rho_{(l)} \phi_{(l)} \alpha_{(l)} u_k}{\partial x_k} &=& \rho_{(l)}\phi_{(l)}\dot{\alpha}_{(l)} \ .
\end{eqnarray}
Here $\alpha_{(l)}$ represents any such history parameter which is advected and evolved with a material as time progresses. A material may have more than one history variable, in which case $\alpha$ represents a vector.

Although it might appear that the model describes solid materials, it will be shown in the next section that fluids can be considered as a special case. It is the equation-of-state for each material that ultimately distinguishes solids from fluids. 
The system allows for the fully-coupled multi-physics solution of problems, which in this work includes the interaction of damageable elastoplastic solids, fluid mixtures and vacuum. The extension to yet more multi-physics applications follows straightforwardly by the inclusion of additional history parameters and closure relations, for example reactive fluid mixtures as shown in \citet{WallisMultiPhysics}.

\subsection{Thermodynamics}
\label{sec:thermodynamics}
The internal energy $\mathscr{E}$ for each material is defined by an equation-of-state that conforms to the general form:
\begin{equation}\label{eq_eos_gen}
\mathscr{E}_{(l)}\left(\rho_{(l)},T_{(l)},\dev{\mathbf{H}^e},\alpha_{(l)}\right) = \mathscr{E}_{(l)}^c\left(\rho_{(l)},\alpha_{(l)}\right)+ \mathscr{E}_{(l)}^t\left(\rho_{(l)},T_{(l)}\right) + \mathscr{E}_{(l)}^s\left(\rho_{(l)},\dev{\mathbf{H}^e},\alpha_{(l)}\right) \ ,
\end{equation}
where
\begin{equation}
\dev{\mathbf{H}^e} = \ln\left(\Vbar\right) \ 
\end{equation}
is the deviatoric\footnotemark \ Hencky strain tensor and $T$ is the temperature. The three terms on the right hand side are the contribution due to cold compression or dilation, $\mathscr{E}_{(l)}^c\left(\rho_{(l)},\alpha_{(l)}\right)$, the contribution due to temperature deviations, $\mathscr{E}_{(l)}^t\left(\rho_{(l)},T_{(l)}\right)$, and the contribution due to shear strain $\mathscr{E}_{(l)}^s\left(\rho_{(l)},\dev{\mathbf{H}^e},\alpha_{(l)}\right)$. The cold compression energy will generally be provided by the specific  closure model for each material. The thermal energy is given by
\begin{eqnarray}
\mathscr{E}_{(l)}^t(\rho_{(l)},T)&=& C_{(l)}^{\text{V}}\left(T-T_{(l)}^0\theta_{(l)}^D\left(\rho_{(l)}\right)\right) \ ,
\end{eqnarray}
where $C_{(l)}^{\text{V}}$ is the heat capacity, $T_{(l)}^0$ is a reference temperature, and $\theta_{(l)}^{\text{D}}(\rho_{(l)})$ is the non-dimensional Debye temperature.
The Debye temperature is related to the Gr\"uneisen function, $\Gamma(\rho_{(l)})$, via
\begin{equation}
\Gamma_{(l)}(\rho_{(l)}) = \frac{\partial \ln\theta_{(l)}^{\text{D}}(\rho_{(l)})}{\partial \ln(1/\rho_{(l)})} = \frac{\rho_{(l)}}{\theta_{(l)}^{\text{D}}(\rho_{(l)})}\frac{\partial \theta_{(l)}^{\text{D}}(\rho_{(l)})}{\partial\rho_{(l)}} \ .
\end{equation}
The specific form of the Gr\"uneisen function for each material must be provided, but it is generally taken to be constant. The shear energy is given by
\begin{equation}
\mathscr{E}_{(l)}^s(\rho_{(l)},\dev{\mathbf{H}^e},\alpha_{(l)}) =  \frac{G_{(l)}\left(\rho_{(l)},\alpha_{(l)}\right)}{\rho_{(l)}}  \mathcal{J}^2\left(\dev{\mathbf{H}^e}\right) \ ,
\end{equation}
where $G(\rho_{(l)},\alpha_{(l)})$ is the shear modulus, and
\begin{equation}
\mathcal{J}^2(\dev{\mathbf{H}^e}) = \tr\left(\dev{\mathbf{H}^e}\cdot\dev{\mathbf{H}^e}^{\text{T}}\right)
\end{equation}
is the second invariant of shear strain. This form is chosen such that the resultant stresses are a non-linear analogy of Hooke's law. \citet{BruhnsStrainEnergy} find that this form provides good empirical agreement for a range of materials and deformations.
It is remarked that other extensions of infinitesimal strain theory to finite deformations have taken the density divisor to be the reference value $\rho^0_{(l)}$. This formulation can be replicated here without further modification to the model or numerical method by setting $G=\left(\frac{\rho_{(l)}}{\rho^0_{(l)}}\right) \overline{G}$, and redefining $\overline{G}$ as the actual shear modulus.
\footnotetext{
For any $N\times N$ matrix $\mathbf{M}$, $\dev{\mathbf{M}}:=\mathbf{M}-\frac{1}{N}\tr(\mathbf{M})\mathbf{I} \ $ denotes the matrix deviator. Here, $\tr(\mathbf{M})$ denotes the trace, and $\mathbf{I}$ denotes the identity matrix.
}

For each component, the Cauchy stress, $\boldsymbol\sigma$, and pressure, $p$, are inferred from the second law of thermodynamics and classical arguments for irreversible elastic deformations:
\begin{eqnarray}
\boldsymbol\sigma_{(l)} &=& p_{(l)}\mathbf{I} + \dev{\boldsymbol\sigma_{(l)}} \\
p_{(l)} &=& \rho^{2}_{(l)} \frac{\partial\mathscr{E}_{(l)}}{\partial \rho_{(l)}} \label{eq:p_energy_derivative}\\ 
\dev{\boldsymbol\sigma_{(l)}} &=& 2G_{(l)}\cdot\dev{\mathbf{H}^e} \ .
\end{eqnarray}
It is reiterated that, although it might appear that the model describes solid materials, fluids can be considered a special case where the shear modulus is zero, resulting in a spherical stress tensor and no shear energy contribution. This formulation lends itself well to diffuse interface modelling where different phases that share the same underlying model can combine consistently in mixture regions.

Mixture rules must be provided to represent the state of regions containing multiple materials in a thermodynamically consistent way. The following mixture rules are applied \cite{Allaire, Barton2019}:
\begin{eqnarray}
1 &=& \sum_{l=1}^N\phi_{(l)}\label{vof_mix_rule} \\
\rho &=& \sum_{l=1}^N \phi_{(l)} \rho_{(l)} \\
\rho \mathscr{E} &=& \sum_{l=1}^N \phi_{(l)} \rho_{(l)} \mathscr{E}_{(l)}\label{eq:energyMixtureRule}\\ 
G &=& \frac{\sum_{l=1}^N\left(  \phi_{(l)} G_{(l)}(\rho_{(l)},\alpha_{(l)})/\Gamma_{(l)}\right)}{\sum_{l=1}^N\left(\phi_{(l)}/\Gamma_{(l)}\right)}\\
c^2 &=& \frac{\sum_{l=1}^N \left( \phi_{(l)} Y_{(l)} c_{(l)}^2 / \Gamma_{(l)} \right)}{\sum_{l=1}^N\left(\phi_{(l)}/\Gamma_{(l)}\right)} \ ,
\end{eqnarray}
where $c$ is the sound speed and $Y_{(l)} = ({\phi_{(l)}\rho_{(l)}})/{\rho} $ is the mass fraction.

\subsection{Closure Models}
\label{sec:ClosureModels}
It can be seen that, by presenting the internal energy in the form outlined, equation \eqref{eq:p_energy_derivative} can be written in the form:
\begin{align}
 p_{(l)} = p_{\text{\scriptsize{ref}},(l)} + \rho_{(l)}\Gamma_{(l)}\left(\mathscr{E}_{(l)} - \mathscr{E}_{\text{\scriptsize{ref}},(l)}\right) \label{eq:MieGruneisenEOS} \ .
\end{align}
Here, $\mathscr{E}_{\text{\scriptsize{ref}},(l)} = \mathscr{E}^c_{(l)} + \mathscr{E}^s_{(l)}$ and $p_{\text{\scriptsize{ref}},(l)} = \rho^{2}_{(l)} \frac{\partial\mathscr{E}_{\text{\scriptsize{ref}},(l)}}{\partial \rho_{(l)}}$. This is the form of the standard Mie-Gr\"uneisen equation-of-state. This encompasses a wide range of different materials, not limited to solids, depending on the choice of the reference curves $\mathscr{E}_{\text{\scriptsize{ref}},(l)}(\rho_{(l)})$, $p_{\text{\scriptsize{ref}},(l)}(\rho_{(l)})$, and $\Gamma_{(l)}(\rho_{(l)})$. These additional freedoms incorporate thermal, compaction and shear effects. For example:
\begin{itemize}
 \item $\mathscr{E}_{\text{\scriptsize{ref}},(l)} = p_{\text{\scriptsize{ref}},(l)} = 0, \ \Gamma(\rho)=\Gamma_0=\gamma-1$, where $\gamma$ is the adiabatic index, gives the ideal gas law used for relatively simple gases such as air.
 \item $p_{\text{\scriptsize{ref}},(l)} = -\gamma p_{\infty}, \  \mathscr{E}_{\text{\scriptsize{ref}},(l)} = e_{\infty}, \ \Gamma(\rho)=\Gamma_0=\gamma-1$ gives the stiffened gas equation of state, used to model denser fluids such as water. 
 \item $\Gamma(\rho)=\Gamma_0=\gamma-1$ and 
 \begin{align}
   p_{\text{\scriptsize{ref}},(l)}(\rho_{(l)}) &= {\cal A}e^{-{\cal R}_1\frac{\rho_0}{\rho_{(l)}}}+{\cal B}e^{-{\cal R}_2\frac{\rho_0}{\rho_{(l)}}} \\
   \mathscr{E}_{\text{\scriptsize{ref}},(l)}(\rho_{(l)}) &= \frac{{\cal A}}{{\cal R}_1\rho_0}e^{-{\cal R}_1\frac{\rho_0}{\rho_{(l)}}}+\frac{{\cal B}}{{\cal R}_2\rho_0}e^{-{\cal R}_2\frac{\rho_0}{\rho_{(l)}}}
 \end{align}
 gives the JWL equation-of-state, widely used for condensed phase explosives or reaction products \cite{MiNi16}.
 \item $\Gamma(\rho)=\Gamma_0=\gamma-1$ and
 \begin{align}
   \mathscr{E}_{\text{\scriptsize{ref}},(l)}(\rho_{(l)}) &= \frac{K_0}{2\rho_{(l)}\alpha^2}\left(\left(\frac{\rho_{(l)}}{\rho_0}\right)^{\alpha}-1\right) + \mathscr{E}^s_{(l)} \\
   G(\rho_{(l)}) &= G_0\left(\frac{\rho_{(l)}}{\rho_0}\right)^{\beta+1} \\
   p_{\text{\scriptsize{ref}},(l)}(\rho_{(l)}) &= \rho^{2}_{(l)} \frac{\partial\mathscr{E}_{\text{\scriptsize{ref}},(l)}}{\partial \rho_{(l)}}
 \end{align}
 gives the Romenskii equation-of-state, used for elastic solids \cite{RomenskiiEOS}.
\end{itemize}

The form of the Mie-Gr\"uneisen equation of state now allows the total pressure to be obtained from the total energy. To achieve this, equation \ref{eq:energyMixtureRule} is combined with the isobaric assumption that forms part of the mechanical equilibrium of the underlying \citet{Allaire} model. Firstly the internal energy is obtained:
\begin{equation}
\rho \mathscr{E} = \rho E - \frac{1}{2}\rho|\mathbf{u}| \ .
\end{equation}
Then the Mie-Gr\"uneisen form of the equation of state is used to rewrite equation \ref{eq:energyMixtureRule}, using the isobaric assumption:
\begin{equation}
\rho \mathscr{E} = \sum_{l=1}^N \phi_{(l)} \rho_{(l)} \left(\mathscr{E}_{\text{\scriptsize{ref}},(l)} + \frac{(p - p_{\text{\scriptsize{ref}},(l)})}{\rho_{(l)}\Gamma_{(l)}}\right) \ .
\end{equation}
Finally, this equation can be rewritten in terms of $p$:
\begin{equation}
p = \frac{\rho \mathscr{E} - \sum_{l} \phi_{(l)}\left(\rho_{(l)}\mathscr{E}_{\text{\scriptsize{ref}},(l)} - \frac{p_{\text{\scriptsize{ref}},(l)}}{\Gamma_{(l)}}\right)}{\sum_{l}\frac{\phi_{(l)}}{\Gamma_{(l)}}} \ .
\end{equation}

\subsubsection{Plasticity}

The introduction of plasticity through the source term $\boldsymbol\Phi$ follows the method of convex potentials (see for example \citet{Ottosen2005}) and is therefore thermodynamically compatible.
In this approach, the von Mises yield criterion forms the scalar potential which leads to the plastic flow rule:
\begin{align}
 \boldsymbol\Phi &= \chi \sqrt{\frac{3}{2}}\frac{\dev{\boldsymbol\sigma}}{\Jnorm{\dev{\boldsymbol\sigma}}} \cdot \Vbar \ .
\end{align}
The plastic flow rate $\chi$ is a closure model and must be suitable for arbitrary mixtures. For a mixture of $N$ materials:
\begin{align}
\chi = \frac{\sum_{l=1}^N\left(  \phi_{(l)} \chi_{(l)}(\rho_{(l)})/\Gamma_{(l)}\right)}{\sum_{l=1}^N\left(\phi_{(l)}/\Gamma_{(l)}\right)} \ .
\end{align}
If any material does not obey a plasticity model, such as fluids, then that material contributes $\chi_{(l)} = 0$.

The form of $\chi$ relates the particular material flow model. This work considers both ideal plasticity, where $\chi$ is a Heaviside function such that the relaxation is non-zero only when the stress exceeds the yield surface $\sigma_Y$:
\begin{align}
 \chi_{(l)} = \chi_{(l)}^0 H\left[ \sqrt{\frac{3}{2}} ||\text{dev}({\boldsymbol\sigma})_{(l)}|| -\sigma_Y \right] \ ,
\end{align}
and the rate sensitive isotropic work-hardening model outlined by \citet{JohnsonCook}:
\begin{equation}
\chi_{(l)} = \chi_{(l)}^0 \exp\left[\frac{1}{c_3}\left(\frac{\sqrt{\frac{3}{2}}||\dev{{\boldsymbol\sigma}_{(l)}}||}{\sigma_Y\left(\varepsilon_{p,(l)}\right)}-1\right)\right], \label{JCPlasticity}
\end{equation}
where ${\chi}_{0}>0$ is the reference plastic strain-rate and the constant $c_3$ controls the rate dependency. In the Johnson and Cook model the yield stress is given by:
\begin{align}
 \sigma_Y\left(\varepsilon_{p,(l)}\right) = \left(c_1+c_2(\varepsilon_{p,(l)})^n\right)\left( 1 - \left(\frac{T-T_0}{T_{\text{melt}}-T_0}\right)^m   \right)\ ,
\end{align}
where $c_1$ is the yield stress, $c_2$ is the strain hardening factor, $n$ is the strain hardening exponent, $T_{\text{melt}}$ is the melting temperature of the material, $T_0 = 298$ K is a reference temperature, and $m$ is the thermal softening exponent. This yield surface at the reference strain rate $\chi_{(l)}^0$ is a function of the accumulated plastic strain, $\varepsilon_p$, making it necessary to add an additional evolution equation to the system to advect and evolve the plastic strain:
\begin{align}
\frac{\partial \rho_{(l)}\phi_{(l)} \varepsilon_{(l),p}}{\partial t} + \frac{\partial \rho_{(l)} \phi_{(l)} \varepsilon_{(l),p} u_k}{\partial x_k} = \rho_{(l)}\phi_{(l)}\chi_{(l)} \ .
\end{align}
Using the Johnson-Cook model in this way results in a viscoplastic flow rule, where plastic deformations can accumulate from the onset of loading. Note however that the parameter $c_3$ is usually small such that $\chi_{(l)}\ll\chi_{(l)}^0$ for stresses much below the characteristic stress.
Indeed, as $c_3 \rightarrow 0$ the plastic flow becomes rate independent and the stress becomes bounded by a yield surface.

\subsection{Damage}
Modelling damage is important if realistic material behaviour is to be captured for solid materials undergoing finite deformation. This work focuses on ductile damage, a significant mode of damage in most metals, where substantial plastic deformation occurs before the onset of fracture, as opposed to brittle fracture, where fracture directly proceeds from elastic loading. 

Physically, damage reduces the load carrying ability of a solid material until, at a critical level, the material loses all strength and fractures. Various branches of physics deal with damage on different scales, depending on the application and type of damage at hand. The increase of damage in a material can be explained on the micro-scale by the nucleation and coalescence of micro-voids in the structure of the material and is intimately linked to plastic effects. At a macro-scale, this presents itself as the formation of visible fractures within a material. 

A continuum damage mechanics (CDM) model is used to simulate ductile fracture. CDM is a thermodynamically consistent way of modelling the effect of damage at the continuum scale, where thermodynamic forces are derived using the second law in order to advect and evolve damage in a given material \cite{MurakamiCDM}. Damage is assumed to be isotropic, and can thus can be quantified and tracked by the addition of a single scalar field for each damageable material, $D_{(l)}$:
\begin{align}
\frac{\partial \rho_{(l)}\phi_{(l)} D_{(l)}}{\partial t} + \frac{\partial \rho_{(l)} \phi_{(l)} D_{(l)} u_k}{\partial x_k} = \rho_{(l)}\phi_{(l)}\dot{D}_{(l)} \ .
\end{align}

It is remarked that other formulations of damage can be derived, including fully anisotropic models such as \citet{BartonAnisotropicDamage} that quantify damage using a tensor quantity. However, the efficacy of the damage model itself is outside the scope of this paper. Rather, the intention is to provide a general framework that demonstrates the capability of the new numerical method, regardless of what specific constitutive material models are employed.

The contributions to the internal energy from cold compression and shear energy for damageable materials are affected by the damage parameter, and taken to be 
\begin{eqnarray}
\mathscr{E}_{(l)}^c\left(\rho_{(l)},D_{(l)}\right) &=& (1-D_{(l)})\mathscr{E}_{(l)}^{c,\text{u}}\left(\rho_{(l)}\right) \\
\mathscr{E}_{(l)}^s\left(\rho_{(l)},\dev{\mathbf{H}^e},D_{(l)}\right) &=& (1-D_{(l)})\mathscr{E}_{(l)}^{s,\text{u}}\left(\rho_{(l)},\dev{\mathbf{H}^e}\right) \ .
\end{eqnarray}
where $\mathscr{E}_{(l)}^{c,\text{u}}$ and $\mathscr{E}_{(l)}^{s,\text{u}}$ are the functions corresponding to the undamaged state, and can be any of the forms highlighted in section~\ref{sec:ClosureModels}.
From this proposition, the Cauchy stress tensor derived from the Clausius-Duhem inequality becomes:
\begin{eqnarray}
\boldsymbol\sigma_{(l)} &=& p_{(l)}\mathbf{I} + \dev{\boldsymbol\sigma_{(l)}} \\
p_{(l)} &=& (1-D_{(l)})\rho^{2}_{(l)} \frac{\partial\mathscr{E}_{(l)}}{\partial \rho_{(l)}}\\ 
\dev{\boldsymbol\sigma_{(l)}} &=& (1-D_{(l)})2G_{(l)}\cdot\dev{\mathbf{H}^e} \ .
\end{eqnarray}
The presence of damage causes a reduction in load carrying area, with $(1-D_{(l)})$ being the reduction factor, leading to the {\it effective-stress} principal first proposed by \citet{kachanov:1958}.
In this approach, damage $D_{(l)}$ is quantified by the fractional area of micro-voids on a plane intersecting a representative volume element.

In a similar way to plasticity, the Clausius-Dunhem inequality gives the form of the evolution equation for damage in terms of its conjugate thermodynamic force, and the method of convex potentials is then used to derive the evolution of the damage parameter:
\begin{align}
 \dot{D}_{(l)} = - \dot{\lambda}_{(l)}\pdv{F^D_{(l)}}{Y_{(l)}} \ ,
\end{align}
where $\lambda_{(l)}$ is the Lagrange multiplier, $Y_{(l)} = {\partial\mathscr{E}_{(l)}}/{\partial D_{(l)}}$ is the elastic energy release rate due to damage (the thermodynamic force conjugate to damage), and $F^D_{(l)}$ is the damage dissipation potential. The associative flow rule for damaged materials is 
\begin{equation}
\dot{\lambda}_{(l)} = \dot{\varepsilon}_{p,(l)}(1-D_{(l)}) \ .
\end{equation}
The damage dissipation potential is taken from \citet{Borona}, who propose the following potential to account for micro-mechanical processes:
\begin{align}
 F^D = \left[ \frac{1}{2}\left(-\frac{Y}{S}\right)\frac{S}{1-D_{(l)}}\right] \frac{(D_{\text{crit},(l)}-D_{(l)})^{\frac{\alpha-1}{\alpha}}}{\varepsilon_p^{\frac{2+n}{n}}} \ .
\end{align}
Under certain assumptions on the equation of state (detailed in the appendix) this gives:
\begin{align}
 \dot{D}_{(l)} = \alpha \frac{(D_{\text{crit},(l)}-D_{0,(l)})^{\frac{1}{\alpha}}}{\ln(\frac{\varepsilon_{\text{crit}}}{\varepsilon_{\text{thresh}}})}  R_t\left(\frac{p}{\sigma_{\text{eq}}}\right)(D_{\text{crit},(l)}-D_{(l)})^{\frac{\alpha-1}{\alpha}}\left(\frac{\dot{\varepsilon}_p}{\varepsilon_p}\right) \ ,
\end{align}
where $D_{\text{crit}}, D_{0}, \alpha, \varepsilon_{\text{crit}}, \varepsilon_{\text{thresh}}$ are material parameters and $R_t$ is the stress triaxiality function:
\begin{equation}
R_t\left(\frac{p}{\sigma_{\text{eq}}}\right) = \frac{2E_{(l)}(1-D_{(l)})^2 \mathscr{E}_{(l)}^{c,\text{u}}\rho_{(l)}}{\sigma^2_{\text{eq}}} + \frac{E_{(l)}}{3G_{(l)}^{\text{u}}} 
\end{equation}
where $G_{(l)}^{\text{u}}, E_{(l)}^{\text{u}}$ are respectively the shear and Young's modulus in the undamaged state. A full derivation of the damage model is provided in the appendix.

\citet{PirondiAndBorona} extended this model to deal with cyclic loading, proposing the following addition:
\begin{align}
 \varepsilon_p^- &= \left\lbrace\mqty{ \varepsilon_p && \text{if: } \ p < 0 \\ 0 && \text{else} \\}\right. \\
 \dot{D} &= \left\lbrace\mqty{ \dot{D} && \text{if: } \ p < 0 \ \text{and} \ \varepsilon_p^- > \varepsilon_{p,\text{thresh}} \\ 0 && \text{else} \\}\right.
\end{align}
which conveys the underlying assumption that damage can only accrue in states of tension, not under compression. Finally, threshold plastic strain $\varepsilon_{p,\text{thresh}}$ is taken to depend on the stress triaxiality using the model from \citet{StressTriaxialityDependence}:
\begin{align}
 \varepsilon_{p,\text{thresh}} = \frac{\varepsilon_{\text{thresh}}^{1/R_t}}{R_t^{1/n}}\left( \frac{\varepsilon_{\text{crit}}^{2n}-\varepsilon_{\text{thresh}}^{2n}}{\varepsilon_{\text{crit}}^{2n/R_t}-\varepsilon_{\text{thresh}}^{2n/R_t}}     \right)^{\frac{1}{2n}} \ ,
\end{align}
where $n$ is the strain hardening exponent.

\section{Numerical Approach}

The model is solved on a Cartesian mesh with local resolution adaptation in space and time. This is achieved using the AMReX software from Lawrence Berkeley National Laboratory \cite{amrex}, which includes an implementation of the structured adaptive mesh refinement (SAMR) method of \citet{berger:1988} for solving hyperbolic systems of partial differential equations of the form of equation \eqref{eq:sys_vec_form}.
In this approach, cells of identical resolution are grouped into logically rectangular sub-grids or `patches'. 
Refined grids are derived recursively from coarser ones, based upon a flagging criterion, to form a hierarchy of successively embedded levels.
All mesh widths on level $l$ are $r_l$-times finer than on level $l-1$, i.e. $\Delta t_l:=\Delta t_{l-1}/r_l$ and $\Delta \mathbf{x}_{l}:=\Delta \mathbf{x}_{l-1}/r_l$ with $r_l\in\mathbb{N}, r_l\ge 2$ for $l>0$ and $r_0=1$. 
The numerical scheme is applied on level $l$ by calling a single-grid update routine in a loop over all patches constituting the level. 
The discretisation of the constitutive models does not differ between patches or levels, so for clarity the method shall be described for a single sub-grid.
Cell centres are denoted by the indices $i,j,k\in\mathbb{Z}$ and each cell $C^{l}_{ijk}$ has the dimensions $\Delta \mathbf{x}^l_{ijk}$.

It is convenient when describing the numerical method to express the system of equations in compact vector form by separating it into various qualitatively different parts: a conservative hyperbolic part for each spatial dimension, non-conservative terms from the volume fraction and stretch tensor updates, a source term due to plastic flow and associated damage accumulation, and a source term to account for geometrical effects. This can be written as:
\begin{equation}
 \frac{\partial \mathbf{U}}{\partial t}+ \frac{\partial \mathbf{F}_k}{\partial x_k} =  \mathbf{s}_{\text{\scriptsize{non-con.}}} + \mathbf{s}_p + \mathbf{s}_g \ . \label{eq:sys_vec_form} 
\end{equation}
Subject to the closure relations previously outlined, this is given by:
\begin{eqnarray}
 \pdv{t}\mqty(	\phi_{(l)} \\
		\phi_{(l)}\rho_{(l)} \\
		\phi_{(l)}\rho_{(l)}D_{(l)} \\
		\phi_{(l)}\rho_{(l)}\varepsilon_{p,(l)} \\
		\rho u_i \\
		\rho E \\
		\Vbar_{ij} \\)  
		+ \pdv{x_k}\mqty(\phi_{(l)}u_k \\
		\phi_{(l)}\rho_{(l)}u_k \\
		\phi_{(l)}\rho_{(l)}D_{(l)}u_k \\
		\phi_{(l)}\rho_{(l)}\varepsilon_{p,(l)}u_k \\
		\rho u_iu_k -\sigma_{ik} \\
		\rho Eu_k - u_i\sigma_{ik} \\
		\Vbar_{ij}u_k - \Vbar_{kj}u_i \\)
		= \mqty(\phi_{(l)}\pdv{u_k}{x_k} \\
		0 \\
		0 \\
		0 \\
		0 \\
		0 \\
		\frac{2}{3}\Vbar_{ij}\pdv{u_k}{x_k} - u_i\pdv{\Vbar_{kj}}{x_k} \\)
		+ \mqty(0 \\
			0 \\
			\phi_{(l)}\rho_{(l)}\dot{D}_{(l)} \\
			\phi_{(l)}\rho_{(l)}\dot{\varepsilon}_{p,(l)} \\
			0 \\
			0 \\
			\boldsymbol\Phi_{ij} \\) + \mathbf{s}_g \ .
\end{eqnarray}
When considering cylindrical symmetry, the conservative variables and geometrical term are given by:
\begin{align}
 \mathbf{U} =\mqty(	\phi_{(l)} \\
		\phi_{(l)}\rho_{(l)} \\
		\phi_{(l)}\rho_{(l)}D_{(l)} \\
		\phi_{(l)}\rho_{(l)}\varepsilon_{p,(l)} \\
		\rho u_r \\
		\rho u_z \\
		\rho E \\
		\Vbar_{ij} \\),		
\ \  \vb{s}_g =-\frac{1}{r} \mqty( 0 \\
		\phi_{(l)}\rho_{(l)} u_r \\
		\phi_{(l)}\rho_{(l)}D_{(l)} u_r \\
		\phi_{(l)}\rho_{(l)}\varepsilon_{p,(l)} u_r \\
		\rho u_r^2 - \sigma_{rr} + \sigma_{\theta\theta} \\
		\rho u_zu_r -\sigma_{rz}\\
		\rho Eu_r-(u_r\sigma_{rr} + u_z\sigma_{zr}) \\
		\frac{1}{3} \Vbar_{ij} u_r - \delta_{i\theta}\Vbar_{ij} u_r ) \ .
\end{align}

In what follows, the underpinning method of \citet{Barton2019} is summarised first before detailing the new modifications that facilitate the resolution of interface slide, separation, and void.

\subsection{Underpinning Method}

Following \citet{Barton2019}, the inhomogeneous system is integrated for time intervals $[t^n,t^{n+1}]$, where the time-step $\Delta t=t^{n+1}-t^n$ is chosen to be a fraction of the global maximum allowable time step required for stability of the hyperbolic update method \cite{CFL}.
For high strain-rate applications, the plastic relaxation can occur over smaller time scales, which can lead to local stiffness.
To address this issue, without resorting to forecasting stiff zones and resolving the time scales of irreversible physical processes, Godunov's method of fractional steps is used, where the hyperbolic part is updated first, followed by serially adding in contributions from each source term:
\begin{align}
\frac{\partial \mathbf{U}}{\partial t}  &= -\frac{\partial \mathbf{F_k}}{\partial x_k} + \mathbf{s}_{\text{\scriptsize{non-con.}}} &\text{IC:} \quad \mathbf{U}^n  \xRightarrow{\Delta t} {\mathbf{U}}^{\hstar}\label{exp_update}\\
\frac{\partial \mathbf{U}}{\partial t}  &= \mathbf{s}_x &\text{IC:} \quad \mathbf{U}^{\hstar} \xRightarrow{\Delta t} {\mathbf{U}}^\bigstar\label{imp_udate_p} \ ,
\end{align}
where the result of the each step is used as the initial condition (IC) for the next. Here $\vb{s}_x=\vb{s}_p, \vb{s}_g$. Once all source terms have been added, the last $\mathbf{U}^{\bigstar}$ becomes $\mathbf{U}^{n+1}$.

\subsubsection{Hyperbolic Update}

Replacing the spatial derivatives with a discretised conservative approximation, the hyperbolic system can be written
\begin{equation}\label{sys_mat_dis}
 \frac{\text{d}}{\text{d}t}{\mathbf{U}}_{ijk}+\mathcal{D}_{ijk}\left({\mathbf{U}}\right) = 0,
\end{equation}
where ${\mathbf{U}}_{ijk}$ represents the vector of conservative variables stored at cell centres,  and 
\begin{eqnarray}
 \mathcal{D}_{ijk} :=&& \frac{1}{\Delta x^x_{ijk}}\left(\widetilde{\mathbf{F}}^x_{i+1/2,jk}-\widetilde{\mathbf{F}}^x_{i-1/2,jk}\right)\nonumber\\
&+&\frac{1}{\Delta x^y_{ijk}}\left(\widetilde{\mathbf{F}}^y_{i,j+1/2,k}-\widetilde{\mathbf{F}}^y_{i,j-1/2,k}\right)\nonumber\\
&+&\frac{1}{\Delta x^z_{ijk}}\left(\widetilde{\mathbf{F}}^z_{ij,k+1/2}-\widetilde{\mathbf{F}}^z_{ij,k-1/2}\right) - \mathbf{s}_{\text{\scriptsize{non-con.}},ijk},\label{eq:spat_op}
\end{eqnarray}
where $\widetilde{\mathbf{F}}^d_{m\pm1/2}$, for $m=i,j,k$, are the cell wall numerical flux functions in the direction $d = x,y,z$. 
The numerical fluxes are computed through successive sweeps of each spatial dimension and summed according to equation \eqref{eq:spat_op}. 

Fluxes are computed using the HLLD solver from \citet{Barton2019} (but with the additional flux-modifiers that shall be set out next). 
To achieve higher order spatial accuracy, the initial conditions for the Riemann solver are taken to be the MUSCL reconstruction of the cell centred primitive variables. It should be noted that this procedure is generally carried out on the conservative variables; the method instead follows \citet{JohnsenColonius}, who showed that, for multi-material problems, reconstruction of the primitive variables leads to fewer oscillations around interfaces. An artificial interface reconstruction is applied to reduce numerical diffusion around interfaces. This is achieved using the Tangent of Hyperbola INterface Capturing (THINC) method: an algebraic interface reconstruction technique that fits a hyperbolic tangent function to variables inside a cell. 
In contrast to \citet{Barton2019}, who used the original THINC method of \citet{XiaoTHINC}, the more recent MUSCL-BVD-THINC scheme of \citet{BVDTHINC} is employed. This more recent scheme provides an additional check to minimise oscillations by comparing the reconstructed state's cell boundary variation with the previously calculated MUSCL reconstruction. THINC-reconstructed states are only accepted when their total boundary variation is lower than that of the MUSCL scheme alone.

The non-conservative source terms that result from the volume fraction and deformation tensor updates are added in the hyperbolic step following a mid-point rule based procedure similar to that used by several authors \cite{MiNi16, Allaire, Barton2019}.

To achieve a higher temporal resolution in the update of the hyperbolic terms, the third order Runge-Kutta time integration scheme is used:
\begin{eqnarray}
\mathbf{U}^{(1)}    &=& \mathbf{U}^{n} - \Delta t  \mathcal{D}\left({\mathbf{U}}^{n}\right)\\
\mathbf{U}^{(2)}    &=& \mathbf{U}^{(1)} - \Delta t  \mathcal{D}\left({\mathbf{U}}^{(1)}\right)\\
\mathbf{U}^{(3)} &=& \frac{3}{4} \mathbf{U}^{n} + \frac{1}{4} \mathbf{U}^{(2)} \\
\mathbf{U}^{(4)}    &=& \mathbf{U}^{(3)} - \Delta t  \mathcal{D}\left({\mathbf{U}}^{(3)}\right)\\
\mathbf{U}^{\hstar} &=& \frac{1}{3} \mathbf{U}^{n} + \frac{2}{3} \mathbf{U}^{(4)} \ .
\end{eqnarray}
It is remarked that the second order Runge-Kutta method could also be used, but in practice it is found that the third-order method offers additional robustness in some cases.

\subsubsection{Plastic and Damage Update}
\label{sec:numericalPlastic}
The plastic update is performed after the hyperbolic step as detailed above and consists of solving the equations: 
\begin{align}
\dot{\varepsilon_p} &= \chi \label{eq:plasticStrainUpdate} \\
\dot{\Vbar} &= - \chi \sqrt{\frac{3}{2}}\frac{\dev{\sigma}}{\Jnorm{\dev{\sigma}}} \Vbar \ . \label{eq:vtensorplasticupdate}
\end{align}

The potentially stiff ODEs governing the plasticity evolution are solved using an analytical technique as detailed by \citet{Barton2019}, which reduces the problem to a single ODE for each material in a given cell. The algorithm can be summarised as follows: 
\begin{equation}\label{eq:eq_plast_qnew}
\left(\Vbar\right)^{\bigstar} = \exp\left(\dev{\mathbf{H}^e}^{\bigstar}\right)
\end{equation}
\begin{equation}\label{eq:eq_plast_hnew}
\dev{\mathbf{H}^e}^{\bigstar} = \frac{\mathcal{J}^{\bigstar}}{\mathcal{J}^\hstar} \dev{\mathbf{H}^e\left({\Vbar}^\hstar\right)}
\end{equation}
\begin{equation}\label{eq:eq_plast_jnew}
\mathcal{J}^{\bigstar} = \frac{\sum_{l=1}^N \left(\phi_{(l)}/\Gamma_{(l)}\right)\mathcal{J}^{\bigstar}_{(l)}}{\sum_{l=1}^N \phi_{(l)}/\Gamma_{(l)}}
\end{equation}
\begin{equation}\label{eq:eq_plast_ojb}
\mathscr{R}\left(\mathcal{J}_{(l)}^{\bigstar}\right) = \mathcal{J}_{(l)}^{\bigstar} - \mathcal{J}_{(l)}^\hstar + \Delta t \sqrt{\frac{3}{2}}\chi_{(l)}\left(\mathcal{J}_{(l)}^{\bigstar},\varepsilon_{p_{(l)}}^{\bigstar}\left(\mathcal{J}_{(l)}^{\bigstar}\right)\right) = 0 \ .
\end{equation}
The last objective equation is solved using a simple bisection algorithm between the limits $\mathcal{J}_{(l)}^{\bigstar}\in[0:\mathcal{J}_{(l)}^\hstar]$.

This algorithm can be evaluated, at least in part, everywhere in a problem irrespective of whether a material adheres to a plasticity law or not. This is because the stretch tensor may lose its symmetry and unimodular properties as a result of numerical error, and the algorithm encompasses the enforcement of  
\begin{equation}\label{eq:eq_symm_rein}
\overline{\mathbf{V}}^e \leftarrow \sqrt{\overline{\mathbf{V}}^e\overline{\mathbf{V}}^{e^\text{T}}} \ ,
\end{equation}
to ensure that the stretch tensor remains symmetric, and the condition $\Vbar=\mathbf{I}$ is true for fluids. However, for tests that do not feature materials that obey a plasticity law, performing this algorithm is not necessarily required, as not enforcing the symmetry condition was not found to significantly affect the results.

The implementation of this algorithm by \citet{Barton2019} employs singular value decomposition (SVD) so that equation \eqref{eq:eq_plast_qnew} and equation \eqref{eq:eq_plast_hnew} are evaluated exactly. 
Since SVD is relatively expensive, this part of the overall numerical scheme constitutes a primary overhead. To alleviate this computational burden, the following modifications are made. The Hencky strain tensor in equation \eqref{eq:eq_plast_hnew} is evaluated using a variant of the approximation by \citet{Bazant1998}:
\begin{equation}
\dev{\mathbf{H}^{e}} \approx \frac{1}{2}\mathbf{H}_B^{e^{(1)}}\left(\Vbar{\Vbar}^{\text{T}}\right) \ ,
\end{equation}
where
\begin{equation}
\mathbf{H}_B^{e^{(m)}}\left(\Vbar\right) = \frac{1}{2m}\left({\Vbar}^m-{\Vbar}^{-m}\right) \ ,
\end{equation}
and the invariants are evaluated from this in turn. The exponential matrix in equation \eqref{eq:eq_plast_qnew} is evaluated using the first diagonal Pad\'{e} approximant:
\begin{equation}
(\Vbar)^{n+1} \approx  \left( 1- \frac{1}{2}\dev{\mathbf{H}^{e}}^{n+1}\right)^{-1}\left(1 + \frac{1}{2}\dev{\mathbf{H}^{e}}^{n+1}\right) \ .
\end{equation}
This approximation of the strain was proposed in \citet{Barton2019} but was used only where stresses are computed for the numerical flux functions. 
Use of the first Pad\'{e} approximate is found to be sufficient compared to second or higher variants since the second norm of $\mathbf{H}^e$ is known from the outset not to exceed sufficiently small values for the materials of interest. These approximations avoid the use of a SVD, and the algorithm reduces to simple matrix operations. The use of these approximations is further discussed in \ref{app:HenckyApproximations}.

Alongside this algorithm, the plastic strain and damage parameters are updated:
\begin{align}
 \dot{\varepsilon}_{p,(l)} &= \chi = \sqrt{\frac{2}{3}}\left(\mathcal{J}_{(l)}^\hstar - \mathcal{J}_{(l)}^{\bigstar}\right)/\Delta t \\
 \dot{D}_{(l)} &= \dot{D}_{(l)}(\dot{\varepsilon}_{p,(l)}) \ .
\end{align}
%At this point, if the damage parameter $D_{(l)}$ for a given material is found to exceed the critical damage threshold, $D_{\text{crit}}$, then the damage holder volume fraction is updated as detailed in section \ref{sec:fracture}.

\subsection{Flux-modifiers}

Since the underpinning model does not account for certain interface boundary conditions, the fluxes in the hyperbolic update are modified in selected cells in such a way that the desired boundary conditions are captured. This is the crux of the new method, and the primary distinction compared to sharp interface Ghost Fluid methods. In Ghost Fluid methods, the same numerical flux is used across the entire domain, meaning ghost-regions of the domain must be filled with states that provide suitable material boundary interaction. By contrast, the new method uses a modified flux at the material interface, encoding the effect of the material boundary condition into the flux method itself, thereby not requiring a ghost-region to hold any additional states. The following sections will therefore explain how appropriate material boundary conditions are included in the modified fluxes.

\begin{figure}
\centering
\includegraphics[width = 0.8\textwidth]{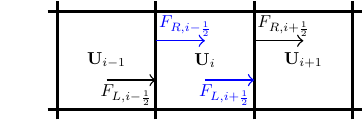}
\caption{The two-flux approach. The fluxes that the cell in the middle `sees' for the purposes of the conservative update formula are highlighted in blue.}
\label{fig:twoFluxApproach}
\end{figure}

In some cases when a flux-modifier is present, two different fluxes are required across a cell interface. This two-flux approach has an approximately equivalent effect to introducing a Ghost Fluid on either side, where each side's flux is based on the cell on one side and its own Ghost Fluid cell on the other side. 
The presence of two fluxes modifies the conservative update formula: what was
\begin{equation}
 \frac{\text{d}}{\text{d}t}{\mathbf{U}}_{ijk}+\frac{1}{\Delta x^x_{ijk}}\left(\widetilde{\mathbf{F}}^x_{i+1/2,jk}-\widetilde{\mathbf{F}}^x_{i-1/2,jk}\right) = 0 \ ,
\end{equation}
becomes
\begin{equation}
 \frac{\text{d}}{\text{d}t}{\mathbf{U}}_{ijk}+\frac{1}{\Delta x^x_{ijk}}\left(\widetilde{\mathbf{F}}^x_{L,i+1/2,jk}-\widetilde{\mathbf{F}}^x_{R,i-1/2,jk}\right) = 0 \ .
\end{equation}
The method is illustrated in Figure \ref{fig:twoFluxApproach}. 

The method sacrifices some degree of strict conservation at interfaces at the benefit of allowing shear and void opening. However, it is emphasised that the fluxes are modified only in a compact region around material interfaces -- when the flux-modifiers are not present the two fluxes at each cell interface are the same, returning the method to the standard conservative update formula.
Moreover, in the case of shear, this flux imbalance applies predominantly to the tangential vector quantities, retaining a greater degree of conservation for scalar quantities such as mass. In the case of void opening, the full effect of this imbalance is not felt as the flux into a void region is of comparatively less significance as it will be over-written by the void seeding routine. A two-flux approach such as this is not without precedent; the volume fraction parts of the original \citet{Allaire} model and the stretch tensor parts of \citet{Barton2019} employ non-conservative update formulae. The extent of these conservation loss effects will be quantified in Section \ref{sec:Validation}. 

The fluxes in the following sections are found by considering the Riemann problem between two states. Various different flux methods are outlined, so the notation $\mathbf{F}^{\bullet}(\mathbf{U}_1,\mathbf{U}_2)$ will denote the flux between states $\mathbf{U}_1$ and $\mathbf{U}_2$ using the $\bullet$ flux method, where $\bullet$ can be substituted for $n$ (normal/standard), $\nu$ (void) or $s$ (shear).
The higher order methods and interface reconstruction of the underpinning method can be included in the usual way in the definition of the initial states $\mathbf{U}_L$, $\mathbf{U}_R$ between which the flux is to be calculated. 

\subsubsection{Shear}

Physically, slip boundary conditions are produced at material interfaces where the lack of adhesion between materials causes them to be able to slide along each other in directions tangential to the interface normal. This is represented mathematically as the tangential components of the interface stress tensor being zero. Vector components normal to the interface and scalar quantities are not affected by this condition, but tangential vector components such as the transverse momentum/velocity are.
A similar approach has been used \citet{FavrieElasticDiffuse} and \citet{FavriePlasticDiffuse} for allowing slip between solid-gas interfaces, but this was not extended to the case of solid-solid interaction. Another successful approach has been led by \citet{VitaliXEM}, called the XEM method.

The underpinning algorithm proceeds until it is required to find the flux across a material interface. Thus the two states between which the flux is to be found, $\mathbf{U}_L,\mathbf{U}_R$ have already been obtained. 

Material interfaces are defined as where the volume fraction for any material `changes sign' between the two cells being considered. The sign of the volume fraction is defined as:
\begin{eqnarray}
 \text{sign}(0.5 - \phi) \ .
\end{eqnarray}
The material for which this is the case is denoted the \textit{mixture material}. When three or more materials are present, there is a chance that several material interfaces can coincide between the two cells being considered. In this case, this method approximates the solution by selecting only one mixture material, taken to be whichever of the volume fractions that change sign has the largest difference between the two cells. This approximation was not found to affect the results, but could be further examined in future.

In the method, a distinction is drawn between the `normal' and `tangential' state variables. The normal state variables consist of all scalar variables (such as density) and the parts of tensorial quantities that are aligned with the interface, and the tangential parts are the remainder. The normal parts of a state $\mathbf{U}$ are denoted $\mathbf{U}^{\parallel}$ and the tangential parts are denoted $\mathbf{U}^{\perp}$. For vector quantities such as momentum and velocity, these different parts are relatively easy to define:
\begin{align}
 \mathbf{u}^{\parallel} &= (\mathbf{u} \cdot \mathbf{\hat{n}})\mathbf{\hat{n}} \\
 \mathbf{u}^{\perp}     &= \mathbf{u} - \mathbf{u}^{\parallel} \ ,
\end{align}
where $\mathbf{\hat{n}}$ is the outward pointing unit normal vector to the mixture interface. For the stretch tensor, however, the process is slightly more involved. It proves easier to rotate the tensor into the frame where the interface normal lies along the $\mathbf{\hat{x}}$ direction (or any other coordinate basis aligned frame). In what follows, states that have been rotated into this reference frame will be denoted with a tilde, $\mathbf{U} \rightarrow \tilde{\mathbf{U}}$. In this frame, the components of the $x-$row and $x-$column, except for the component on the diagonal, are the tangential parts, and the rest are the normal parts:
\begin{align}
 \mathbf{\tilde{V}} &= \mathbf{R}\mathbf{\Vbar}\mathbf{R}^{\text{T}} \\
 \tilde{\mathbf{V}}^{\parallel} &= \mqty(\tilde{V}_{xx} && 0 && 0 \\ 0 && \tilde{V}_{yy} && \tilde{V}_{yz} \\ 0 && \tilde{V}_{zy} && \tilde{V}_{zz}) \\
 \tilde{\mathbf{V}}^{\perp} &= \mqty(0 && \tilde{V}_{xy} && \tilde{V}_{xz} \\ \tilde{V}_{yx} && 0 && 0 \\ \tilde{V}_{zx} && 0 && 0) \ .
\end{align}
The normal and tangential parts in the original reference frame are then found by reversing the rotation:
\begin{align}
 \mathbf{\Vbar}^{\parallel} &= \mathbf{R}^{\text{T}}\tilde{\mathbf{V}}^{\parallel}\mathbf{R} \\
 \mathbf{\Vbar}^{\perp}     &= \mathbf{R}^{\text{T}}\tilde{\mathbf{V}}^{\perp}\mathbf{R} \ .
\end{align}

Here $\mathbf{R}$ is the rotation matrix which maps the normal vector $\mathbf{\hat{n}}$ to the $x$-axis $\mathbf{\hat{x}}$. In two dimensions, $\mathbf{R}$ is a simple rotation around the $z$-axis, given in terms of the components of the normal vector by:
\begin{align}
 \mathbf{R} &= \mqty( n_x && n_y && 0 \\ -n_y && n_x && 0 \\ 0 && 0 && 1 ) \ .
\end{align}
In three dimensions, the definition of $\mathbf{R}$ is more involved and not unique. The problem comes down to finding a rotation matrix which will map a vector $\mathbf{a}$ ($\mathbf{\hat{n}}$) to another vector $\mathbf{b}$ ($\mathbf{\hat{x}}$). Two different methods were trialled. The first relies on noting that $\mathbf{a}$ can be rotated onto $\mathbf{b}$ by rotating by $\pi$ around the axis $\hat{\mathbf{c}} = \frac{\mathbf{a}+\mathbf{b}}{|\mathbf{a}+\mathbf{b}|}$. Using Rodrigues' rotation formula, it is then found that:
\begin{align}
 \mathbf{R} &= 2\left(\hat{\mathbf{c}}\otimes\hat{\mathbf{c}}\right) - \mathbf{I} \ .
\end{align}
This method has the advantage of being very simple to define, but has several disadvantages. Firstly, it will on average involve some unnecessary twisting, as in general when $\mathbf{a}=\mathbf{b}$, $\mathbf{R} \neq \mathbf{I}$. Secondly, this method fails when $\mathbf{a} = -\mathbf{b}$, where $|\mathbf{a}+\mathbf{b}| = 0$, so error checking must be implemented around this pole.
Another viable method rotates around the axis defined by the vector normal to the plane defined by $\mathbf{a}$ and $\mathbf{b}$. This vector, $\hat{\mathbf{d}}$, is orthogonal to both $\mathbf{a}$ and $\mathbf{b}$ and can be found with $\hat{\mathbf{d}} = \frac{\mathbf{a}\times\mathbf{b}}{|\mathbf{a}\times\mathbf{b}|}$. It then follows again from Rodrigues' rotation formula that:
\begin{align}
 \mathbf{R} &= \mathbf{I} + \frac{|\mathbf{a}\times\mathbf{b}|}{|\mathbf{a}||\mathbf{b}|}[\hat{\mathbf{d}}]_{\times}+\left(1-\frac{\mathbf{a}\cdot\mathbf{b}}{|\mathbf{a}||\mathbf{b}|}\right)[\hat{\mathbf{d}}]_{\times}^2 \ , \\
 \nonumber \text{where: } \\
 [\hat{\mathbf{d}}]_{\times} &= \mqty(0 && -d_3 && d_2 \\ d_3 && 0 && -d_1 \\ -d_2 && d_1 && 0 \\ ) \ .
\end{align}
This version also features a pole where error checking must be implemented at $\mathbf{a} = \mathbf{b}$, but involves much less unnecessary twisting.

%\subsubsection{Shear Flux Modifier} %%%%%%%%%%%%%%%%%%%%%%%%%%%%%%%%%%%%%%%%%
%%%%%%%%%%%%%%%%%%%%%%%%%%%%%%%%%%%%%%%%%%%%%%%%%%%%%%%%%%%%%%%%%%%%%%%%%%%%%%%
%%%%%%%%%%%%%%%%%%%%%%%%%%%%%%%%%%%%%%%%%%%%%%%%%%%%%%%%%%%%%%%%%%%%%%%%%%%%%%%

The flux-modifier then proceeds as follows: \\[0.5cm]
\framebox[\textwidth]{\begin{minipage}{0.9\textwidth}{
The shear flux-modifier: 
\begin{enumerate}
 \item First, the outward pointing interface normal, $\mathbf{\hat{n}}$, is estimated using Young's method \cite{YoungsNormal} applied to the mixture material volume fraction, and averaged between the two cells.
 \item $\mathbf{U}_L$ and $\mathbf{U}_R$ are rotated into the reference frame where the normal $\mathbf{\hat{n}}$ points along $\mathbf{\hat{x}}$.
 \item Two new states are constructed: $\tilde{\mathbf{U}}_L^s,\tilde{\mathbf{U}}_R^s$. These contain the normal parts of one side, combined with the tangential parts of the other side:
  \begin{eqnarray*}
  \tilde{\mathbf{U}}_L^s & = & \tilde{\mathbf{U}}_L^{\perp} + \tilde{\mathbf{U}}_R^{\parallel} \\
  \tilde{\mathbf{U}}_R^s & = & \tilde{\mathbf{U}}_R^{\perp} + \tilde{\mathbf{U}}_L^{\parallel} \ .
  \end{eqnarray*}
  %\item For the Rieman-problem-based method, these contain the star-states of the $\tilde{\mathbf{U}}_L$,$\tilde{\mathbf{U}}_R$ Riemann problem, found with the HLLD solver using a slip boundary condition between the cells:
  %\begin{eqnarray*}
  %\tilde{\mathbf{U}}_L^s & = & \tilde{\mathbf{U}}_L^{**} \\
  %\tilde{\mathbf{U}}_R^s & = & \tilde{\mathbf{U}}_R^{**} \ .
  %\end{eqnarray*}
 %\end{itemize}
 \item The appropriate stress tensor is then calculated in these new states, where the tangential stress components are zero. This can be done in two ways, see Section \ref{sec:StressFreeVsStrainFree}.
 \item All the aforementioned cells are then rotated back into the original frame of reference:
 \begin{eqnarray*}
\mathbf{U} = \mathbf{R}^{\text{T}}\tilde{\mathbf{U}}\mathbf{R} \ .
\end{eqnarray*}
 \item The standard flux is then calculated between each pair of states: 
\begin{eqnarray*}
\mathbf{F}^{s}_{L} &=& \mathbf{F}^{n}(\mathbf{U}_L,\mathbf{U}_L^s), \\  
\mathbf{F}^{s}_{R} &=& \mathbf{F}^{n}(\mathbf{U}_R^s,\mathbf{U}_R).
\end{eqnarray*}
\end{enumerate}
}\end{minipage}}\\[0.5cm]
This procedure predominantly affects the tangential components of any state. However, in two and three dimensions, the interface will not generally be coordinate basis aligned, so the procedure should modify components that are normal to a given hyperbolic sweep direction, as they will have some component pointing along the interface normal.

%\subsubsection{Shear Seeding Routine} %%%%%%%%%%%%%%%%%%%%%%%%%%%%%%%%%%%%%%%%
%%%%%%%%%%%%%%%%%%%%%%%%%%%%%%%%%%%%%%%%%%%%%%%%%%%%%%%%%%%%%%%%%%%%%%%%%%%%%%%
%%%%%%%%%%%%%%%%%%%%%%%%%%%%%%%%%%%%%%%%%%%%%%%%%%%%%%%%%%%%%%%%%%%%%%%%%%%%%%%

As it stands, the shear flux-modifier alone would not be sufficient. If only this flux-modifier were applied, the tangential components would not move with the interface normal velocity. Before every time step, a seeding routine is first performed. This routine propagates the correct tangential velocity components into mixed cell regions. Mixed cells are defined as those either next to a material interface or having a volume fraction for any material that is in the range $0.5 < \phi < 1.0 - \delta$ where delta is some small amount, taken in this work to be $1\times 10^{-4}$. The volume fraction for which this inequality applies is then the \textit{mixture material}. The following procedure is then employed:\\[0.5cm]
\framebox[\textwidth]{\begin{minipage}{0.9\textwidth}{
The shear seeding routine:
\begin{enumerate}
 \item The outward pointing interface normal, $\mathbf{\hat{n}}$, is calculated using Young's method \cite{YoungsNormal}.
 \item A probe is sent out along the normal direction using the method described in Section \ref{sec:probeCalculation}, and the material values are interpolated at that point to give a new state $\mathbf{U}_{\text{Interp}}$.
 \item  If the interpolated mixture material volume fraction is higher than in the original cell, both the original and the interpolated cell are rotated so the normal lies in the $\mathbf{\hat{x}}$ direction, giving $\tilde{\mathbf{U}}$ and $\tilde{\mathbf{U}}_{\text{Interp}}$.
 \item A new state is then created containing the tangential components of the interpolated cell and the normal components of the original cell, in the same way as the flux-modifier:
 \begin{equation*}
  \tilde{\mathbf{U}}^s  = \tilde{\mathbf{U}}^{\parallel} + \tilde{\mathbf{U}}_{\text{Interp}}^{\perp} \ .
 \end{equation*}
 \item This new state is then rotated back into the original frame of reference and replaces the original state.
 \begin{equation*}
  \mathbf{U} \leftarrow \mathbf{U}^s \ .
 \end{equation*}
\end{enumerate}
}\end{minipage}}\\[0.5cm]
The procedure is found to adequately seed the mixture regions with the suitable tangential components. If correctly applied, this procedure should only affect mixture regions, and with the application of interface reconstruction, these regions are generally small and the process remains localised.

\subsubsection{Strain-free and Stress-free Tangential Conditions}
\label{sec:StressFreeVsStrainFree}
To calculate the correct interface stress tensor in the shear flux-modifier states $\tilde{\mathbf{U}}_L^s,\tilde{\mathbf{U}}_R^s$ there are two options. The first is to leave the stretch tensor unchanged and simply set the tangential components of the stress tensor to zero. This corresponds to the stress-free condition. In this intuitive approximation, the strain-free nature of the interface will emerge naturally as the simulation progresses. The second option is to calculate the strain-free state that will result in a stress-free interface. This approach is somewhat more involved, but more physically grounded. Here, a new stretch tensor is calculated that results in the tangential stress being zero when it is calculated in the ordinary way. 
This strain-free state is calculated in a similar way to how the stretch tensor is reconstructed during the plastic update. The deviatoric Hencky strain is obtained, its tangential components are set to zero, and then it is reconstructed into the stretch tensor, following the procedure laid out in Section \ref{sec:numericalPlastic}.
In the course of this work, it was found that both techniques produce satisfactory results, but the second, more physical method is preferred.

\subsubsection{Void}

To enable interaction with voids such as those produced when two solid materials separate, an additional scalar variable, the void volume fraction, $\nu$, is introduced. The void volume fraction is advected in exactly the same way as material volume fractions in the standard flux, by solving
\begin{align}
\frac{\partial \nu }{\partial t} + \frac{\partial \nu u_k}{\partial x_k} = \nu \frac{\partial u_k}{\partial x_k}  \ .
\end{align}
The void flux is then modified in two scenarios:
\begin{enumerate}
 \item Either cell $\mathbf{U}_L$, $\mathbf{U}_R$ already contains some void -- the `pre-existing void' case.
 \item The cells are at a material interface or are critically damaged and are coming apart. The `void generation' case.
\end{enumerate}

In the case where the void already exists and doesn't need to be generated, the stress tensor in each cell is weighted by the void volume fraction:
\begin{equation} 
\boldsymbol\sigma \leftarrow (1-\nu)\boldsymbol\sigma
\end{equation}
and the standard flux can be used. This in itself is insufficient to model realistic void interaction and a void seeding routine is employed in a similar way to the shear flux method. This will be outlined below.

In addition, this work presents two options for the void generation flux-modifier. The first is a `simple' method, akin to the original Ghost Fluid method of \citet{FedkiwGFM}, where quantities in the ghost cells are simply copied across from the real material. The second is a Riemann-solver-based method akin to the later Ghost Fluid Methods of \citet{RiemannGFM} and \citet{rGFM}. Both methods are found to work satisfactorily, with the simple method being cheaper and generally more robust, but the Riemann-problem-based method retaining greater conservation for strong shocks.

Other methods with which this work can be compared and contrasted include \citet{BarlowVoidOpening} and  \citet{BarlowVoidOpening2}

%\subsubsection{Void Flux Modifier} %%%%%%%%%%%%%%%%%%%%%%%%%%%%%%%%%%%%%%%%%
%%%%%%%%%%%%%%%%%%%%%%%%%%%%%%%%%%%%%%%%%%%%%%%%%%%%%%%%%%%%%%%%%%%%%%%%%%%%%%%
%%%%%%%%%%%%%%%%%%%%%%%%%%%%%%%%%%%%%%%%%%%%%%%%%%%%%%%%%%%%%%%%%%%%%%%%%%%%%%%

The void generation flux-modifier is applied in all interfacial cells that are coming apart. Material interfaces are flagged in the same way as the shear flux-modifier, namely when a volume fraction `changes sign' between the two cells. Damage interfaces are defined as where either cell between which the flux is to be calculated is critically damaged (this fracture criterion is explained in detail in the next section). Interfaces are deemed to be moving apart when they are under tension:
\begin{align}
\mathbf{\hat{n}}^{\text{T}} \cdot \left(\boldsymbol\sigma_R +\boldsymbol\sigma_L\right) \cdot \mathbf{\hat{n}} > 0 \ .
\end{align}
In this case, the following procedure is carried out:\\[0.5cm]
\framebox[\textwidth]{\begin{minipage}{0.9\textwidth}{
The void flux-modifier:
\begin{enumerate}
 \item The interface normal $\mathbf{\hat{n}}$ is calculated using Young's method \cite{YoungsNormal}.
 \item Two new states are constructed: $\mathbf{U}_L^{\nu},\mathbf{U}_R^{\nu}$. 
  \begin{itemize}
  \item For the `simple' method, these states are directly copied from their counterpart:
  \begin{eqnarray*}
  \mathbf{U}_L^{\nu} & = & \mathbf{U}_R \\
  \mathbf{U}_R^{\nu} & = & \mathbf{U}_L \ .
  \end{eqnarray*}
  \item For the Riemann-problem-based method, these contain the star-states of the interaction $\mathbf{U}_L$ and $\mathbf{U}_R$ with void, found with the HLLD solver using a void boundary condition:
  \begin{eqnarray*}
  \mathbf{U}_L^{\nu} & = & \mathbf{U}_L^{**} \\
  \mathbf{U}_R^{\nu} & = & \mathbf{U}_R^{**} \ .
  \end{eqnarray*}
 \end{itemize}
 \item The void volume fraction in these new states is then set to 1, and the appropriate weighted stress tensor is then calculated.
 \item The standard flux is then calculated between each pair of states: 
\begin{eqnarray*}
\mathbf{F}^{\nu}_{L} &=& \mathbf{F}^{n}(\mathbf{U}_L,\mathbf{U}_L^{\nu}),\\
\mathbf{F}^{\nu}_{R} &=& \mathbf{F}^{n}(\mathbf{U}_R^{\nu},\mathbf{U}_R).
\end{eqnarray*}
\end{enumerate}
}\end{minipage}}\\[0.5cm]

%\subsubsection{Void Seeding Routine} %%%%%%%%%%%%%%%%%%%%%%%%%%%%%%%%%%%%%%%%%
%%%%%%%%%%%%%%%%%%%%%%%%%%%%%%%%%%%%%%%%%%%%%%%%%%%%%%%%%%%%%%%%%%%%%%%%%%%%%%%
%%%%%%%%%%%%%%%%%%%%%%%%%%%%%%%%%%%%%%%%%%%%%%%%%%%%%%%%%%%%%%%%%%%%%%%%%%%%%%%

To enable the correct interaction with void states, the thermodynamic variables in areas with high void volume fraction present are seeded with values that mimic the interaction of the flow with void. This is somewhat akin to the Ghost Fluid method's filling of the ghost cells with the material boundary condition. The following procedure is used: \\[0.5cm]
\framebox[\textwidth]{\begin{minipage}{0.9\textwidth}{
The void seeding routine:
\begin{enumerate}
 \item In cells with a void volume fraction above a specified threshold (generally this work takes $\nu = 0.9$, but a range of values is found to work), the void normal $\mathbf{\hat{n}}_{\nu}$ is calculated using Young's method \cite{YoungsNormal}.
 \item A probe is sent out along the normal direction using the method described in Section \ref{sec:probeCalculation} and the material values are interpolated at that point to give a new state $\mathbf{U}_{\text{Interp}}$.
 \item If the interpolated void volume fraction is lower than the void volume fraction of the original cell, a new state $\mathbf{U}^{\nu}$ is calculated:
 \begin{itemize}
  \item For the `simple' method, this new state is directly copied from the interpolated cell: 
  \begin{equation*}
  \mathbf{U}^{\nu} = \mathbf{U}_{\text{Interp}}.
  \end{equation*}
  \item For the Riemann-problem-based method, this new state is the star-state of the interaction of $\mathbf{U}_{\text{Interp}}$ with void. This is found as before by rotating $\mathbf{U}_{\text{Interp}}$ into the direction where the normal $\mathbf{\hat{n}}_{\nu}$ points along $\mathbf{\hat{x}}$, and then using the HLLD solver with a void boundary condition. This star-state then becomes the new state:
  \begin{equation*}
  \tilde{\mathbf{U}}^{\nu} = \tilde{\mathbf{U}}^{**} \ .
  \end{equation*}
  This state $\tilde{\mathbf{U}}^{\nu}$ is then rotated back to the original reference frame.
 \end{itemize}
 \item The new state is then linearly combined with the old state, but keeping history variables such as damage, plastic strain and volume fractions constant:
  \begin{equation*}
  \mathbf{U} \leftarrow (1-\nu)\mathbf{U} + \nu \mathbf{U}^{\nu} \ .
  \end{equation*}
\end{enumerate}
}\end{minipage}}\\[0.5cm]
Again this seeding routine only affects void mixture regions, and with interface reconstruction this procedure is kept localised.

%\begin{figure}
%\centering
%\includegraphics[width = 0.8\textwidth]{VoidWaveFan.pdf}
%\caption{The material-void Riemann problem.}
%\label{fig:voidInteraction}
%\end{figure}

\subsection{Probe Calculation}
\label{sec:probeCalculation}
\begin{figure}
\centering
\includegraphics[width = 0.5\textwidth]{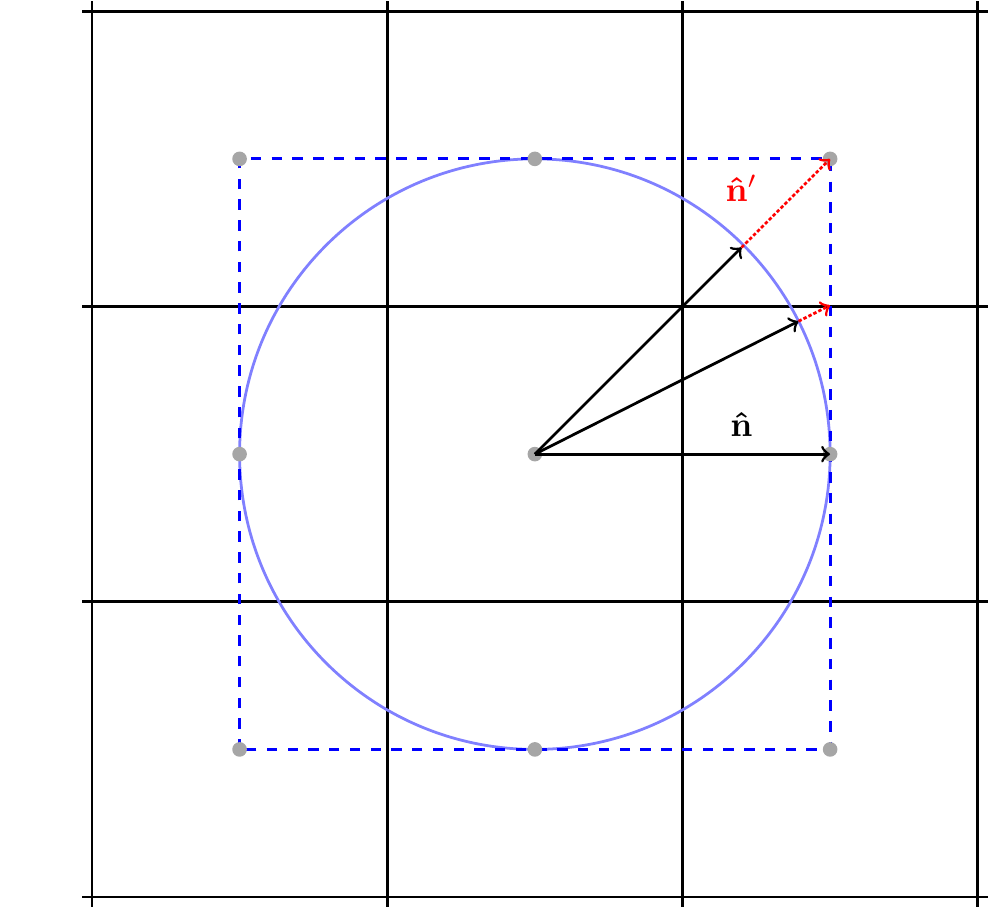}
\caption{Mapping the unit circle to a square with side length 2. This stretch ensures that the probe can reach exterior cell centres in diagonal directions.}
\label{fig:circleToSquare}
\end{figure}

The seeding routines require probes to be sent out in different directions in order to interpolate material values at different points. The calculation of these probes is explained in this section. In order to start, the current cell centre position $\mathbf{c}$, the interface normal (probe direction) $\mathbf{\hat{n}}$, and vector of cell dimensions $\mathbf{\dd x}$, are assumed to be known. In two and three dimensions, if the normal does not point along a coordinate basis direction, the distance a probe has to travel to reach another cell is greater, as the probe travels along the diagonal of the cell. Therefore, in order to ensure that probe directions along the diagonal can still reach suitable material values outside the current cell, the probe vector is scaled depending on the normal direction. This is performed in a dimension-independent way by stretching the unit sphere defined by all normal directions to fill a cube with side length 2 (in two dimensions this corresponds to stretching the unit circle to fill a square with side length 2). This stretching is denoted $\mathbf{\hat{n}}\rightarrow\mathbf{\hat{n}}'$. Figure \ref{fig:circleToSquare} demonstrates this mapping in two dimensions. 

To perform the mapping, firstly the component of $\mathbf{\hat{n}}$ with the largest absolute value (either positive or negative) is denoted $n_{\text{max}}$. The components of $\mathbf{\hat{n}}'$ are then given by:
\begin{align}
 n_i' = \frac{n_i}{|n_{\text{max}}|} \ .
\end{align}

Once this stretch has been performed, the final probe position is given by:
\begin{equation}
\mathbf{p} = \mathbf{c} + l \cdot\left( \mathbf{\hat{n}}' \odot \mathbf{\dd x}\right) \ ,
\end{equation}
where $\odot$ is element-wise multiplication. The factor $l$ modifies the probe length, and values between 1 and 1.5 were found to work well for the problems presented.

\subsection{Fracture}
\label{sec:fracture}
One advantage of the new method is that fracture is handled naturally by the same routines that mediate interface separation. Critically damaged cells, defined as any cell where:
\begin{equation}
D_{(l)} \ge D_{\text{crit},(l)} \ ,
\end{equation}
are simply treated as any other material interface where the void generation flux-modifier can be applied. No additional algorithm needs to be applied to initiate fracture. In this case, the interface normal can be taken as the current sweep direction.
This approach also removes the need for the highly non-conservative approach of deleting critically damaged cells to form fracture, as is commonplace in finite element and level set based codes \cite{ParticleLevelSetDamage}, \cite{UdaykumarDamage}, \cite{BartonLevelSetDamage}. This also avoids the issue of potentially having to redistribute the conserved quantities such a mass and momentum to the neighbouring cells around the deleted cells. 

However, this method does come with its own associated challenges; the diffusive nature of all Eulerian codes means that damaged cells may undergo artificial healing. As an initially critically damaged material is advected, diffusion can `heal' the damage by causing it to fall under the critical damage threshold. This causes a number of issues:
\begin{itemize}
 \item Regions that should be able to come apart and form a crack are prevented from separating.
 \item The transition to critically damaged and back is discontinuous; quantities such as the shear modulus and yield strength go suddenly to zero.
 \item When a history-dependent plasticity model such as strain-hardening is employed, the plastic strain remains unchanged through healing. This means a material can be damaged, heal, and then continue to accrue plastic strain, leading to unphysical increased hardening. 
 \item It prevents the use of a stochastic initial damage profile to mimic the effect of defects in the structure of real materials, as high-frequency modes in the initial distribution are quickly smeared out. 
\end{itemize}

Previous works, such as \citet{VitaliTransportDiffusion} and \citet{BartonLevelSetDamage}, have employed semi-Lagrangian updates to avoid these issues. This technique could also be applied to the current method, but is beyond the scope of the current study. In this work, THINC reconstruction is applied to the damage field and this is found to be sufficient to allow the large strain rate ductile fracture in problems at hand. Additionally, this work does not apply a stochastic initial condition to the damage profile to mimic material defects and ageing. Rather, the aim is to demonstrate the validity of the underlying fracture mechanics, while acknowledging that the method can be extended with techniques similar to \citet{VitaliTransportDiffusion} and \citet{BartonLevelSetDamage} if so desired.  

%This work proposes a simpler method to avoid some of these issues. An additional scalar history variable, $\delta$, called the \emph{Damage Holder}, is advected with the flow for each damageable material:
%\begin{align}
%\frac{\partial \delta }{\partial t} + \frac{\partial \delta u_k}{\partial x_k} = \delta \frac{\partial u_k}{\partial x_k}  \ .
%\end{align}
%The damage holder is set to zero initially. When a given material, $(l)$, becomes critically damaged in a cell, the damage holder field is reinitialised locally as:
%\begin{align}
% \delta = \text{max}\left(\delta,\phi_{(l)}\right) \ .
%\end{align}
%When the damage is needed for a thermodynamic operation, the value in a given cell is calculated as:
%\begin{align}
% D_{\text{actual},(l)} = D_{(l)}(1-\delta)+\delta D_{\text{crit},(l)} \ .
%\end{align}
%This avoids the issues laid out above. Regions that have been damaged are now tracked; previously, one could not tell if a cell with sub-critical damage was ever damaged or not. The diffusion has been, in a sense, `transferred' to a volume fraction field rather than the damage parameter. As THINC sharpening can now be applied to the damage holder field, the diffusion is greatly reduced. 
%This method avoids the first three problems laid out above, but as the damage holder field is only introduced for critically damaged cells it does not enable the use of stochastic initial conditions.

\section{Validation and Verification}
\label{sec:Validation}
This section provides both numerical and experimental comparison for the new method. In two- and three-dimensional tests, materials are delineated from void by the $\nu = 0.5$ volume fraction contour, with void areas shown in white for clarity.

\begin{table}
\begin{center}
    \begin{tabular}[t]{|l|l|l|l|l|l|l|}
      \hline
      Material & $\rho_0$ \ kgm$^{-3}$ & $K_0$ \ GPa & $G_0$ \ GPa & $\alpha$ & $\beta$ & $\Gamma_0$ \\
      \hline
      Aluminium     & 2703.0 & 76.3  & 26.36& 0.627& 2.288& 1.484 \\
      Al 5083 H32   & 2670.0 & 72.2  & 25.8 & 0.63 & 2.29 & 1.48  \\
      Copper        & 8930.0 & 136.45& 39.38& 1.0  & 3.0  & 2.0   \\
      CuBe          & 8370.0 & 131.3 & 53.6 & 1.0  & 3.0  & 2.0   \\
      Steel         & 8030.0 & 156.2 & 77.2 & 0.569& 2.437& 1.563 \\
      Polycarbonate & 1193.0 & 5.52  & 1.18 & 1.49 & ~*   & 0.61  \\
      AerMet        & 7940.0 & 156.2 & 77.2 & 0.569& 2.437& 1.84  \\
      \hline
    \end{tabular}
\end{center}
\caption{Romenskii equation of state material parameters. *A constant shear modulus variant was used for this material.}
\label{tab:MaterialParameters}
\end{table}

\subsection{Solid-Solid Riemann Problem}\label{sec:section_ivp1}

Firstly, the shear algorithms are verified in one dimension. This test is a Riemann problem taken from \citet{Barton2019} involving the collision of a stressed piece of aluminium with a stressed piece of copper. The test is challenging, representing a highly unrealistic initial stress configuration. Moreover, the problem features 7 different waves: 2 longitudinal, 4 transverse and one contact. The problem has an exact solution with which this work can be compared to find an estimate for the convergence; further details on the exact solution are provided in \ref{app:ExactSolution}. The initial conditions are given by:
\begin{align}
&\mbox{Left:  } \ \mathbf{u} = \mqty( 2000 \\ 0 \\ 100) \mbox{ ms}^{-1}, \quad F = \mqty(1 & 0 & 0 \\ -0.01 & 0.95 & 0.02 \\ -0.015 & 0 & 0.9 )\\
&\mbox{Right:  } \ \mathbf{u} = \mqty( 0 \\ -30 \\ -10) \mbox{ ms}^{-1}, \quad F = \mqty(1 & 0 & 0 \\ 0.015 & 0.95 & 0 \\ -0.01 & 0 & 0.9 )\ ,
\end{align}
where $F$ is the deformation tensor. Note that the stretch tensor, $\Vbar$, can be found from $F$ using a singular value decomposition. A full, exact singular value decomposition can be used for this task as this is only required once at the start of a simulation.
The test is run at a resolution of 500 cells on a domain $x=[0,0.01]$ m with an interface at $x=0.005$ m, for a time of 0.5 $\mu$s, using a CFL of 0.9.
For this initial test it is assumed that the materials behave purely elastically and both materials are governed by the Romenskii equation of state used by Barton \cite{Barton2019} (see section \ref{sec:ClosureModels}), using the material parameters laid out in Table \ref{tab:MaterialParameters}.

\begin{figure}
\centering
\includegraphics[width = 0.95\textwidth]{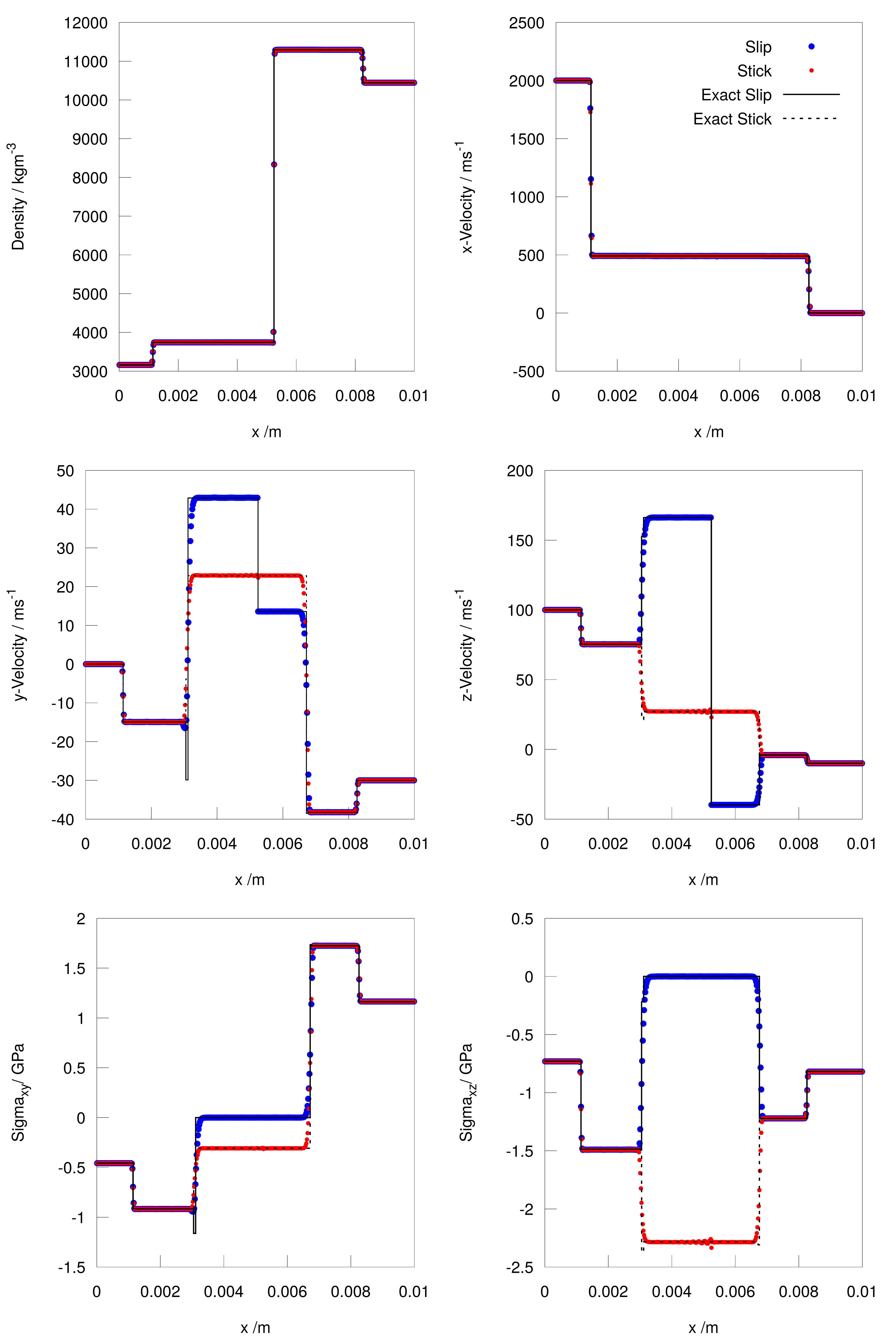}
\caption{The solid-solid Riemann problem. This test is used to provide one-dimensional shear verification, comparing the case with the shear flux-modifier and the case without to their exact solutions. Excellent agreement is observed in both cases for this strenuous test.}
\label{fig:Barton1}
\end{figure}

The results are shown in Figure \ref{fig:Barton1}, where comparison is drawn between both slip and stick boundary conditions. It can be seen that the method performs well in this tough test, with a clean tangential velocity discontinuity, smooth zero tangential stress, and unaffected longitudinal waves. The slip boundary condition also removes the small THINC-induced oscillations in the $xz$-stress noticed by \citet{Barton2019}. 

Table \ref{tab:SolidConvergence} shows the convergence for both stick and slip cases for the density and velocities. In both cases, the density and $x$-velocity show a roughly first order convergence, as expected for such a shock-dominated problem. This also verifies that the flux-modifier does not affect the convergence in the normal direction. The convergence for the two transverse velocities are slightly lower, but importantly it is better in the slip case where the flux-modifier is applied, even despite the additional contact wave that must be resolved. This suggests that flux-modifiers do not degrade the convergence of the underlying solution.

\begin{table}
  \begin{center}
    \begin{tabular}{c|c|ll|ll|ll|ll}%
    \hline
    \multirow{2}{*}{Method}  & \multirow{2}{*}{Resolution} & \multicolumn{2}{c|}{Density}   & \multicolumn{2}{c|}{$x$-velocity} & \multicolumn{2}{c|}{$y$-velocity} & \multicolumn{2}{c}{$z$-velocity}\\
    ~                        & ~                           & $L_1$ Error  & Order           & $L_1$ Error    & Order            & $L_1$ Error & Order              & $L_1$ Error & Order    \\\hline
    \multirow{7}{*}{Stick}    & ~                           & ~            & ~               & ~              & ~                & ~           & ~                  & ~           &          \\
    & 100 &  0.05653 & ~        &  0.01548 &  ~       &  0.00143 &  ~       &  0.00116 &  ~       \\ 
    & 200 &  0.02428 &  1.21917 &  0.00793 &  0.96479 &  0.00086 &  0.74022 &  0.00084 &  0.46520 \\ 
    & 400 &  0.01286 &  0.91657 &  0.00427 &  0.89176 &  0.00050 &  0.78438 &  0.00054 &  0.64268 \\ 
    & 800 &  0.00710 &  0.85778 &  0.00198 &  1.11011 &  0.00031 &  0.69452 &  0.00033 &  0.70210 \\ 
    &1600 &  0.00343 &  1.05106 &  0.00120 &  0.71821 &  0.00020 &  0.60264 &  0.00022 &  0.61283 \\ 
    &3200 &  0.00205 &  0.74423 &  0.00058 &  1.06062 &  0.00014 &  0.54894 &  0.00014 &  0.66939 \\  
    \multirow{7}{*}{Slip}    & ~                           & ~            & ~               & ~              & ~                & ~           & ~                  & ~           &          \\
    &  100 &  0.05656 &  ~       &  0.01551 &  ~       &  0.00240 &  ~       &  0.00304 &  ~        \\ 
    &  200 &  0.02427 &  1.22096 &  0.00787 &  0.97837 &  0.00135 &  0.82735 &  0.00163 &  0.90084 \\ 
    &  400 &  0.01282 &  0.92021 &  0.00418 &  0.91192 &  0.00078 &  0.78909 &  0.00086 &  0.92802 \\ 
    &  800 &  0.00722 &  0.82829 &  0.00204 &  1.03288 &  0.00051 &  0.61001 &  0.00047 &  0.86087 \\ 
    & 1600 &  0.00339 &  1.09041 &  0.00111 &  0.87972 &  0.00031 &  0.71911 &  0.00027 &  0.77833 \\ 
    & 3200 &  0.00205 &  0.72749 &  0.00057 &  0.95303 &  0.00019 &  0.74642 &  0.00017 &  0.66865 \\   \hline
    \end{tabular}
  \end{center}
\caption{Convergence for the solid-solid Riemann problem. The proposed flux-modifier does not degrade the convergence of the underlying method.}
\label{tab:SolidConvergence}
\end{table}

\begin{figure}
\centering
\includegraphics[width = 0.85\textwidth]{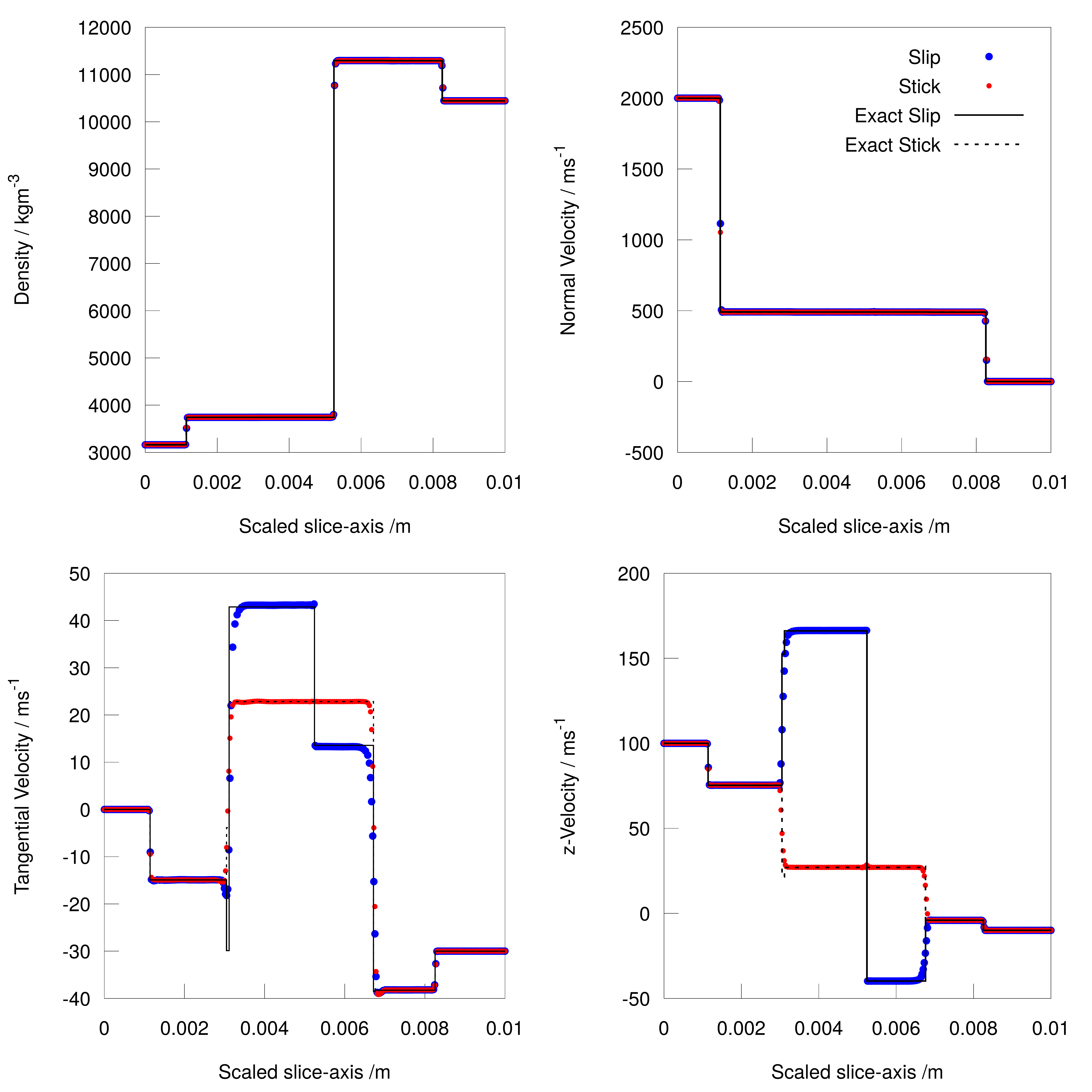}
\caption{The rotated solid-solid Riemann problem. This test is run in 2D using an interface normal of $(x,y) = (1,1)/\sqrt{2}$ to test the shear flux-modifier in the non-grid-aligned case. A slice is then taken through the domain to compare to the 1D case and the exact solution (Figure \ref{fig:Barton1}), and the results match well.}
\label{fig:Barton1_rot} 
\end{figure}

Additionally, this test can be run in two dimensions with the interface normal rotated by and angle of $45^{\circ}$ to verify that the algorithm can equally cope with non-grid-aligned interfaces. The initial conditions for the test are the same, except the velocity and deformation tensors are rotated to account for the rotated interface. The test is then run using a CFL of 0.6 and resolution of $250\times250$ cells, with one layer of AMR of refinement factor 2. Figure \ref{fig:Barton1_rot} shows this rotated test, where a cut has been taken through the domain along the line $(x,y) = (1,1)$ and the spatial axis has been scaled by a factor of $\sqrt{2}$ to make the results comparable with the exact solution. 

\subsection{Solid-Vacuum Riemann Problem}\label{sec:section_ivp2}

The interaction of solids with vacuum in one dimension is now examined. This is done using a test from \citet{Barton2019} that features a high-speed stressed block of aluminium in contact with vacuum. The initial conditions are given by:
\begin{align}
&\mbox{Left:  } \ \mathbf{u} = \mqty( 2000 \\ 0 \\ 100) \mbox{ ms}^{-1}, \quad F = \mqty(1 & 0 & 0 \\ -0.01 & 0.95 & 0.02 \\  -0.015 & 0 & 0.9 \\ ), \quad \nu = 0\\
&\mbox{Right:  } \ \mathbf{u} = \mqty( 0 \\ 0 \\ 0) \mbox{ ms}^{-1}, \quad F = \mqty(1 & 0 & 0 \\ 0 & 1 & 0 \\ 0 & 0 & 1 ) , \quad \nu = 1\ .
\end{align}
The problem features a rarefaction wave, 2 transverse shocks in the solid, and a material-vacuum contact wave. The test is run for 0.6 $\mu$s on a domain $x = [0:10]$ mm using a CFL 0.9 with the initial discontinuity at $x = 5$ mm.

\begin{figure}
\centering
\includegraphics[width = 0.80\textwidth]{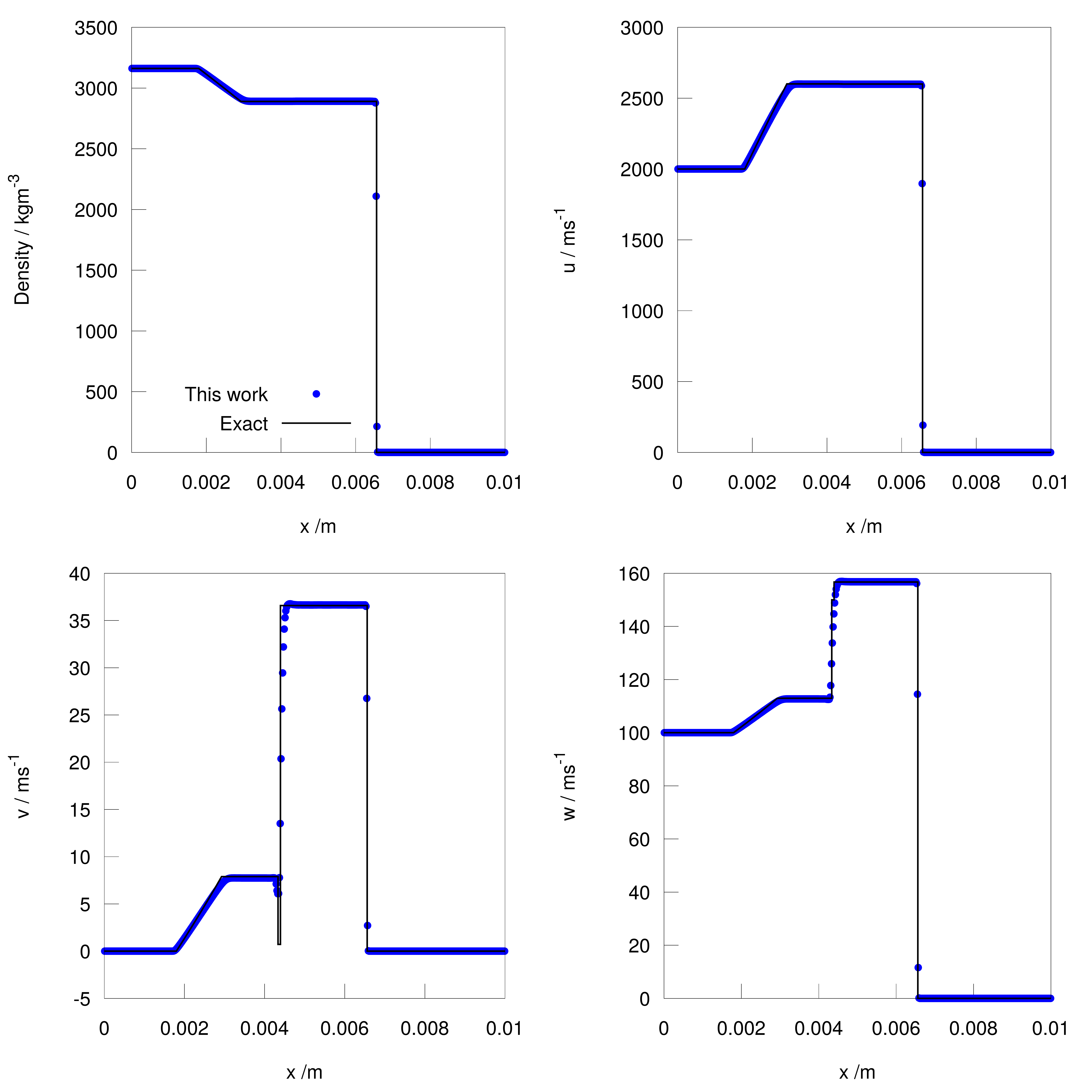}
\caption{The solid-vacuum Riemann problem. This problem tests the interaction of a stressed solid with an area of pre-existing void. The work at hand is compared to the exact solution, and matches very well, keeping a sharp material-void interface while resolving the multiple waves within the material.}
\label{fig:Barton3}
\end{figure}

\begin{figure}
\centering
\includegraphics[width =0.95\textwidth]{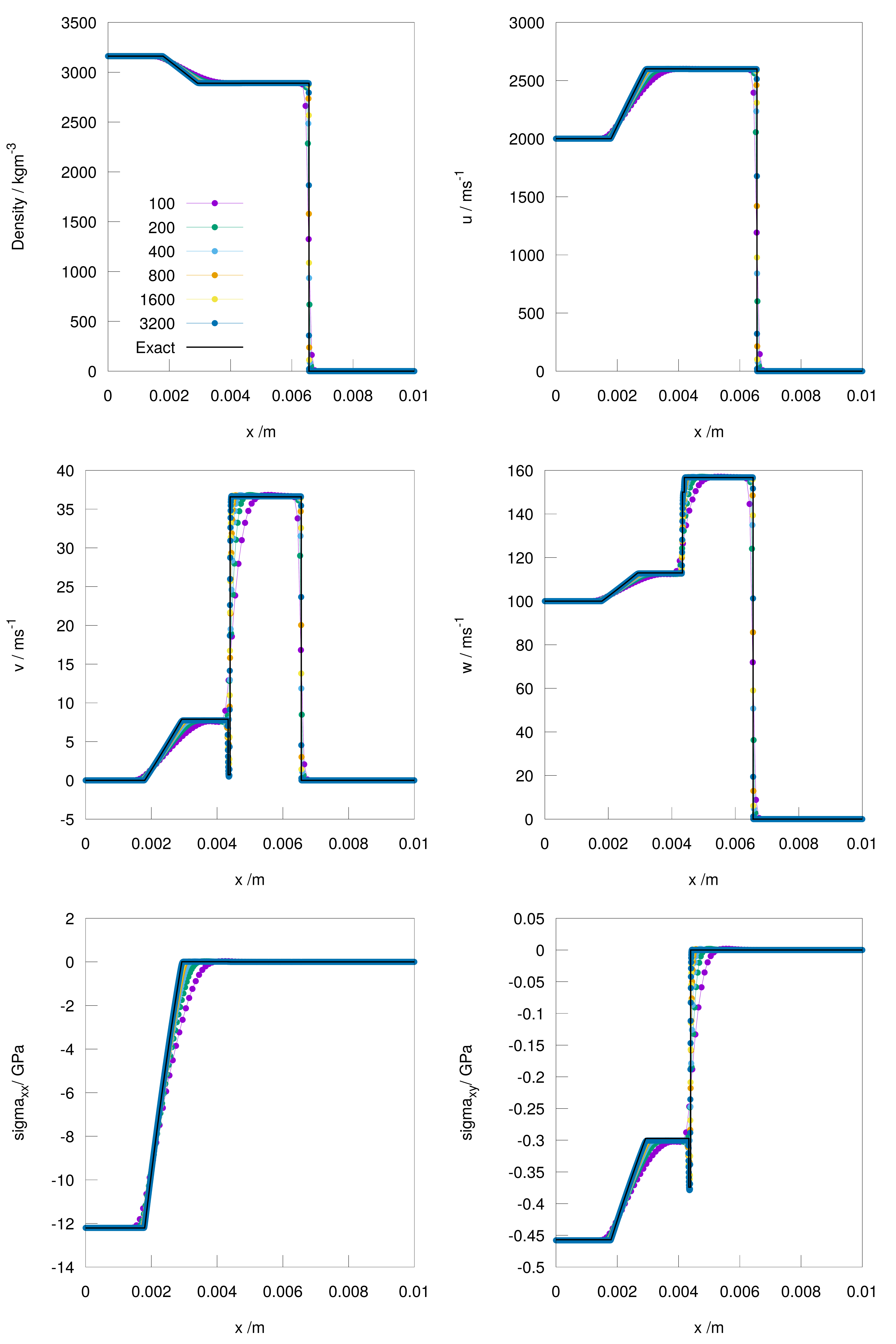}
\caption{The solid-vacuum Riemann problem shown for various different resolutions. For practical resolutions, the method matches the exact solution extremely well.}
\label{fig:Barton3Convergence}
\end{figure}

The results are shown in Figure \ref{fig:Barton3}. Again, the results are compared with the exact solution. The method handles the strong velocity discontinuities across the material interface well, for both normal and tangential components. This test also matches the results in \citet{Barton2019} well, where the void was instead replaced with air. Due to the large material differences between air and aluminium it is expected that the wave structure inside the solid is very similar in both the air- and vacuum-case, which is observed in the results of this test. This is an example of where this method can be particularly useful - the solution inside the air has little effect on the wave structure in the solid, and replacing the air with void saves computational resources.
The effect of resolution is also examined, as shown in Figure \ref{fig:Barton3Convergence} and Table \ref{tab:Barton3Convergence}. Figure \ref{fig:Barton3Convergence} compares multiple resolutions to the exact solution, derived using the method laid out in \ref{app:ExactSolution}. For practical resolutions, it can be seen that the method matches the exact solution very well. Additionally, Table \ref{tab:Barton3Convergence} shows the void algorithms also maintain the convergence to the exact solution, even in this strenuous test.

\begin{table}
  \begin{center}
    \begin{tabular}{c|ll|ll|ll|ll}%
    \hline
    \multirow{2}{*}{Resolution} & \multicolumn{2}{c|}{Density}   & \multicolumn{2}{c|}{$\sigma_{xx}$}      & \multicolumn{2}{c|}{$x$-velocity} & \multicolumn{2}{c}{$y$-velocity} \\
    ~                           & $L_1$ Error  & Order           & $L_1$ Error  & Order              & $L_1$ Error    & Order            & $L_1$ Error & Order               \\\hline
    ~                           & ~            & ~               & ~            & ~                  & ~              & ~                & ~           & ~                   \\
     100 &  0.02609 &  ~       &  0.24831 & ~        &  0.03114 &  ~       &  0.00112 &  ~       \\ 
     200 &  0.01024 &  1.34911 &  0.12300 &  1.01344 &  0.01312 &  1.24658 &  0.00053 &  1.06342 \\ 
     400 &  0.00554 &  0.88558 &  0.05958 &  1.04568 &  0.00698 &  0.91040 &  0.00032 &  0.74531 \\ 
     800 &  0.00329 &  0.75150 &  0.02677 &  1.15449 &  0.00396 &  0.81883 &  0.00019 &  0.72190 \\ 
    1600 &  0.00221 &  0.57811 &  0.01210 &  1.14490 &  0.00253 &  0.64459 &  0.00013 &  0.60614 \\ 
    3200 &  0.00110 &  0.99971 &  0.00593 &  1.02980 &  0.00134 &  0.91885 &  0.00009 &  0.48300 \\ 
\hline
    \end{tabular}
  \end{center}
\caption{Convergence for the solid-vacuum Riemann problem.}
\label{tab:Barton3Convergence}
\end{table}

\subsection{Gas-Vacuum Riemann Problem}

The void components are validated in one dimension using an ideal gas vacuum Riemann problem, with a $\gamma = 1.4$ ideal gas on the left and vacuum on the right. This is a notoriously strenuous test of the material-void interaction which many approximate Riemann solvers struggle to capture, especially in regards to the interface velocity and internal energy of the gas. The test features only two waves: a strong rarefaction back into the material and the material-void contact discontinuity. This test is difficult because the density of the gas at the material interface is incredibly small, while its velocity is large. For an ideal gas, the test has an exact solution with which this method can be compared to find an estimate for the convergence rate. The test is non-dimensional and is run until a time of $t = 0.1$ over a domain of $x = [0:1.5]$, with varying resolution using a CFL of 0.9. The initial conditions for the problem are:
\begin{align}
&\mbox{Left:  } \   (x<x_0)  \quad \rho_L = 1, \quad p_L = 2.5, \quad \mathbf{u}_L = \mathbf{0}, \quad \nu_L = 0\\
&\mbox{Right:  }\   (x>x_0)  \quad \rho_R = 0, \quad p_R = 0, \quad \mathbf{u}_R = \mathbf{0}, \quad \nu_R = 1 \ ,
\end{align}
where the interface $x_0$ is taken to be 0.3. The exact solution is given by:
\begin{align}
\mqty(\rho && \mathbf{u} && p ) &= \left\lbrace\mqty{ \mqty(\rho_L && \mathbf{u}_L && p_L ) && \text{if}\quad \frac{x-x_0}{t} < -c_0 \\
						     \mqty(\rho^* && \mathbf{u}^* && p^* ) && \text{if}\quad \frac{x-x_0}{t} < \frac{2c_0}{\gamma-1} \\
						     \mqty(\rho_R && \mathbf{u}_R && p_R ) && \text{if}\quad \frac{x-x_0}{t} > \frac{2c_0}{\gamma-1}\\}\right. \ , \\
\nonumber \text{where: } \\						     
\mathbf{u}^* &= 2\frac{\frac{x-x_0}{t} + c_0}{\gamma-1} \\
\rho^* &= \left(\left(\mathbf{u}^*-\frac{x-x_0}{t}\right)^2\left(\frac{\rho_L^{\gamma}}{\gamma p_L}\right)\right)^{\frac{1}{\gamma-1}} \\
p^* &= p_L\left(\frac{\rho^*}{\rho_L}\right)^{\gamma} \ .
\end{align}
\begin{figure}
\centering
\includegraphics[width = \textwidth]{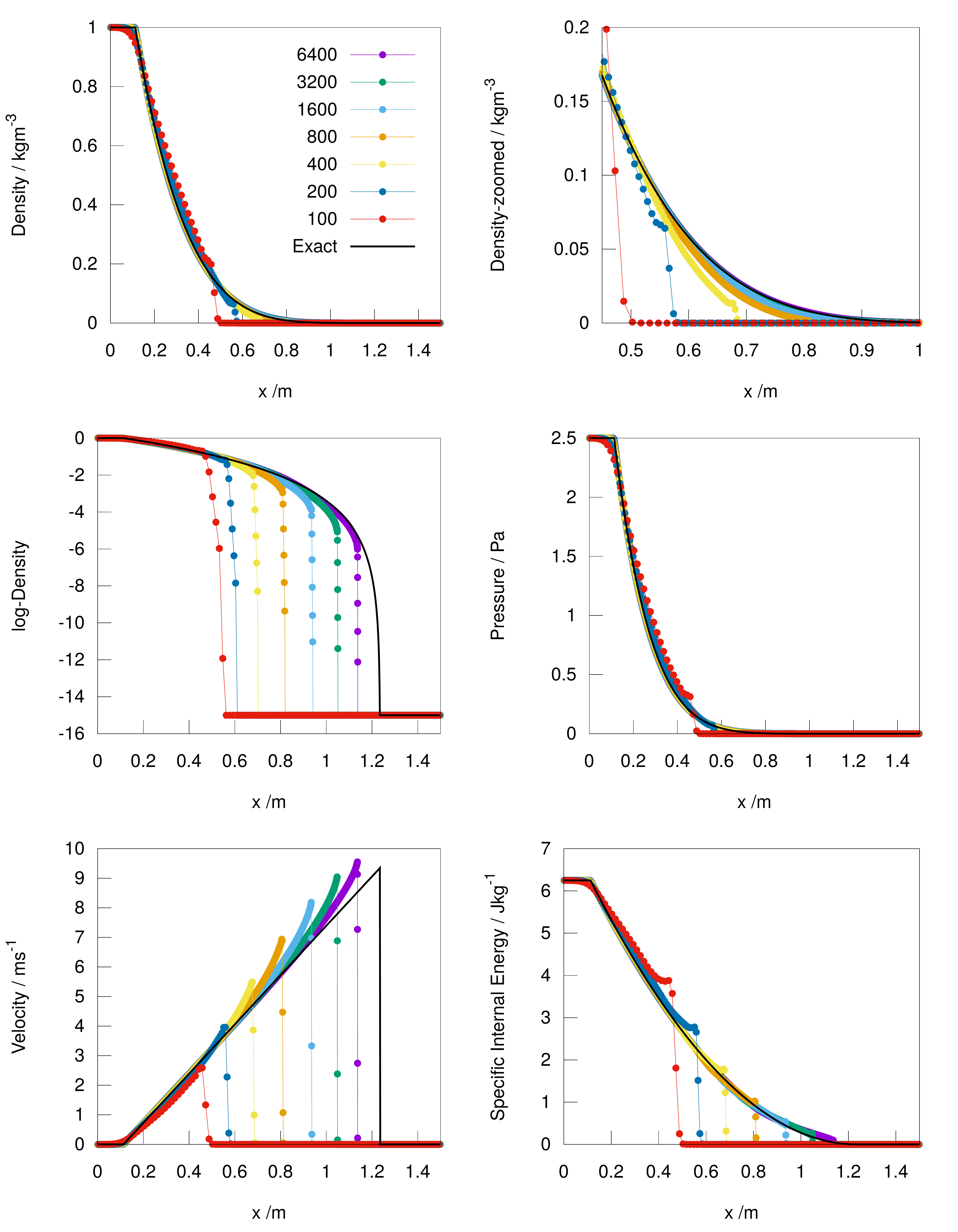}
\caption{The gas-vacuum Riemann problem. This difficult test examines the interaction of a fluid with a pre-defined area of void, comparing this method to the analytical solution, and good agreement is seen in the density and pressure profiles. As with all the tests presented here, THINC reconstruction is applied to the diffuse gas-vacuum boundary to keep it sharp.}
\label{fig:VRP_diffuse}
\end{figure}

%\begin{figure}
%\centering
%\includegraphics[width = \textwidth]{VRP_sharp.pdf}
%\caption{The gas-vacuum Riemann problem for a Ghost Fluid Method.}
%\label{fig:VRP_sharp}
%\end{figure}

\begin{figure}
\centering
\includegraphics[width = 0.5\textwidth]{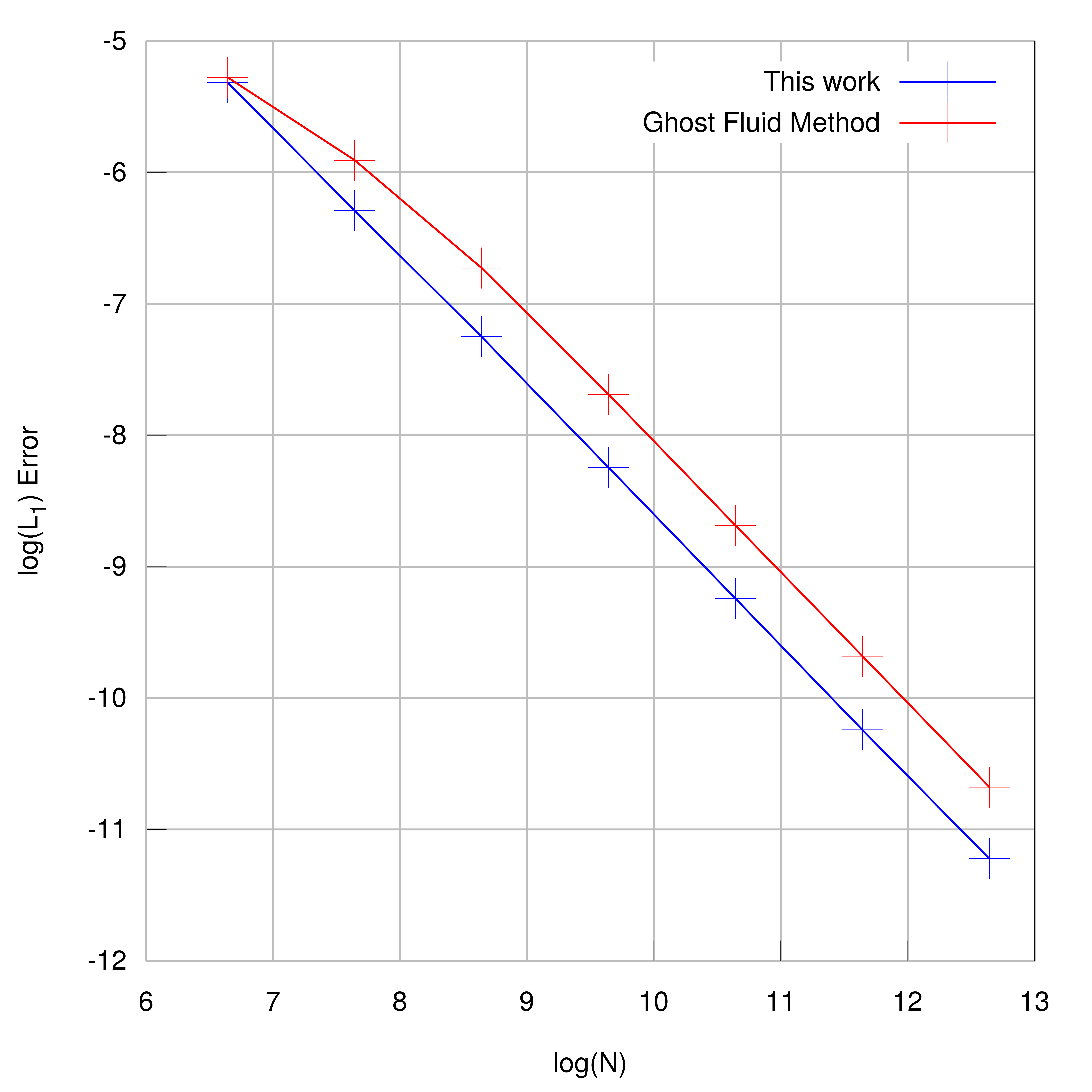}
\caption{The density convergence rate for the gas-vacuum Riemann problem, comparing the work at hand to a Ghost Fluid method. Both methods show first order convergence. $N$ is the number of cells used in the test.}
\label{fig:VRPConvergence}
\end{figure}

\begin{table}
  \begin{center}
    \begin{tabular}{c|c|ll|ll|ll}%
    \hline
    \multirow{2}{*}{Method}  & \multirow{2}{*}{Resolution} & \multicolumn{2}{c|}{Density}   & \multicolumn{2}{c|}{Pressure} & \multicolumn{2}{c}{Internal Energy} \\
    ~      &  ~         & $L_1$ Error & Order           & $L_1$ Error & Order                & $L_1$ Error & Order                   \\\hline
\multirow{8}{*}{This work}      &     ~    &          ~       &          ~       &      ~       &         ~        &          ~       &          ~       \\
  &   100 &  0.02567 &  ~       &  0.04737 &  ~       &  0.56130 &  ~ \\ 
  &   200 &  0.01344 &  0.93393 &  0.02226 &  1.08937 &  0.37705 &  0.57400 \\ 
  &   400 &  0.00656 &  1.03531 &  0.01034 &  1.10613 &  0.20617 &  0.87096 \\ 
  &   800 &  0.00329 &  0.99276 &  0.00521 &  0.99035 &  0.09981 &  1.04659 \\ 
  &  1600 &  0.00165 &  0.99761 &  0.00262 &  0.99298 &  0.04236 &  1.23633 \\ 
  &  3200 &  0.00083 &  0.99907 &  0.00131 &  0.99588 &  0.02160 &  0.97203 \\ 
  &  6400 &  0.00041 &  0.99931 &  0.00066 &  0.99649 &  0.01725 &  0.32432 \\ 
\multirow{8}{*}{Ghost Fluid}    &     ~    &          ~       &          ~       &      ~       &          ~       &          ~       &                  \\ 
  &      100 &          0.02578 &          ~ &            0.04528 &          ~ &                0.55784 &          ~ \\ 
  &      200 &          0.01666 &          0.62938 &      0.02666 &          0.76402 &          0.44393 &          0.32953 \\ 
  &      400 &          0.00944 &          0.82010 &      0.01390 &          0.93990 &          0.33020 &          0.42699 \\ 
  &      800 &          0.00485 &          0.96168 &      0.00668 &          1.05690 &          0.21812 &          0.59821 \\ 
  &     1600 &          0.00243 &          0.99756 &      0.00329 &          1.02356 &          0.13510 &          0.69105 \\ 
  &     3200 &          0.00122 &          0.99393 &      0.00164 &          1.00159 &          0.08101 &          0.73792 \\ 
  &     6400 &          0.00061 &          0.99681 &      0.00082 &          0.99978 &          0.04695 &          0.78708 \\  \hline
    \end{tabular}
  \end{center}
\caption{The gas-vacuum Riemann problem test convergence. Here, the method at hand is compared to a level set Ghost Fluid method, and matches the convergence well.}
\label{tab:VRPConvergence}
\end{table}

This test is shown in Figure \ref{fig:VRP_diffuse}. This method is also compared with a Ghost Fluid method, and Figure \ref{fig:VRPConvergence} and Table \ref{tab:VRPConvergence} demonstrate both of these methods show first order convergence to the exact solution in the density and pressure, as would be expected from this kind of interface problem. However, the internal energy converges slower for both methods, and the velocity will eventually overshoot the exact solution in both cases. It should be stressed that detailed fluid-void interaction studies are not the intended purpose of this method; other exact methods may be more appropriate in these scenarios. Despite this, this test shows that the method performs very capably compared to existing sharp interface methods in a strenuous test, maintaining its convergence.

\subsection{Sliding Blocks Test}

\begin{figure}
\centering
\includegraphics[width = 0.95\textwidth]{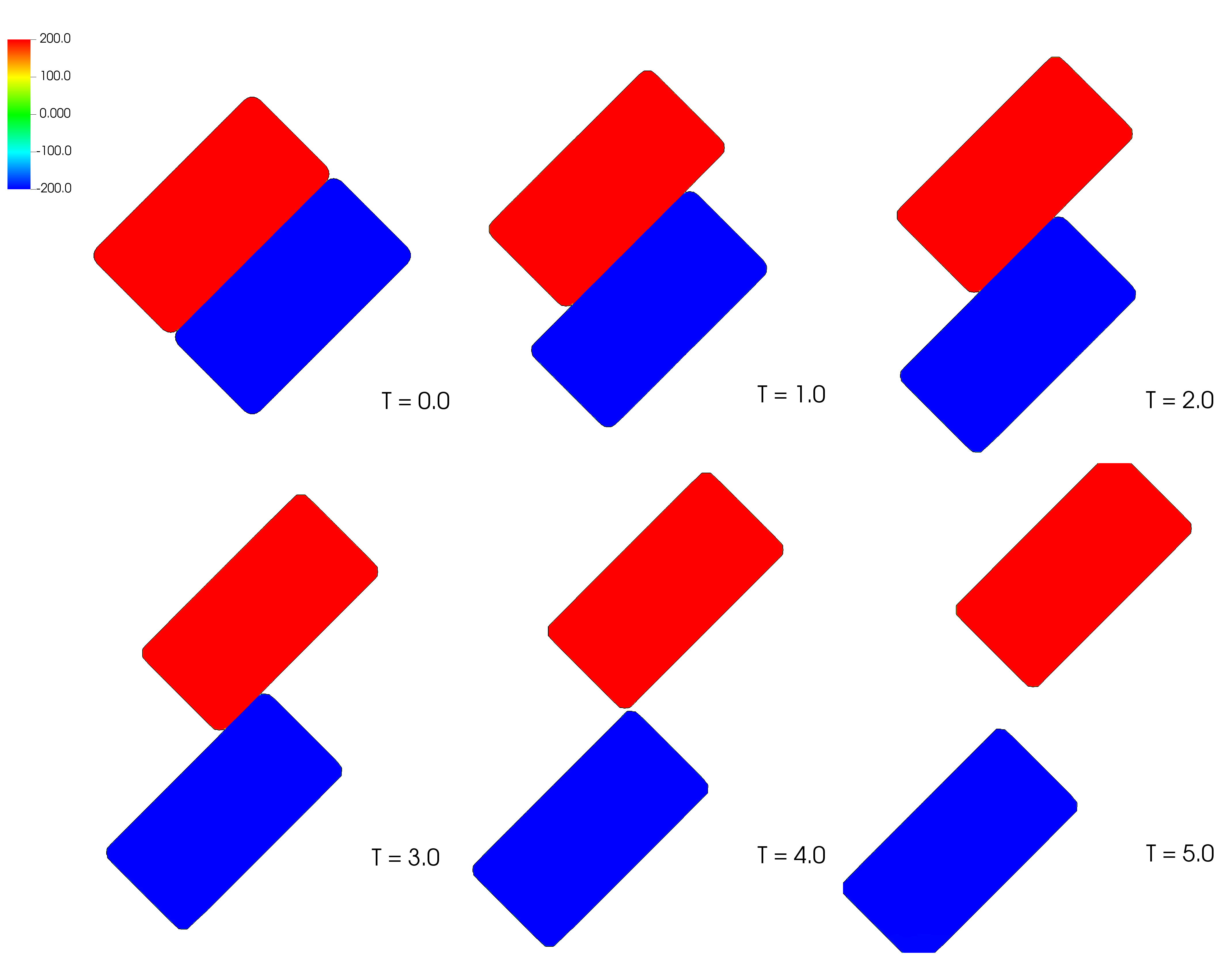}
\caption{The sliding blocks test. Times are shown in $\mu$s. The blocks are surrounded by void, and the shear flux-modifier allows them to slide past each other, separating cleanly at the end.}
\label{fig:SlidingBlocks}
\end{figure}

\begin{figure}
\centering
\begin{subfigure}{.49\textwidth}
  \centering
 \includegraphics[width = \textwidth]{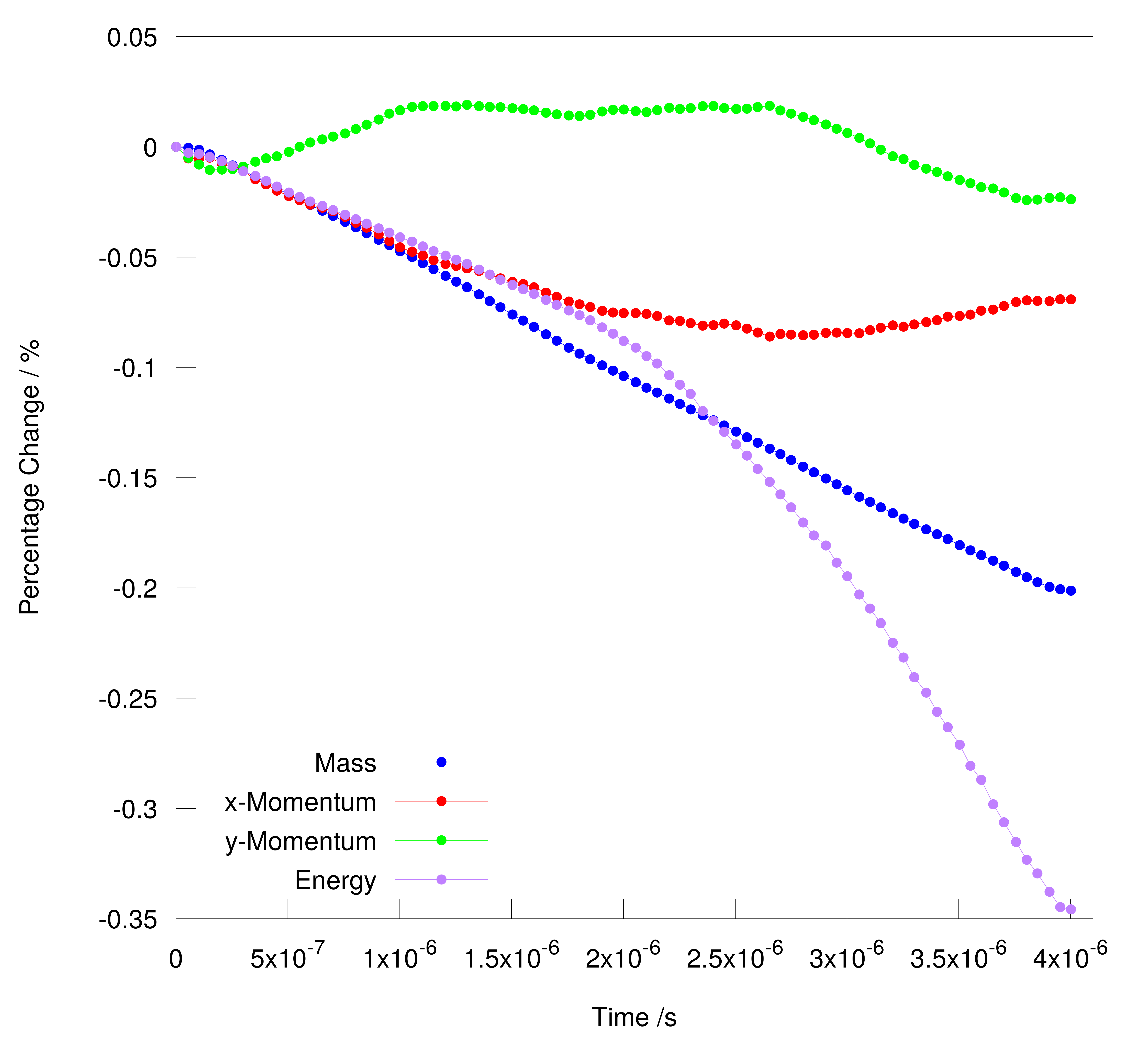}
 \caption{The conservation error for the sliding blocks test, showing mass, momentum and energy loss over the course of the simulation. The method at hand performs very well in terms of conservation, even at this relatively low resolution.}
 \label{fig:SlidingBlocks_Conservation}
\end{subfigure}\hspace{1mm}
\begin{subfigure}{.49\textwidth}
  \centering
 \includegraphics[width = \textwidth]{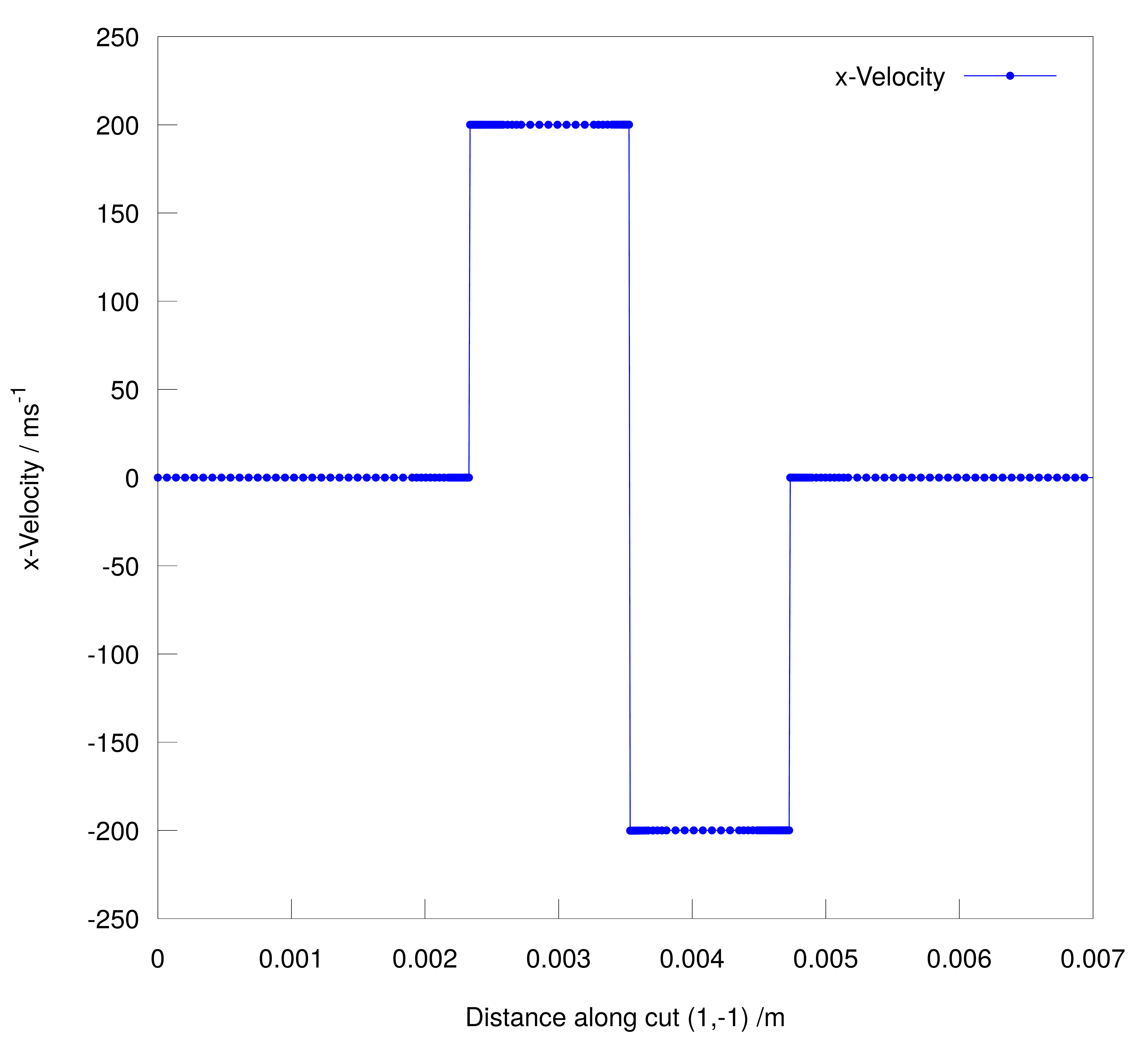}
 \caption{A cut of the $x$-velocity along a line normal to the interface in the sliding blocks test, taken at a time of 1.25 $\mu$s. The plot shows the THINC reconstruction and shear flux-modifier keeping the velocity discontinuities sharp, despite the diffuse interface representation.}
 \label{fig:SlidingBlocks_VelocityCut}
\end{subfigure}
\caption{Further detail on the sliding blocks test.}
\label{fig:test}
\end{figure}

For straightforward two dimensional shear verification, a simple sliding blocks test was considered. In this test, two copper blocks slide past each other, surrounded by vacuum. The velocities of the two blocks are taken along the diagonal of the domain to verify there is no grid-dependence in the shear algorithm. The blocks are both 1.18 mm $\times$ 2.36 mm with a chamfer of 0.17 mm and are given initial velocities:
\begin{align}
\mathbf{u} = \mqty(\pm 200 \\ \pm 200 \\ 0) \ \text{ms}^{-1} \ .
\end{align}
The test is run using a base resolution of 100 $\times$ 100 cells with 3 layers of AMR, using a CFL of 0.6, for a time of 5 $\mu$s. Results are shown in Figure \ref{fig:SlidingBlocks}. The blocks can be seen to slide smoothly along each other, despite the diagonal interface, and separate cleanly. Figure \ref{fig:SlidingBlocks_Conservation} depicts the conservation error over the course of this test. The error remains low, and is predominantly generated by the quasi-non-conservative void volume fraction update, rather than the flux-modifier. Figure \ref{fig:SlidingBlocks_VelocityCut} shows the $x$-velocity along a cut normal to the interface, where the THINC reconstruction and flux-modifiers help to keep the interfaces sharp.

\subsection{Bouncing Blocks and Cylinders}

\begin{figure}
\centering
\includegraphics[width = 0.95\textwidth]{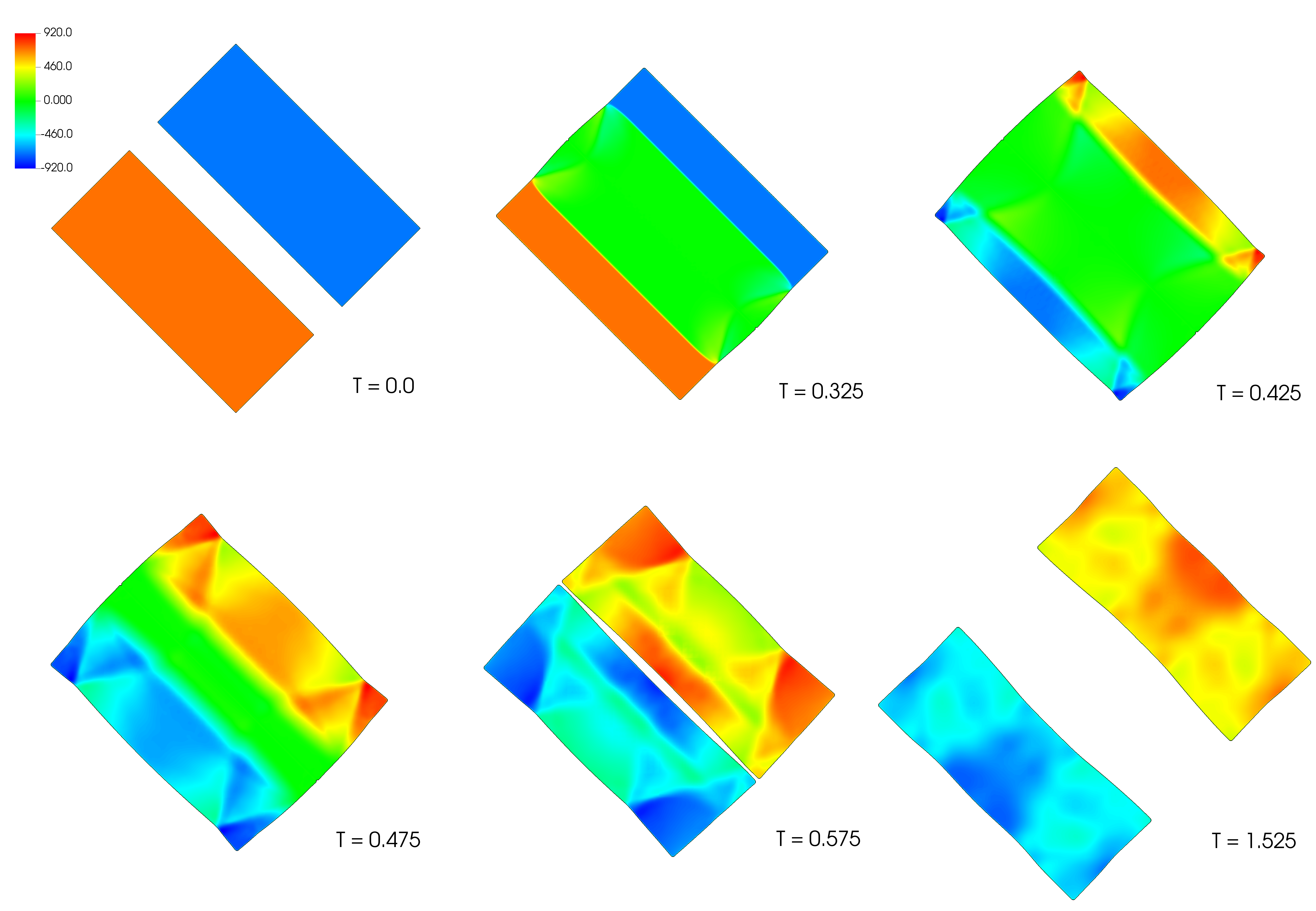}
\caption{The bouncing blocks test, depicting the diagonal velocity. The times shown are in $\mu$s. This test examines the void-generation flux-modifier; as the blocks collide and bounce in vacuum, the interface between closes and opens cleanly.}
\label{fig:BouncingBlocksPictures}
\end{figure}

\begin{figure}
\centering
\includegraphics[width = 0.95\textwidth]{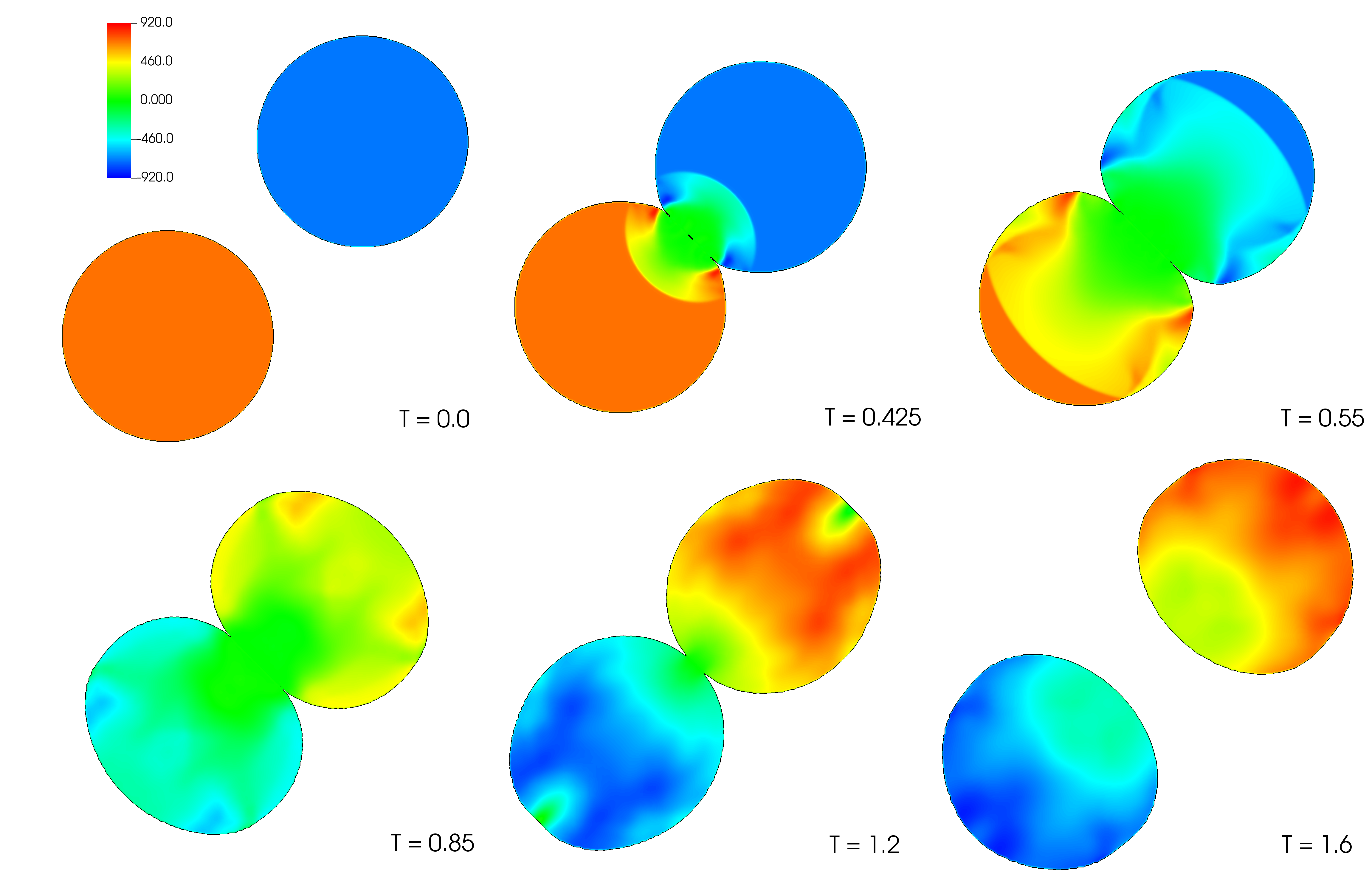}
\caption{The bouncing cylinders test, depicting the diagonal velocity. The times shown are in $\mu$s. This test examines the void-generation flux-modifier; as the cylinders collide and bounce in vacuum, the interface between closes and opens cleanly. This test additionally demonstrates the ability of the method to handle non-planar interface configurations.}
\label{fig:BouncingBallsPictures}
\end{figure}

For a simple two dimensional verification of the void-generation parts of the algorithm, two elastic solid impact tests were considered; a bouncing blocks test and a bouncing cylinders test. These straightforward tests features two purely elastic aluminium materials governed by the Romenskii equation of state colliding and separating in a vacuum. The velocities of the two bodies in either case are taken along the diagonal of the domain to verify there is no grid-dependence in the void-generation algorithm, and the two different geometries test both the planar and non-planar interface separation capabilities. The blocks are both 1 mm $\times$ 2.35 mm, with an initial separation of 0.35 mm. The cylinders both have radii of 0.83 mm, with their centres having an initial separation of 1.53 mm. In both tests, the two bodies are given initial velocities:
\begin{align}
\mathbf{u} = \mqty(\pm 500 \\ \pm 500 \\ 0) \ \text{ms}^{-1} \ .
\end{align}
The tests are run using a base resolution of 100 $\times$ 100 cells with 3 layers of AMR, using a CFL of 0.6, for a time of 1.6 $\mu$s. As can be seen in Figures \ref{fig:BouncingBlocksPictures} and \ref{fig:BouncingBallsPictures}, the two bodies separate cleanly in both cases. It is in this kind of test that the void components of this algorithm give large computational savings. This test would otherwise have to be performed in air, with the simulation having to capture all of the complex but insignificant waves patterns in the air. Moreover, there would likely be large pressure spikes in the air that would be trapped between the blocks, leading to time step restrictions. Both of these problems are avoided with the work at hand.

\begin{figure}
\centering
\begin{subfigure}{.49\textwidth}
  \centering
 \includegraphics[width = \textwidth]{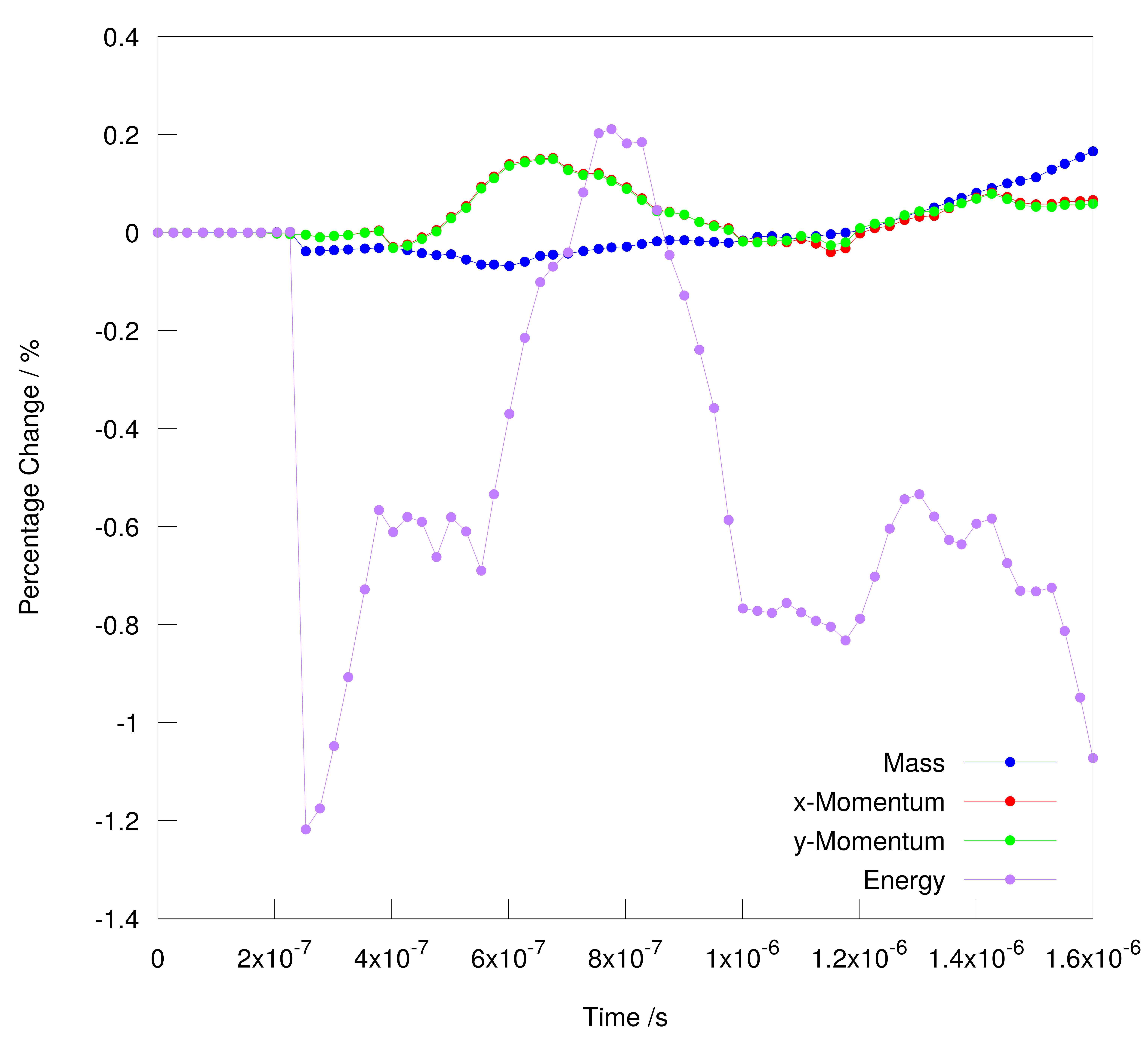}
 \caption{Bouncing blocks.}
 \label{fig:Blocks_Conservation}
\end{subfigure}\hspace{1mm}
\begin{subfigure}{.49\textwidth}
  \centering
 \includegraphics[width = \textwidth]{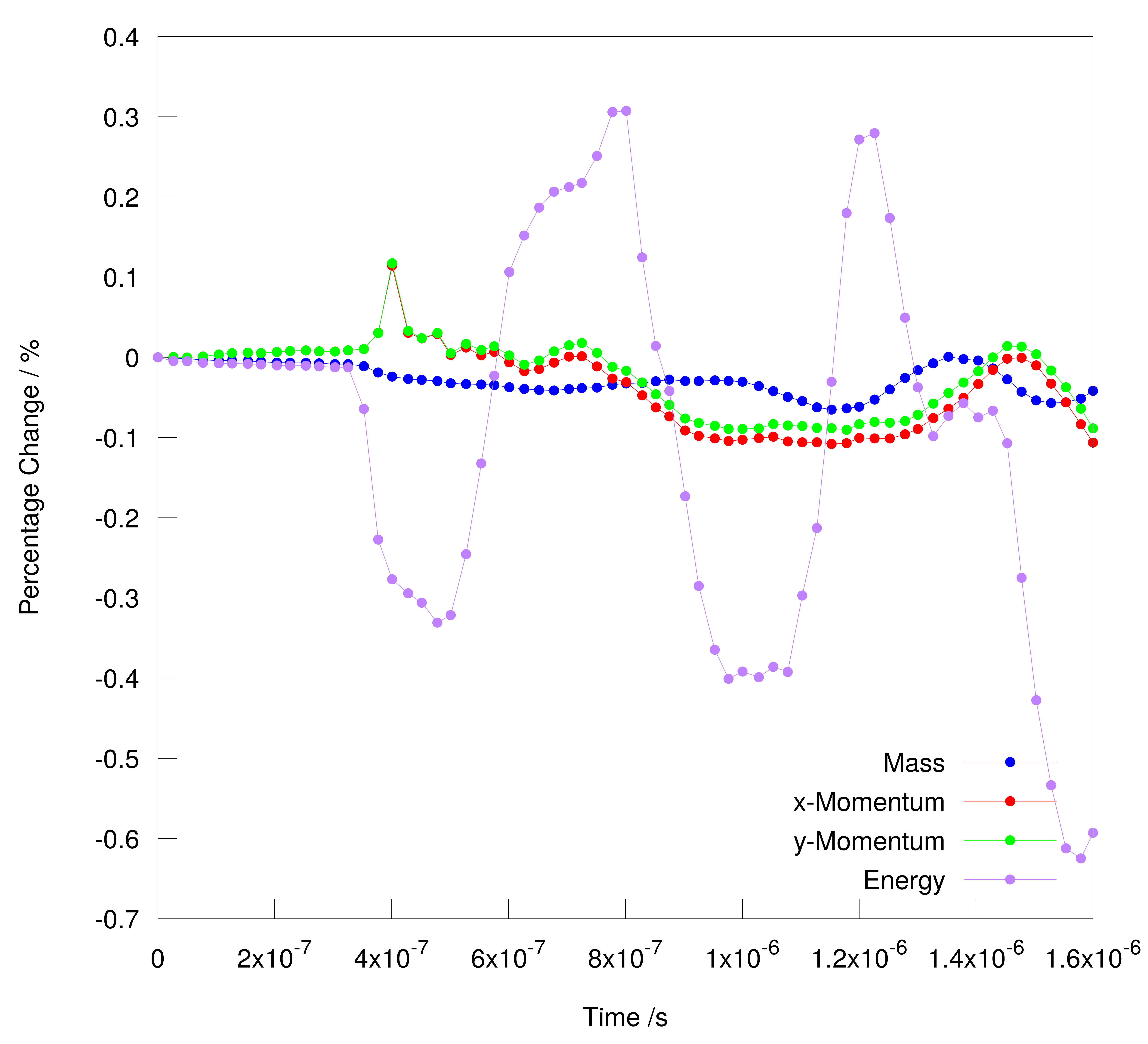}
 \caption{Bouncing cylinders.}
 \label{fig:Balls_Conservation}
\end{subfigure}
\caption{The conservation error for both the bouncing blocks and bouncing cylinders tests, showing mass, momentum and energy loss over the course of the simulation. The conservation error for these tests remains low even despite the low resolution, with the bouncing blocks case showing larger error due to the larger interface area, and correspondingly increased use of the flux-modifier.}
\label{fig:test}
\end{figure}

Figures \ref{fig:Blocks_Conservation} and \ref{fig:Balls_Conservation} depict the conservation errors of this scheme over the course of the two simulations. In both cases, the mass and momentum conservation is very good, with neither test going above or below a 0.2\% change. The total energy conservation is more erratic, but well within the expected bounds for this kind of Eulerian scheme. The total energy conservation error is larger in the case of the bouncing blocks both because there is a larger contact surface, giving increased use of the flux-modifiers, and the geometry for the blocks results in a stronger shock wave hitting the back of the blocks compared to the circular case.

\begin{table}
\begin{center}
    \begin{tabular}[t]{|l|l|l|l|l|l|l|}
      \hline
      Material & $c_1$  GPa & $c_2$  GPa & $c_3$ & $n$ & $m$ & $T_{\text{melt}}$  K \\
      Copper   & 0.09 & 0.292 & 0.025 & 0.31  & 1.09 & 1356.0  \\
      CuBe     & 1.041 & 0.0 & 0.025 & 0.31   & 1.09 & 1286.0  \\
      \hline
    \end{tabular}
\end{center}
\caption{Johnson Cook plasticity material parameters.}
\label{tab:PlasticParameters}
\end{table}

\begin{table}
\begin{center}
    \begin{tabular}[t]{|l|l|l|l|l|l|}
      \hline
      Material & $D_\text{crit}$ & $\epsilon_{p,\text{thresh}}$ & $\epsilon_{p,\text{crit}}$ & $\alpha$ & $\nu$ \\
      \hline
      Copper    & 0.85 & 0.34 & 1.04 & 0.631 & 0.355 \\
      CuBe      & 0.85 & 0.08 & 0.16 & 0.631 & 0.324 \\
      AerMet    & 0.2 & 0.129& 0.64& 0.22  & 0.302 \\
      \hline 
    \end{tabular}
\end{center}
\caption{Damage model parameters.}
\label{tab:DamageParameters}
\end{table}

\subsection{Tribological Pair Test}

\begin{figure}
\centering
\begin{subfigure}{0.4\textwidth}
\centering
\includegraphics[width = 0.9\textwidth]{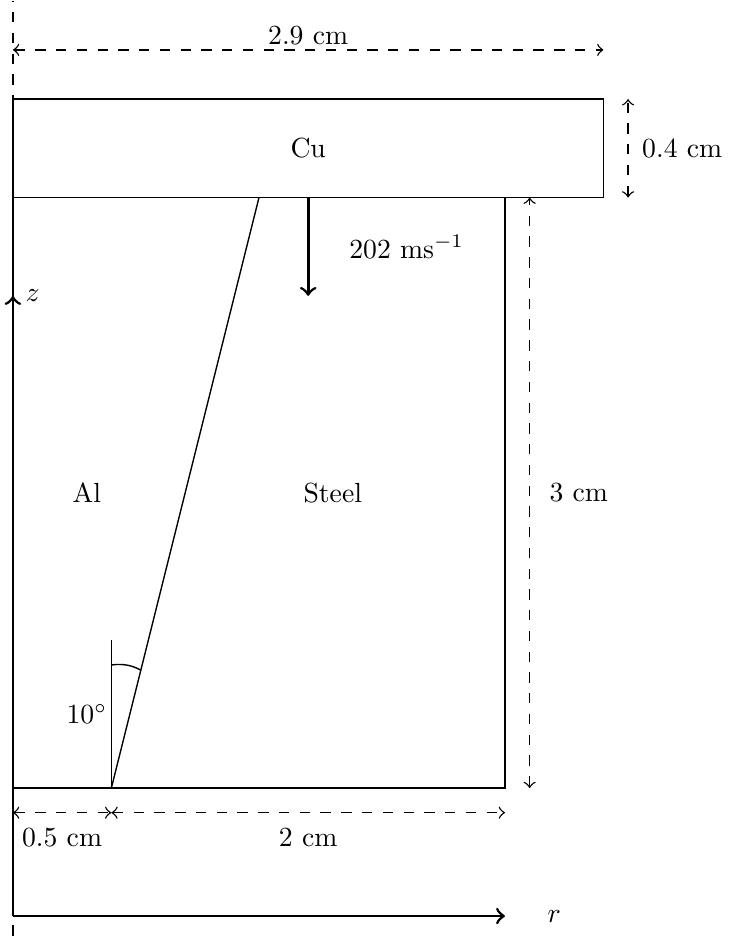}
\caption{Initial conditions for the tribological pair test.}
\label{fig:TriboPairInitialConditions}
\end{subfigure}\hfill
\begin{subfigure}{0.6\textwidth}
\centering
\includegraphics[width = 0.9\textwidth]{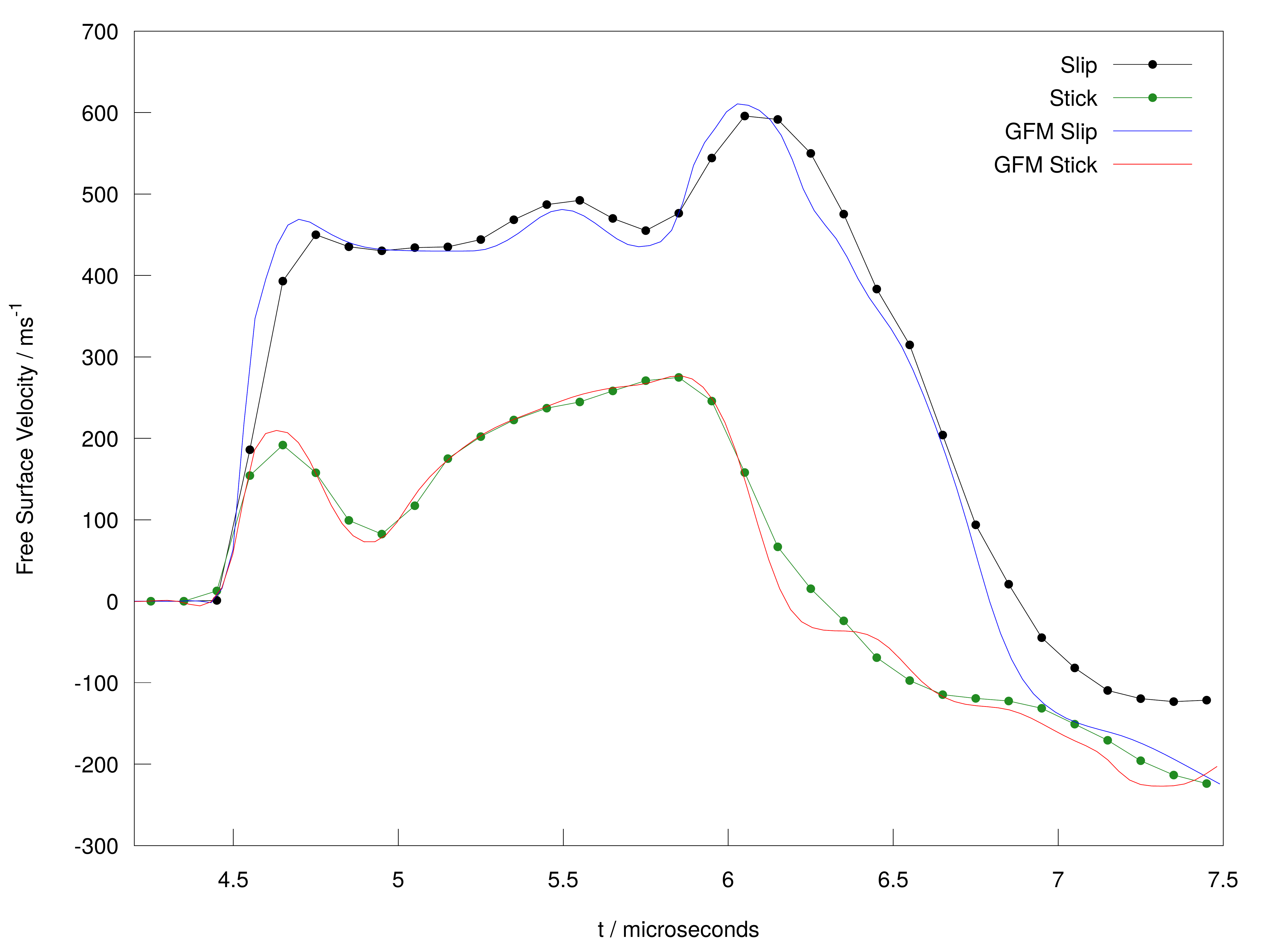}
\caption{A comparison of the free surface velocity profile in the tribological pair test between this method and \citet{BartonSliding} for both slip and stick boundary conditions. Excellent agreement is observed.}
\label{fig:TriboPairVelocityProfile}
\end{subfigure}
\caption{Details of the tribological pair test.}
\end{figure}

For more complex two dimensional shear validation, the tribological pair test from \citet{triboPairExperiment} was considered. The tribological pair test is commonly used to assess friction between different solid materials and allows for a rigorous examination of the ability of the model to support multi-dimensional, oblique sliding. As outlined by \citet{BartonSliding}, this test demonstrates a marked difference between different material boundary conditions. The test consists of a flyerplate colliding with a cylinder comprised of two different materials, as shown in Figure \ref{fig:TriboPairInitialConditions}. Owing to the different material parameters of the inner and outer parts of the cone, the shock wave produced by the collision of the flyerplate travels down the cone at different speeds in the different materials. This produces a tangential velocity discontinuity at the angled interface in the case of the slip boundary conditions. When the wave reaches the bottom of the cone, the slip conditions allow for the centre of the cone to protrude from the bottom, where the initial shock wave arrives first, whereas the stick conditions do not. These differences are captured in the free-surface velocity profile of the test in Figure \ref{fig:TriboPairVelocityProfile}.

In this test, the work at hand is compared to the Ghost Fluid method of \citet{BartonSliding}. To match the test performed by \citet{BartonSliding}, the materials are assumed to behave completely elastically. The initial conditions for the tribological pair test are shown in Figure \ref{fig:TriboPairInitialConditions}. The test is modelled in 2D using cylindrical symmetry with a reflective boundary condition on the $z$-axis. All materials are modelled using the Romenskii equation of state, with parameters outlined in Table \ref{tab:MaterialParameters}. The solid materials are surrounded by void. The copper flyerplate has a downward velocity of 202 ms$^{-1}$. The computational domain spans $r = [0:3.2] \text{ cm}, \ z = [0:4.28] \text{ cm}$. The test is run at a resolution of 80 $\times$ 100 cells with one level of AMR with a refinement factor 2 in order to match the simulation of \citet{BartonSliding}. The test is run at a CFL of 0.6, for 7.5 $\mu$s.

\begin{figure}
\centering
\includegraphics[width = 0.9\textwidth]{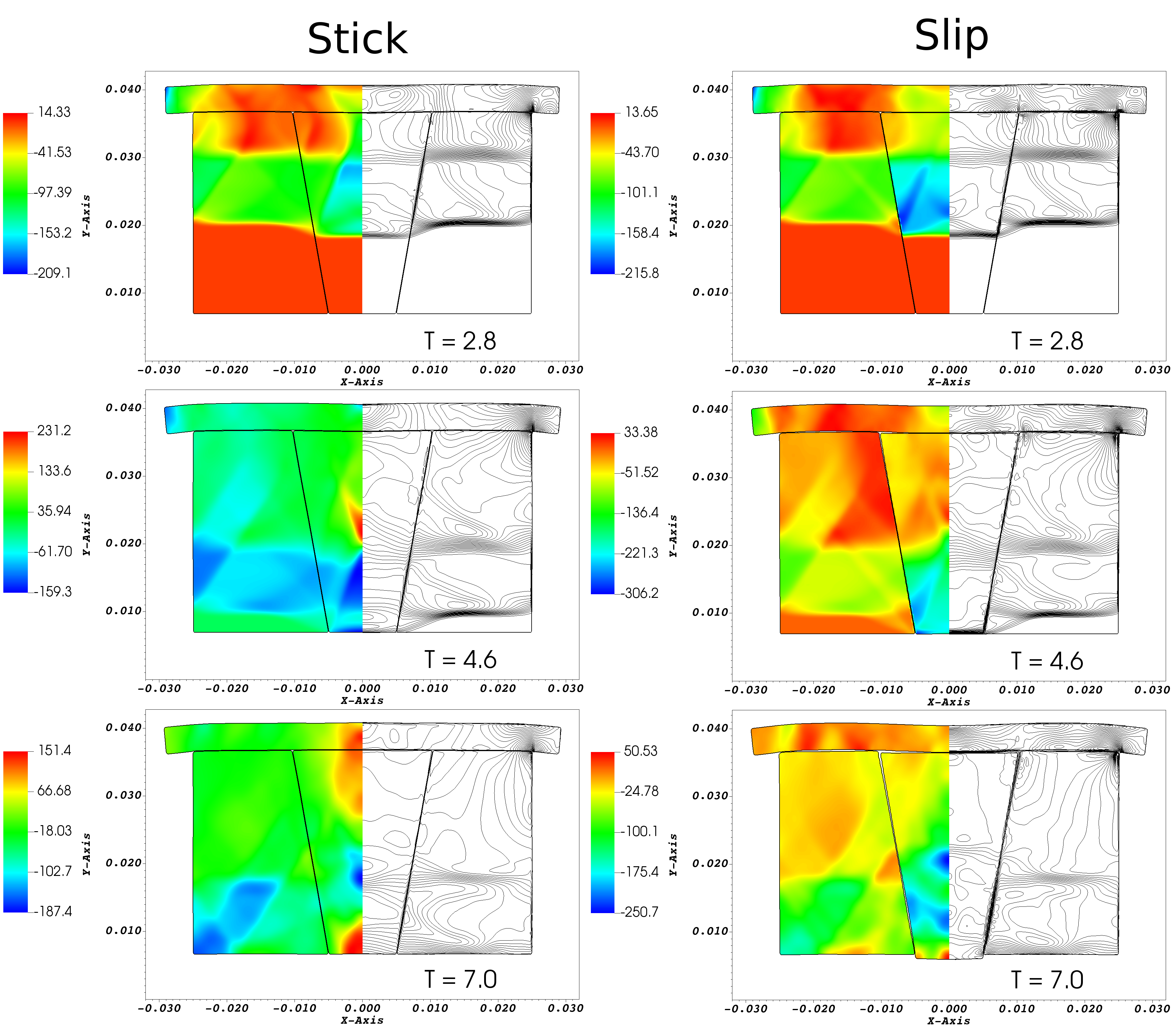}
\caption{Comparison of the stick and slip boundary conditions for the tribological pair test. In each frame, the left hand side depicts the y-velocity and the right hand side shows the pressure contours. The left column is with stick boundary conditions and the right column is with slip boundary conditions. Times shown are in $\mu$s. The method works well in both cases, with the slip boundary condition correctly predicting the protruding cone and separating interfaces - features that cannot be demonstrated without the suitable boundary conditions.}
\label{fig:TriboPairPictures}
\end{figure}

\begin{figure}
\centering
\includegraphics[width = 0.9\textwidth]{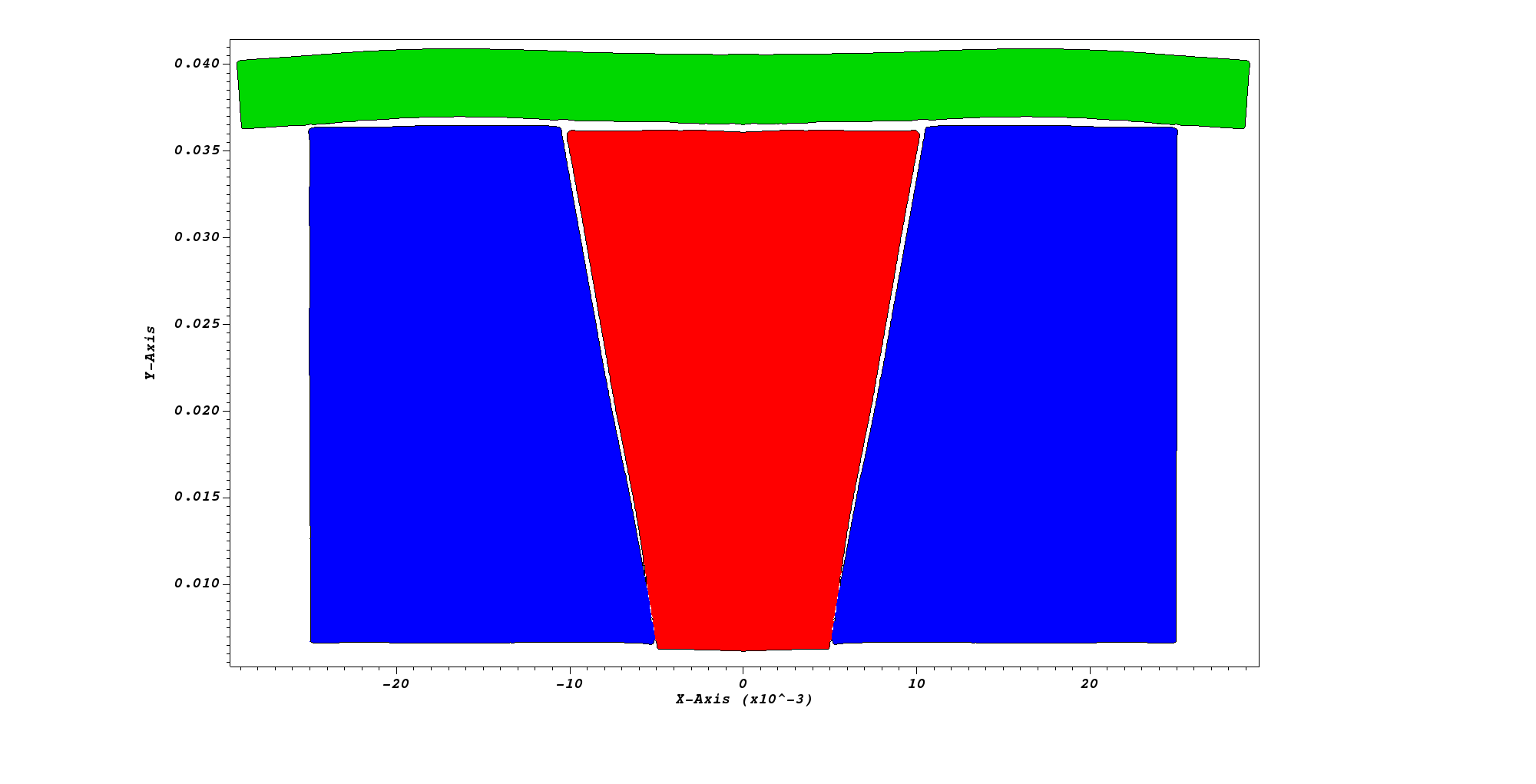}
\caption{The tribological pair test with slip boundary conditions at a higher resolution at a time of $13 \ \mu$s, showing the interface separation between materials. This test was run without plasticity, meaning the interface separation is reduced compared to experiment. Nonetheless, it validates the correct working of the void-generation flux-modifier.}
\label{fig:TriboPairSeparation}
\end{figure}

The results of the test are shown in figure \ref{fig:TriboPairPictures}, where both stick and slip conditions are shown for comparison. The results are qualitatively examined using the free surface velocity profile, taken on the axis at the bottom of the cylinder. Figure \ref{fig:TriboPairVelocityProfile} shows this velocity profile, and it can be seen that the results of this method match those of \citet{BartonSliding} very well. This demonstrates the method's ability to handle multi-material, multi-dimension oblique shear well, matching an advanced elastoplastic Ghost Fluid method. This test also has the added benefit of validating the material void interaction and interface separation. The collision of the strong shock with the void interface at the end of the cone is handled stably and the material is allowed to protrude in the case of slip boundary conditions. Moreover, the test is repeated with an additional layer of AMR, and run for a time of 13 $\mu$s to show how the different materials separate from each other at late times, as shown in Figure \ref{fig:TriboPairSeparation}.

\subsection{Fracture Tests with Experimental Comparison}
\subsubsection{One Dimensional Spallation}

\begin{figure}
\centering
\includegraphics[width = 0.6\textwidth]{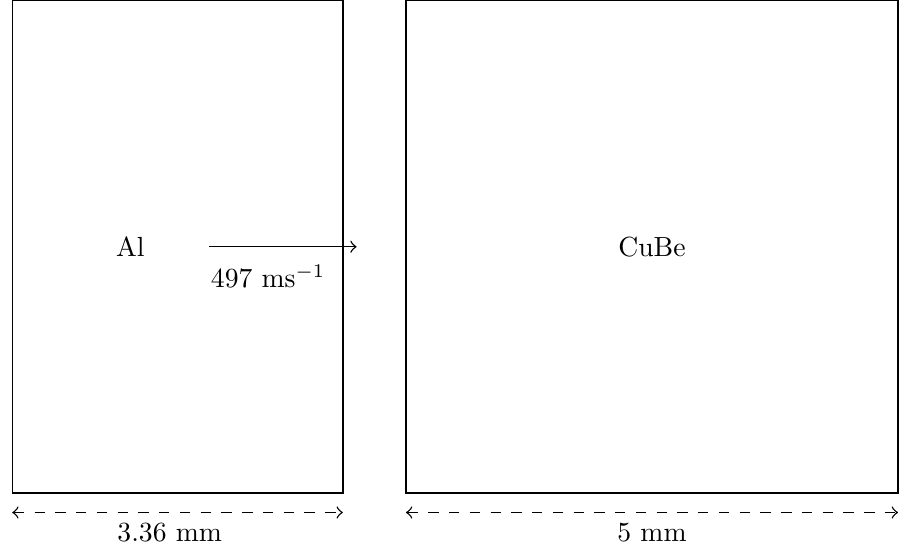}
\caption{The 1D spallation test initial conditions.}
\label{fig:1DSpallationIntialConditions}
\end{figure}

The fracture components of the method are validated in one dimension using an experimental spallation test from \citet{1DSpallationPaper}. In this test, a block of aluminium alloy 5083 H32 collides with a block of copper alloy CuBe TF00, causing the copper to undergo spallation fracture. The test is run for 5 $\mu$s on a domain $x = [0:15]$ mm using a CFL 0.9 with a varying number of cells. The initial conditions for the test are shown in Figure \ref{fig:1DSpallationIntialConditions}. Both materials have plasticity, with the copper using the Johnson Cook model with parameters laid out in Table \ref{tab:PlasticParameters}, and the aluminium using ideal plasticity with a yield stress of 0.275 GPa. The copper is damageable, with parameters laid out in Table \ref{tab:DamageParameters}.

\begin{figure}
\centering
\includegraphics[width = 0.8\textwidth]{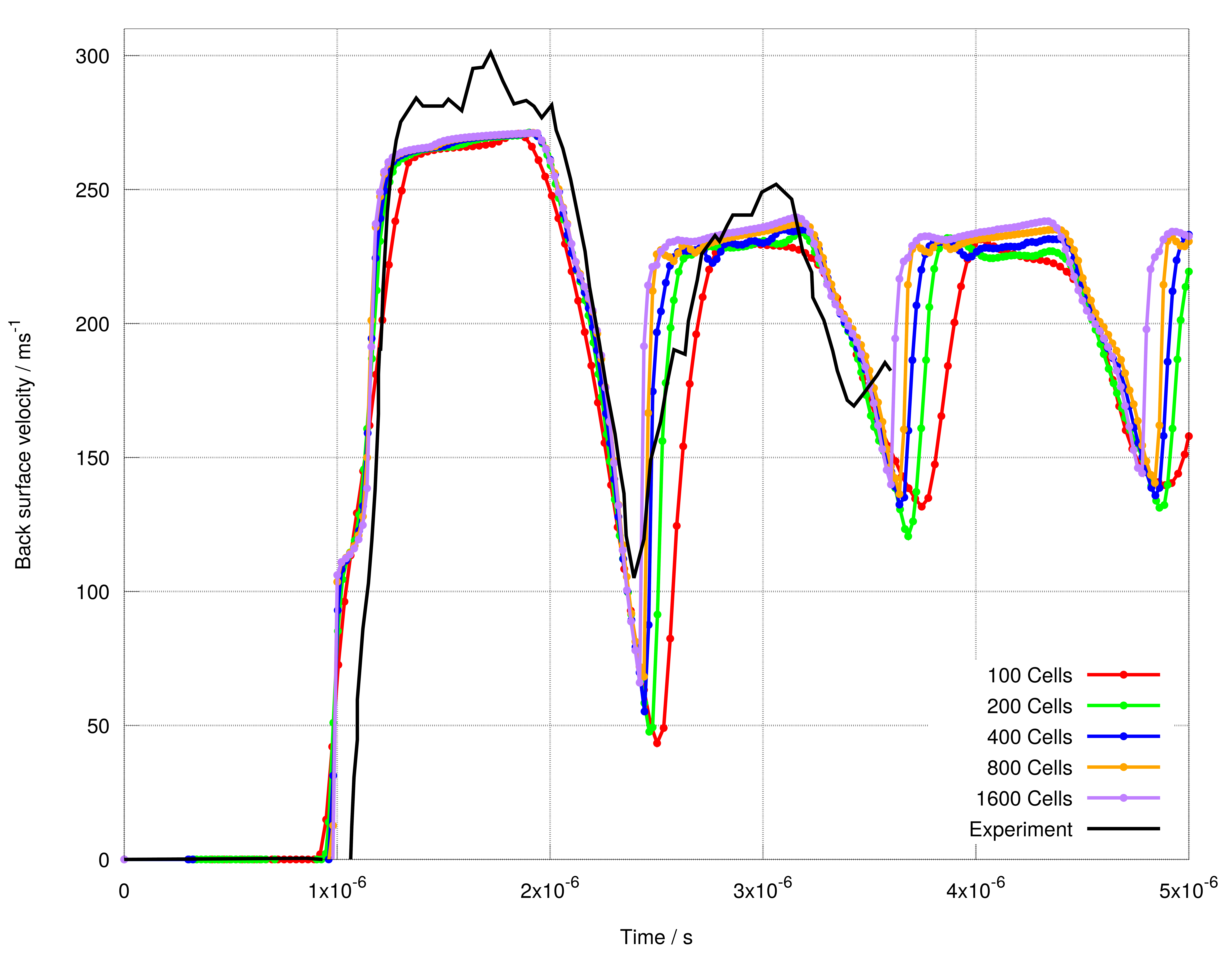}
\caption{The free-surface velocity profile for the 1D spallation test, comparing to experiment. This test demonstrates combination of damage, fracture and void generation working well in one dimension. A range of resolutions is shown here, demonstrating the method's convergence towards the experimental data.}
\label{fig:1DSpallation}
\end{figure}

The results are compared to experiment using the back surface velocity profile of the copper block, shown in Figure~\ref{fig:1DSpallation}. It can be seen that this method matches the experiment well. However, it should be noted that the test is very sensitive to the material damage parameters chosen.

\subsubsection{Plate Impact Spallation}

\begin{figure}
\centering
\includegraphics[width = 0.8\textwidth]{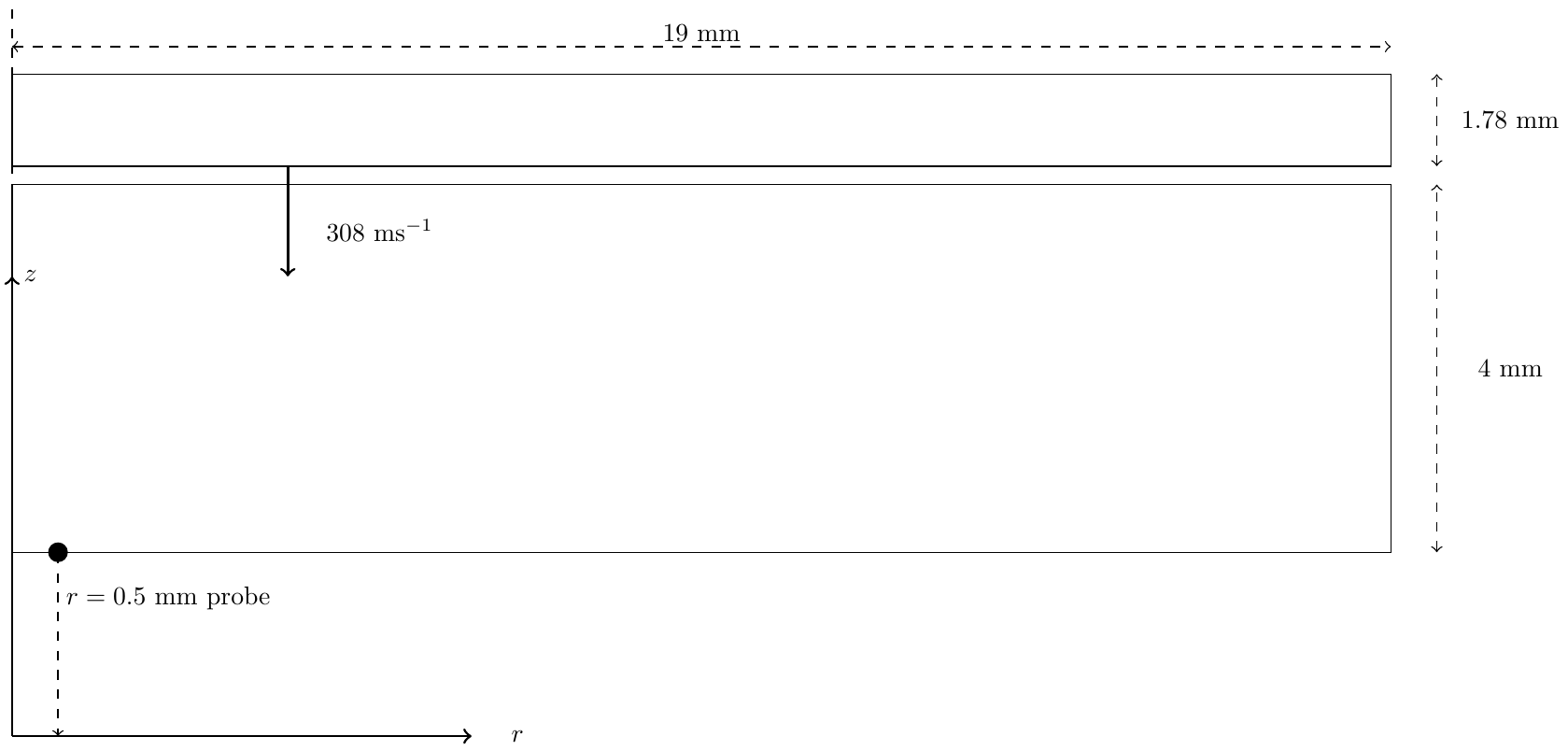}
\caption{The initial conditions for the plate impact spallation test.}
\label{fig:PlateImpactInitialConditions}
\end{figure}

To simultaneously validate all the components of the method, a test containing material shear, separation and fracture is required. Such a test is the plate impact spallation test from \citet{ExperimentalPlateImpact}. This test features a copper flyer-plate impacting another block of copper, the target, with sufficient force that a spallation crack appears in the target. Spall fracture is induced by the interaction of two strong tensile waves inside the target. As has been mentioned, spall damage is important for any fracture model, as it requires void-generation completely inside one material, which cannot physically be modelled by the artificial injection of a low density gas. The initial set-up for this test is shown in Figure \ref{fig:PlateImpactInitialConditions}.
The flyer-plate and target are treated as separate materials, both obeying the Romenskii equation of state for copper. Both materials are damageable, with the damage parameters given in Table \ref{tab:DamageParameters}.

\begin{figure}
\centering
\includegraphics[width = 1.0\textwidth]{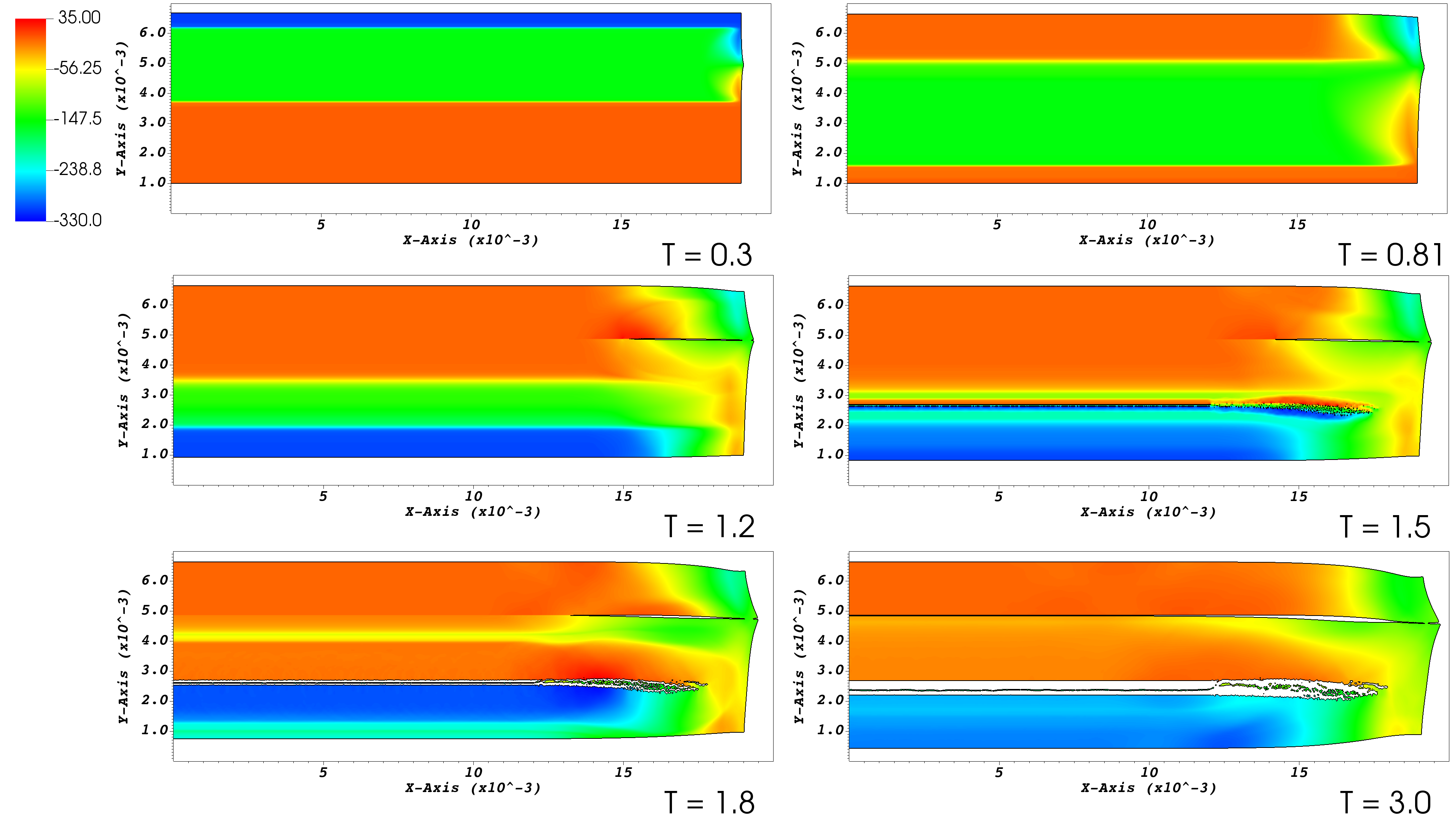}
\caption{The 2D plate impact spallation test. The plots show the downward velocity in $m\cdot s^{-1}$, with the times given in $\mu$s. The test demonstrates the shear, void-generation and fracture algorithms all working together to enable this tough test to be modelled. }
\label{fig:UPSPictures}
\end{figure}

The computational domain spans $r = [0:20 \text{mm}], \ z = [0:7 \text{mm}]$, and the test is run for 3 $\mu$s at a CFL of 0.6. Figure \ref{fig:UPSPictures} shows this test, where a resolution of 500 $\times$ 180 cells with 2 levels of AMR is employed, giving an effective resolution of 10 $\mu$m. The results show the method can handle this large spallation fracture, combined with the slide and separation of the two materials. 

\begin{figure}
\centering
\includegraphics[width = 0.7\textwidth]{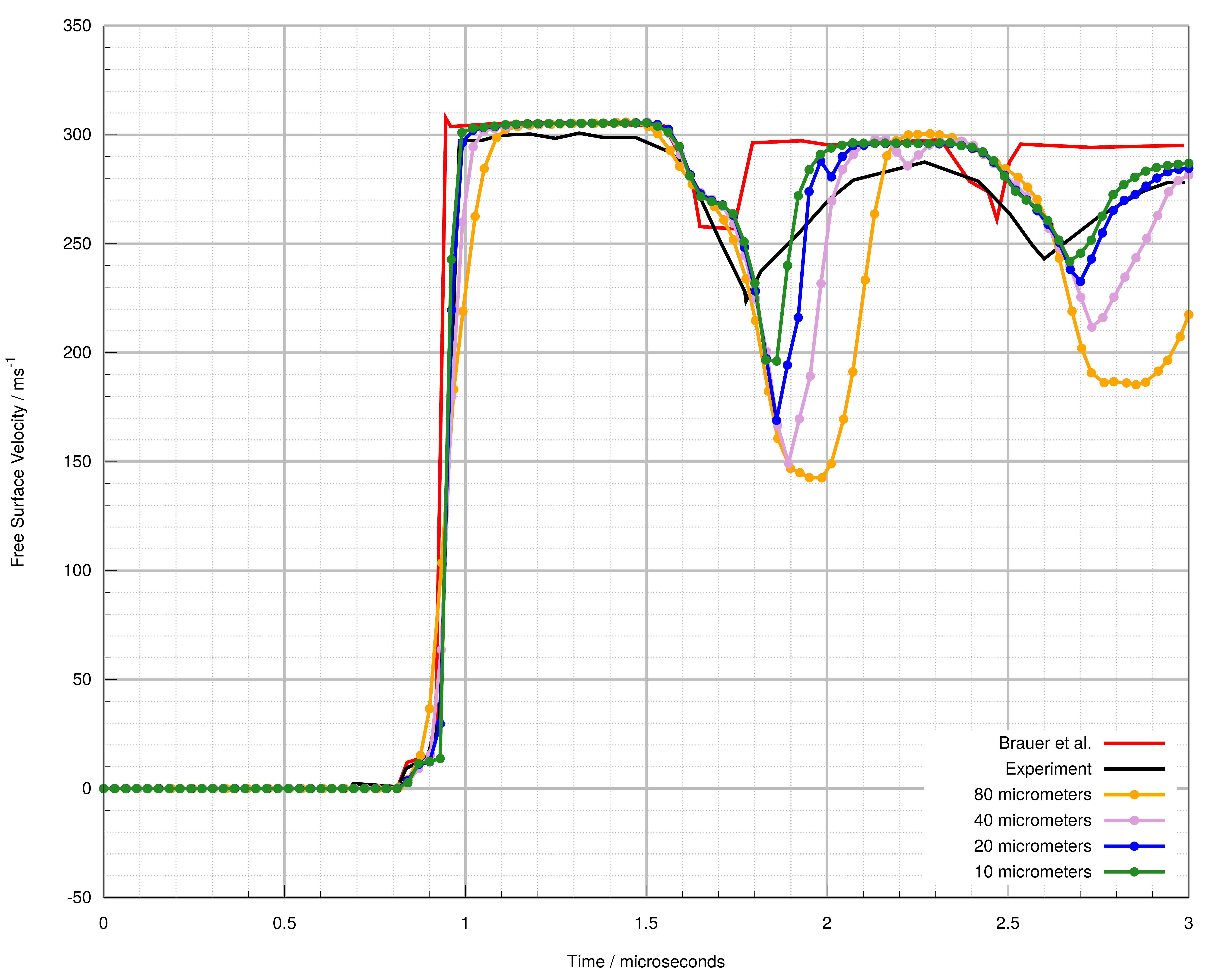}
\caption{The plate impact spallation test free surface velocity profile. Four different resolutions are compared to experiment and previous numerical results and are found to match well, demonstrating good convergence to the experimental data.}
\label{fig:UPSVelocity}
\end{figure}

Similar to the 1D spallation case, the experimental validation for this test comes from the free surface velocity profile, this time of the scab that is formed after spallation. To match experiment, the profile is taken at $r = 0.5$ mm. Comparison is also made with the Ghost Fluid method of \citet{UdaykumarDamage}, who performed the same test with a variety of damage models, using a level set based approach. Four different resolutions are considered: 80, 40, 20 and 10 (Figure \ref{fig:UPSPictures}) $\mu$m. These velocity profiles are shown in Figure \ref{fig:UPSVelocity}. It can be seen that this work matches the experimental oscillation frequency well, and improves with resolution. Moreover, this work matches the qualitative shape of the material separation and the damaged region found by \citet{UdaykumarDamage} with their level set based approach.

For other similar spallation tests on different materials, see for example \citet{KanelSpallation} and \citet{CzarnotaSpallation}.

\subsubsection{Tube Expansion Test}

\begin{figure}
\centering
\includegraphics[width = \textwidth]{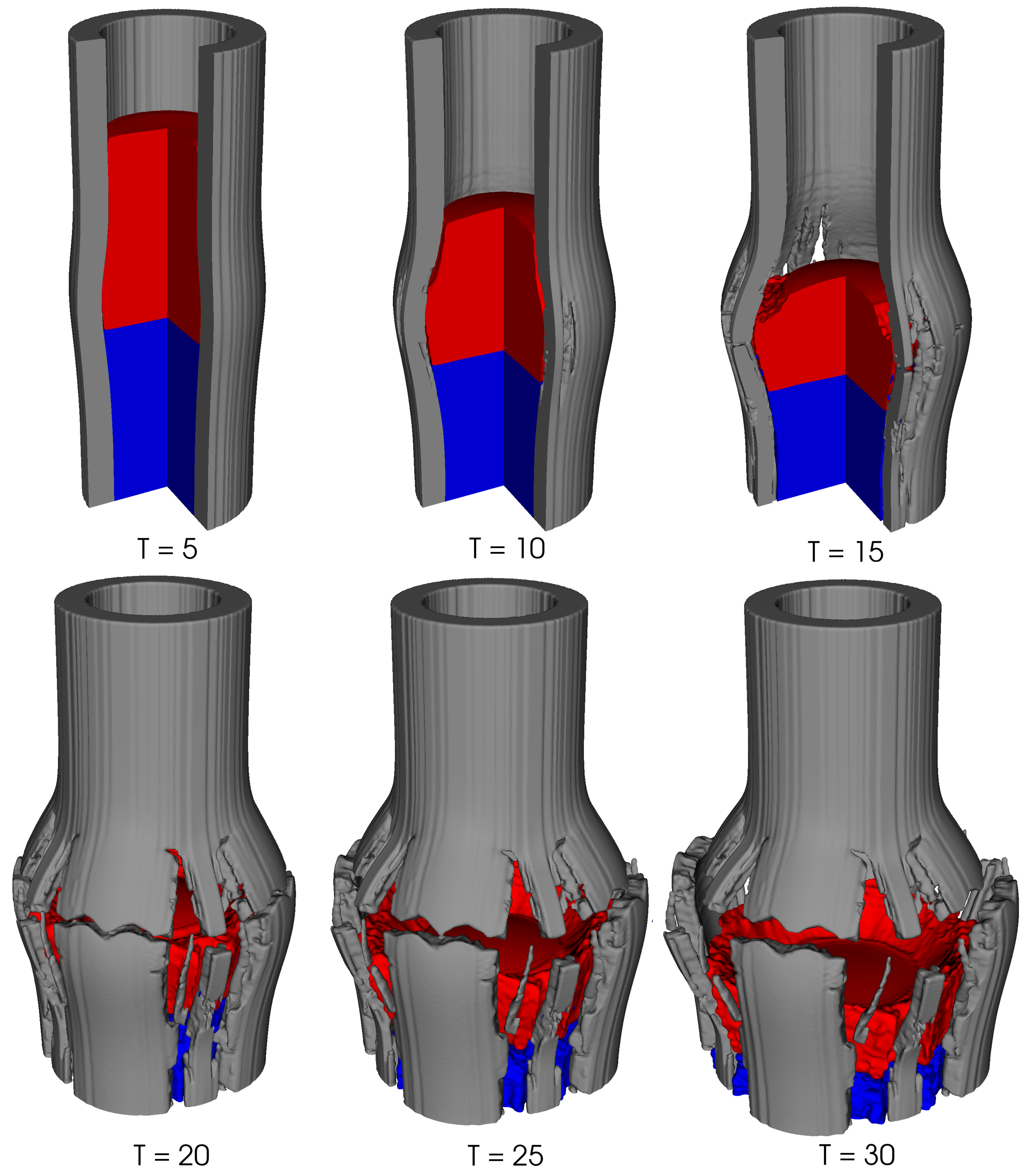}
\caption{The 3D tube expansion test. Times shown are in $\mu$s. The steel tube is depicted in grey and the two polycarbonate blocks are depicted in red and blue. The test examines the three-dimensional capabilities of the method. The work at hand proves to be a practical method to facilitate this kind of large, anisotropic fracture test.}
\label{fig:TubeExpansionPictures}
\end{figure}

\begin{figure}
\centering
\includegraphics[width = \textwidth]{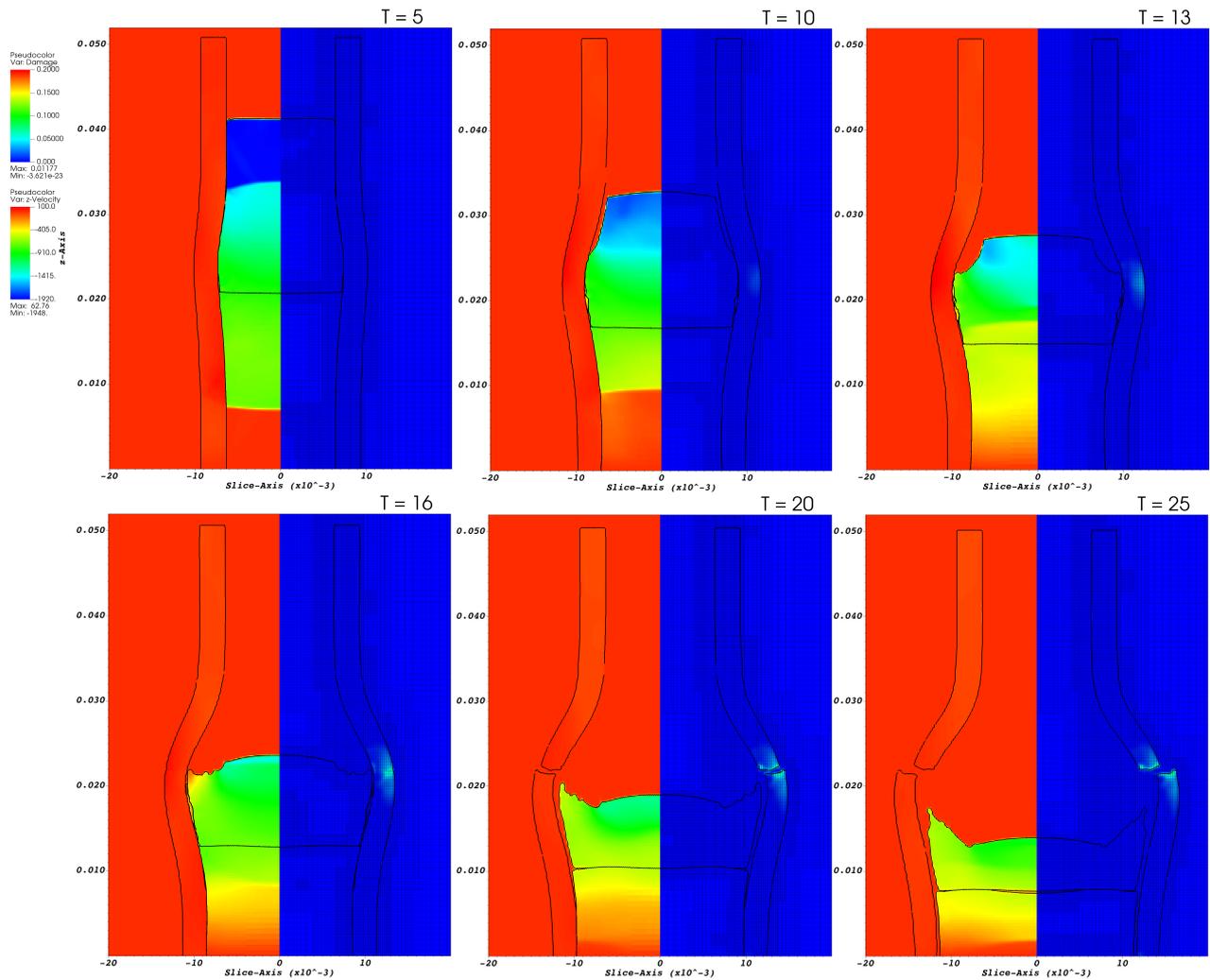}
\caption{A slice of the 3D tube expansion test. Each frame shows the (\textit{left}) $z-$velocity, (\textit{right}) damage evolution and (\textit{both}) the AMR mesh. Times shown are in $\mu$s. This slice demonstrates the flux-modifiers working well to allow the blocks to slide and separate inside the tube, as well as depicting the formation of damage within the steel.}
\label{fig:TubeExpansionSlice}
\end{figure}

\begin{figure}
\centering
 \includegraphics[width = 1.0\textwidth]{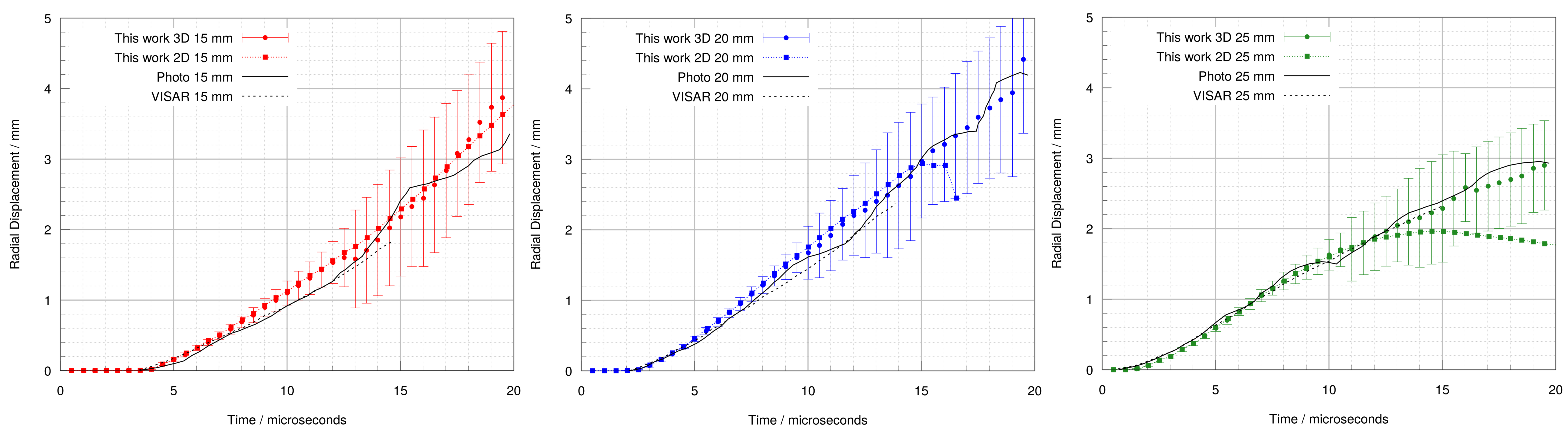}
\caption{Experimental comparison for the tube expansion test. The radial displacement profiles of the tube wall are compared in 2D and 3D versions of this test. The 3D profiles are closer to the experimental results, reflecting the anisotropic nature of the fracture in this problem and demonstrating the necessity of a three dimensional algorithm for fracture studies.}
\label{fig:2D/3DTubeExpansionComparison}
\end{figure}

To validate the method in three dimensions, the tube expansion test from \citet{oldOgiveTest} was considered. The test provides validation of all the components of the algorithm, utilising shear, void and fracture. The test consists of a tube containing a target plug and a projectile. As the projectile hits the target, the tube around them expands and fractures. Both the projectile and target consist of a cylinder of polycarbonate material, with the tube being made of AerMet steel. The tube has a length of 50.8 mm, an inner radius of 6.35 mm and a wall thickness of 3 mm. The two cylinders are identical, both having a radius of 6.35 mm and a length of 25.4 mm. The upper cylinder is given an initial downward velocity of 1.92 kms$^{-1}$. The material parameters for the test are laid out in Tables \ref{tab:MaterialParameters} and \ref{tab:DamageParameters}. Both the polycarbonate and steel are assumed to obey ideal plasticity, with yield stresses of 0.08 and 2.56 GPa respectively. Exploiting the symmetry of the problem, a quarter of the tube is modelled, with reflective boundaries on the $x-$, $y-$ and $z-$axes. The test is run on a domain of $x = [0:20]$ mm, $y = [0:20]$ mm, $z = [0:52]$ mm, using a CFL 0.3 for a time of 30 $\mu$s. A resolution of $32\times32\times80$ cells with 2 levels of AMR, each of refinement factor 2, is used.

The full test is shown in Figure \ref{fig:TubeExpansionPictures}, where each material is delineated by its $\phi = 0.5$ volume fraction surface, with the steel tube shown in grey, and the polycarbonate blocks in red and blue. Figure \ref{fig:TubeExpansionSlice} shows the $z-$velocity, damage evolution and AMR mesh on a vertical slice of the domain. Taking a slice both allows the damage accumulation to be observed, and demonstrates the slide and void opening required for the polycarbonate blocks.

The experimental comparison for this test is the radial displacement profile measured by \citet{oldOgiveTest}, who give results up to 20 $\mu$s at three different positions along the length of the tube using both VISAR probes and photography. This work compares these measured values with those obtained here in Figure \ref{fig:2D/3DTubeExpansionComparison}, and good agreement is observed. The 3D results are obtained by averaging the radial displacement over 20 different positions around the tube. Moreover, these results demonstrate the ability of the method to facilitate three dimensional simulations. This is especially important as many tests often exhibit significant differences when modelled using a fully three dimensional approach, compared to when modelled in two dimensions with radial source terms. This effect is shown in Figure \ref{fig:2D/3DTubeExpansionComparison}, where it can be seen that the 3D profiles match the experimental results better than running the same test using the 2D model with radial symmetry. The ease with which the method can be extended to three dimensions enables this difference to be captured. At an even more fundamental level, a 2D approach can never capture the strain- and fracture-localisation that a 3D model can allow.

\section{Conclusions}

This work has outlined a new method to enable interface slide, void-opening, and material fracture in reduced-equation diffuse interface models. The model has been validated in one, two and three dimensions against experiment and previous simulations. 
The new model facilitates strenuous multi-material, multi-physics simulations in a pragmatic, scalable fashion. This is achieved by a set of flux-modifiers and interface seeding routines that enable material boundary conditions to be applied across diffuse interfaces. The model is general, capable of handling an arbitrary number of materials, in three dimensions, with a broad range of flux methods. 
The method is broadly applicable to any diffuse interface conservative update scheme, and it is not limited to the system of equations outlined here -- the method could be suitably extended to any multi-phase system.
Future work will focus on the inclusion of different boundary conditions such as static and dynamic rigid bodies and frictional boundaries, and the inclusion of additional multi-physics effects such as reactive fluids.

\appendix

\section{Damage Model Derivation}

From the Clausius-Duhem inequality, irreversible damage requires 
\begin{equation}
- Y_{(l)}\dot D_{(l)} \ge 0 \ .
\end{equation}
Using the definitions for the internal energies, the elastic energy release rate is defined as
\begin{equation}
-Y_{(l)}  =  \mathscr{E}_{(l)}^{c,\text{u}} + \frac{1}{\rho_{(l)}}G_{(l)}^{\text{u}}\mathcal{J}^2 \ .
\end{equation}
The equivalent stress can be expressed as 
\begin{equation}
\sigma_{eq}^2 = \frac{3}{2}||\text{dev}\boldsymbol\sigma||^2 = 6 G^2_{(l)} \mathcal{J}^2 = \left(1-D_{(l)}\right)^2{\sigma}_{eq}^{\text{u,2}} \ ,
\end{equation}
where ${\sigma}_{eq}^{\text{u}}$ is the {\it effective} (undamaged) equivalent stress.
Substituting this into the above:
\begin{equation}
-Y_{(l)}  =  \frac{\sigma_{eq}^2}{\left(1-D_{(l)}\right)^2}\frac{1}{2E_{(l)}^{\text{u}}}\left( \frac{2E_{(l)}^{\text{u}}\left(1-D_{(l)}\right)^2 {\mathscr{E}}_{(l)}^{c,\text{u}}}{\sigma_{eq}^2} + \frac{E_{(l)}^{\text{u}}}{3\rho_{(l)}G_{(l)}^{\text{u}}}\right) \ ,
\end{equation}
where the Youngs modulus $E_{(l)}^{\text{u}}$ has been introduced without loss of generality.

It is not immediately apparent that this matches the well-recognised expressions for $Y$ from the original works of \citet{Borona,lemaitre:2005} and others, which assume the St. Venant-Kirchhoff model for the strain-energy density function, so here proof is provided.

Assume the undamaged cold energy is taken to be the logarithmic equation of state from \cite{poirier:2000}, which closely resembles the St. Venant-Kirchhoff model:
\begin{equation}
{\mathscr{E}}_{(l)}^{c,\text{u}} = \frac{K_{(l)}^{\text{u}}}{2\rho^{(l)}_0} \left[\ln \left(\frac{\rho_{(l)}}{\rho_0^{(l)}}\right)\right]^2, \qquad {p}_{(l)}^{c,\text{u}} = K_{(l)}^{\text{u}} \left(\frac{\rho_{(l)}}{\rho_0^{(l)}}\right) \ln \left(\frac{\rho_{(l)}}{\rho_0^{(l)}}\right) \ ,
\end{equation}
so that 
\begin{equation}
{\mathscr{E}}_c^{(l)} = \frac{1}{2K}\frac{\rho_0^{(l)}}{\rho_{(l)}^2} {{p}_{(l)}^{c,\text{u}}}^2 \ .
\end{equation}
Finally, using the relationships between elastic moduli
\begin{equation}
\frac{E_{(l)}^{\text{u}}}{K_{(l)}^{\text{u}}} = 3\left(1-2\nu_{(l)}^{\text{u}}\right),\qquad \frac{E_{(l)}^{\text{u}}}{2G_{(l)}^{\text{u}}} = 1+ \nu_{(l)}^{\text{u}} \ ,
\end{equation}
and substituting all of the above, the elastic damage energy can be written
\begin{equation}
-Y_{(l)} = \frac{1}{\rho} \frac{\sigma_{eq}^2}{\left(1-D_{(l)}\right)^2}\frac{1}{2E_{(l)}^{\text{u}}} R_t\left(\frac{p}{\sigma_{eq}}\right), \qquad R_t\left(\frac{p}{\sigma_e}\right) = \left( 3\left(1-2\nu_{(l)}^{\text{u}}\right)\frac{\rho^{(l)}_0}{\rho_{(l)}}\left(\frac{p}{\sigma_{eq}}\right)^2 + \frac{2}{3}\left(1+\nu_{(l)}^{\text{u}}\right)\right)
\end{equation}
as required. 

The variable $R_t$ is commonly referred to as the stress-triaxiality function. Note that $R_t=1$ for uniaxial loading where $p/\sigma_e=\pm1/3$.
This result proves that the formulation conforms to the strain equivalence hypothesis \cite{lemaitre:2005}, which states that the constitutive model for strain in the damaged state is equivalent to that for undamaged material except the stress is replaced with the effective stress, and implies the linear degradation of shear and bulk moduli:
\begin{equation}
{G}_{(l)}=(1-D_{(l)}){G}_{(l)}^{\text{u}},\qquad {K}_{(l)}=(1-D_{(l)}){K}_{(l)}^{\text{u}} \ .
\end{equation}

\section{Approximations for the Hencky Strain}
\label{app:HenckyApproximations}
Throughout the course of the algorithm presented here, such as in Section \ref{sec:numericalPlastic}, it is necessary to calculate the strain from the deformation, and reconstruct the deformation from the strain. In this work the Hencky strain measure is employed. Although this measure admits a lot of benefits (as described by \citet{Bazant1998} and \citet{PadeHenckyStrain}), its primary downside is its numerical complexity. This derives from the requirement to solve matrix exponential/matrix logarithm problems.

The author of \citet{Barton2019} employs the approximation of \citet{Bazant1998} to calculate the matrix logarithm required to evaluate the strain:
\begin{equation}
\dev{\mathbf{H}^{e}} = \ln\left(\Vbar\right) \approx \frac{1}{2}\mathbf{H}_B^{e^{(1)}}\left(\Vbar{\Vbar}^{\text{T}}\right) \ ,
\end{equation}
where
\begin{equation}
\mathbf{H}_B^{e^{(m)}}\left(\Vbar\right) = \frac{1}{2m}\left({\Vbar}^m-{\Vbar}^{-m}\right) \ .
\end{equation}
To calculate the plastic update \citet{Barton2019} employs a singular value decomposition (SVD), thus avoiding the need to calculate the matrix exponential. This gives the numerically exact solution, but is computationally expensive.

This work also uses the approximation of \citet{Bazant1998} to calculate the strain, but employs a $(1,1)$ Pad\'{e} approximation to the matrix exponential to perform the plastic update:
\begin{equation}
\Vbar = \exp\left(\dev{\mathbf{H}^{e}}\right) \approx \left( \mathbf{I}- \frac{1}{2}\dev{\mathbf{H}^{e}}\right)^{-1}\left(\mathbf{I} + \frac{1}{2}\dev{\mathbf{H}^{e}}\right) \ .
\end{equation}
For simulations involving plasticity, this is found to be sufficient since the norm $||\dev{\mathbf{H}^{e}}||$ is known not to exceed sufficiently small values for the materials of interest in physically realistic scenarios \cite{MatrixExponential}.

If it is required to perform simulations without the physical limitations of plasticity, thus allowing the norm $||\dev{\mathbf{H}^{e}}||$ to grow unchecked, the combination of approximations outlined here may be insufficient, and will lead to the accumulation of errors. In this case, either the SVD can be used in conjunction with the approximation of \citet{Bazant1998}, or if the computational burden of a SVD is still to be avoided, then this work suggests the use of the $(2,2)$ diagonal Pad\'{e} approximation for both the calculation of the strain and the reconstruction of the deformation. To reconstruct the deformation, this work follows \citet{MatrixExponential} in taking:
\begin{align}
 \Vbar \approx \left(\mathbf{I}-\frac{1}{2}\dev{\mathbf{H}^{e}}+\frac{1}{12}\dev{\mathbf{H}^{e}}^2\right)^{-1}\left(\mathbf{I}+\frac{1}{2}\dev{\mathbf{H}^{e}}+\frac{1}{12}\dev{\mathbf{H}^{e}}^2\right) \ .
\end{align}
To calculate the strain, this work follows \citet{PadeHenckyStrain} in taking:
\begin{align}
 \dev{\mathbf{H}^{e}} \approx \frac{3}{2}\left(\mathbf{B}^2-\mathbf{I}\right)\left(\mathbf{B}^2+4\mathbf{B}+\mathbf{I}\right)^{-1} \ ,
\end{align}
where $\mathbf{B} = \Vbar {\Vbar}^{T}$ is the left Cauchy-Green tensor. 
These higher order approximations are suitable for all of the tests presented here, even those which do not employ plasticity.

\citet{Bazant1998} also presents another alternative method for calculating higher order approximations to the matrix logarithm, by taking linear combinations of different approximations:
\begin{align}
\dev{\mathbf{H}^{e}} &\approx \frac{1}{2}\mathbf{H}_B^{e^{(1,2)}}\left(\Vbar{\Vbar}^{\text{T}}\right) = \frac{1}{2}\left(1.307\mathbf{H}_B^{e^{(1)}}-0.307\mathbf{H}_B^{e^{(2)}}\right) \\
\dev{\mathbf{H}^{e}} &\approx \frac{1}{2}\mathbf{H}_B^{e^{(1/2,1)}}\left(\Vbar{\Vbar}^{\text{T}}\right) = \frac{1}{2}\left(1.326\mathbf{H}_B^{e^{(1/2)}}-0.326\mathbf{H}_B^{e^{(1)}}\right) \ .
\end{align}
The latter of these was found to give results similar to the $(2,2)$ Pad\'{e} approximation, but the Pad\'{e} approximations are preferred so as to avoid the use of a matrix square root.

\section{Exact Solutions for Elastic Initial Value Problems}
\label{app:ExactSolution}
Exact solutions to the initial value problems in Sections \ref{sec:section_ivp1} and \ref{sec:section_ivp2} are found using the method proposed in \citet{Barton} and \citet{BartonSliding}. 

From the initial left and right states, an initial estimate of the states between the waves (a.k.a star-states) is determined using numerical solutions. 
The initial solution is used to determine the wave types (distinguishing shocks from rarefactions by checking for entropy jumps and converging or diverging characteristic wave speeds), and then the six wave speeds $S_{j}$ for $1\le j\le6$ (from the Rankine-Hugoniot conditions for shocks, or the eigenvalue evaluated at the tail of the wave for a rarefaction). 
New star-states either side of the contact can then be determined exactly by solving across the three waves on the left, from left to right, then the three waves on the right, from right to left.
In the case of shocks, the solution is found from the Rankine-Hugoniot conditions:
\begin{equation}\label{residuals_shock}
\mathbf{F}(\mathbf{U}^+)-\mathbf{F}(\mathbf{U}^-)-{S}\left(\mathbf{U}^+-\mathbf{U}^-\right)=\mathbf{0}.
\end{equation}
where $\pm$ indicate upstream/downstream states. For rarefaction waves the solution is found by evaluating
\begin{equation}\label{rarefactionODE}
\frac{\partial \mathbf{W}}{\partial \xi}=\frac{\mathbf{r}_j(\mathbf{W})}{\mathbf{r}_j(\mathbf{W})\cdot \nabla_{\mathbf{W}}\lambda_j(\mathbf{W})},
\end{equation}
where $\lambda_j$ denotes the $j$-th eigenvalue, the speed $\xi$ is defined within the range $\lambda_j(\mathbf{W}^-)\le\xi=x/t\le\lambda_j(\mathbf{W}^+)$, $\nabla_{\mathbf{W}}$ denotes the gradient operator with respect to components of the vector of primitive variables, $\mathbf{W}$, and 
$\mathbf{r}_j$ denotes the $j$-th eigenvector.
Full details of how to solve the wave equations can be found in \cite{Barton}

From the star-states next to the contact wave, the residual errors, $\mathscr{R}$, in the required boundary conditions across the contact wave give a measure of the error in the wave speeds (since the number of boundary conditions across the contact equals the number of waves in the Riemann fan). The exact solution method follows an iterative procedure which, given the initial left and right states, seeks the exact wave speeds so as to minimise $\mathscr{R}$. 
In the case of a solid/solid problem with stick conditions the required conditions are equal velocity, normal stress and traction. In the case of slip it is equal normal velocity and stress, and zero traction.
In the case of the solid/vacuum problem there are only three waves, and the boundary conditions are zero normal stress and traction.
The exact solution is determined by solving the non-linear system $\mathscr{R}(S_1,S_2,\ldots,S_6)=\mathbf{0}$ for the wave speeds using the Newton-Raphson method.
For each new guess, the above procedure is repeated to determine a new value for the residuals until they fall below the threshold, which was here set to $10^{-9}$.

\section*{Acknowledgements}

This work was funded by AWE PLC. Tim Wallis acknowledges additional support by a grant from the UK Engineering and Physical Sciences Research Council (EPSRC) EP/L015552/1 for the Centre for Doctoral Training (CDT) in Computational Methods for Materials Science. The authors would like to acknowledge the valuable help of Dr Philip Blakely throughout the development of the method.

\section*{References}

\bibliography{SABERbib}

\begin{thebibliography}{63}
\providecommand{\natexlab}[1]{#1}
\providecommand{\url}[1]{\texttt{#1}}
\expandafter\ifx\csname urlstyle\endcsname\relax
  \providecommand{\doi}[1]{doi: #1}\else
  \providecommand{\doi}{doi: \begingroup \urlstyle{rm}\Url}\fi

\bibitem[Allaire et~al.(2000)Allaire, Clerc, and Kokh]{Allaire}
Gr\'{e}goire Allaire, S\'{e}bastien Clerc, and Samuel Kokh.
\newblock A five-equation model for the numerical simulation of interfaces in
  two-phase flows.
\newblock \emph{Comptes Rendus de l'Academie des Sciences Series I
  Mathematics}, 331\penalty0 (12):\penalty0 1017--1022, 2000.
\newblock ISSN 0764-4442.

\bibitem[Baer and Nunziato(1986)]{BaerNunziato}
M.R. Baer and J.W. Nunziato.
\newblock A two-phase mixture theory for the deflagration-to-detonation
  transition ({DDT}) in reactive granular materials.
\newblock \emph{International Journal of Multiphase Flow}, 12\penalty0
  (6):\penalty0 861--889, 1986.
\newblock ISSN 0301-9322.

\bibitem[Barlow et~al.(2018)Barlow, Klima, and Shashkov]{BarlowVoidOpening2}
Andrew Barlow, Matej Klima, and Mikhail Shashkov.
\newblock Constrained optimization framework for interface-aware sub-scale
  dynamics models for voids closure in lagrangian hydrodynamics.
\newblock \emph{Journal of computational physics}, 371:\penalty0 914--944,
  2018.
\newblock ISSN 0021-9991.

\bibitem[Barlow et~al.(2016)Barlow, Maire, Rider, Rieben, and
  Shashkov]{BarlowALE}
Andrew~J Barlow, Pierre-Henri Maire, William~J Rider, Robert~N Rieben, and
  Mikhail~J Shashkov.
\newblock Arbitrary lagrangian–eulerian methods for modeling high-speed
  compressible multimaterial flows.
\newblock \emph{Journal of computational physics}, 322\penalty0 (C):\penalty0
  603--665, 2016.
\newblock ISSN 0021-9991.

\bibitem[Barton(2016)]{BartonAnisotropicDamage}
Philip~T. Barton.
\newblock An {Eulerian} method for finite deformation anisotropic damage with
  application to high strain-rate problems.
\newblock \emph{International Journal of Plasticity}, 83:\penalty0 225--251,
  2016.
\newblock ISSN 0749-6419.

\bibitem[Barton(2018)]{BartonLevelSetDamage}
Philip~T Barton.
\newblock A level-set based {Eulerian} method for simulating problems involving
  high strain-rate fracture and fragmentation.
\newblock \emph{International Journal of Impact Engineering}, 117:\penalty0
  75--84, 2018.
\newblock ISSN 0734-743X.

\bibitem[Barton(2019)]{Barton2019}
Philip~T. Barton.
\newblock An interface-capturing {Godunov} method for the simulation of
  compressible solid-fluid problems.
\newblock \emph{Journal of Computational Physics}, 2019.
\newblock ISSN 0021-9991.

\bibitem[Barton and Drikakis(2010)]{BartonSliding}
Philip~T Barton and Dimitris Drikakis.
\newblock An {Eulerian} method for multi-component problems in non-linear
  elasticity with sliding interfaces.
\newblock \emph{Journal of Computational Physics}, 229\penalty0 (15):\penalty0
  5518--5540, 2010.
\newblock ISSN 0021-9991.

\bibitem[Barton et~al.(2009)Barton, Drikakis, Romenski, and Titarev]{Barton}
P.T. Barton, D.~Drikakis, E.~Romenski, and V.A. Titarev.
\newblock Exact and approximate solutions of {Riemann} problems in non-linear
  elasticity.
\newblock \emph{Journal of Computational Physics}, 228\penalty0 (18):\penalty0
  7046--7068, 2009.
\newblock ISSN 0021-9991.

\bibitem[Barton et~al.(2011)Barton, Obadia, and Drikakis]{barton:2011}
P.T. Barton, B.~Obadia, and D.~Drikakis.
\newblock A conservative level-set based method for compressible solid/fluid
  problems on fixed grids.
\newblock \emph{Journal of Computational Physics}, 230\penalty0 (21):\penalty0
  7867 -- 7890, 2011.
\newblock ISSN 0021-9991.
\newblock \doi{https://doi.org/10.1016/j.jcp.2011.07.008}.
\newblock URL
  \url{http://www.sciencedirect.com/science/article/pii/S002199911100413X}.

\bibitem[Ba\v{z}ant(1998)]{Bazant1998}
Zden\v{e}k~P. Ba\v{z}ant.
\newblock Easy-to-compute tensors with symmetric inverse approximating {Hencky}
  finite strain and its rate.
\newblock \emph{Journal of Engineering Materials and Technology}, 120\penalty0
  (2):\penalty0 131--136, 04 1998.
\newblock ISSN 0094-4289.
\newblock \doi{10.1115/1.2807001}.
\newblock URL \url{https://doi.org/10.1115/1.2807001}.

\bibitem[Benson(1992)]{benson:1992}
David~J. Benson.
\newblock Computational methods in {Lagrangian} and {Eulerian} hydrocodes.
\newblock \emph{Comput. Methods Appl. Mech. Eng.}, 99\penalty0
  (2–3):\penalty0 235–394, September 1992.
\newblock ISSN 0045-7825.
\newblock \doi{10.1016/0045-7825(92)90042-I}.
\newblock URL \url{https://doi.org/10.1016/0045-7825(92)90042-I}.

\bibitem[Berger and Colella(1989)]{berger:1988}
M.J Berger and P~Colella.
\newblock Local adaptive mesh refinement for shock hydrodynamics.
\newblock \emph{Journal of Computational Physics}, 82\penalty0 (1):\penalty0
  64--84, 1989.
\newblock ISSN 0021-9991.

\bibitem[Bonora(1997)]{Borona}
N.~Bonora.
\newblock A nonlinear {CDM} model for ductile failure.
\newblock \emph{Engineering Fracture Mechanics}, 58\penalty0 (1):\penalty0 11
  -- 28, 1997.
\newblock ISSN 0013-7944.
\newblock \doi{https://doi.org/10.1016/S0013-7944(97)00074-X}.
\newblock URL
  \url{http://www.sciencedirect.com/science/article/pii/S001379449700074X}.

\bibitem[Bonora et~al.(2009)Bonora, Ruggiero, Esposito, and
  Iannitti]{StressTriaxialityDependence}
N.~Bonora, A.~Ruggiero, L.~Esposito, and G.~Iannitti.
\newblock Damage development in high purity copper under varying dynamic
  conditions and microstructural states using continuum damage mechanics.
\newblock \emph{AIP Conference Proceedings}, 1195\penalty0 (1):\penalty0
  107--110, 2009.
\newblock \doi{10.1063/1.3294988}.
\newblock URL \url{https://aip.scitation.org/doi/abs/10.1063/1.3294988}.

\bibitem[Brauer et~al.(2018)Brauer, Rai, Nixon, and Udaykumar]{UdaykumarDamage}
A.~Brauer, N.~K. Rai, M.~E. Nixon, and H.~S. Udaykumar.
\newblock Modeling impact‐induced damage and debonding using level sets in a
  sharp interface {Eulerian} framework.
\newblock \emph{International Journal for Numerical Methods in Engineering},
  115\penalty0 (9):\penalty0 1108--1137, 2018.
\newblock ISSN 0029-5981.

\bibitem[Bruhns et~al.(2001)Bruhns, Xiao, and Meyers]{BruhnsStrainEnergy}
O.~T. Bruhns, H.~Xiao, and A.~Meyers.
\newblock Constitutive inequalities for an isotropic elastic strain-energy
  function based on {Hencky}'s logarithmic strain tensor.
\newblock \emph{Proceedings of the Royal Society A: Mathematical, Physical and
  Engineering Sciences}, 457\penalty0 (2013):\penalty0 2207--2226, 2001.
\newblock ISSN 1364-5021.

\bibitem[Chen et~al.(2010)Chen, Wang, and Liu]{ParticleLevelSetDamage}
Qianyi Chen, Jingtao Wang, and Kaixin Liu.
\newblock Improved {CE/SE} scheme with particle level set method for numerical
  simulation of spall fracture due to high-velocity impact.
\newblock \emph{Journal of Computational Physics}, 229\penalty0 (19):\penalty0
  7503--7519, 2010.
\newblock ISSN 0021-9991.

\bibitem[Courant et~al.(1928)Courant, Friedrichs, and Lewy]{CFL}
R.~Courant, K.~Friedrichs, and H.~Lewy.
\newblock Über die partiellen differenzengleichungen der mathematischen
  physik.
\newblock \emph{Mathematische Annalen}, 100\penalty0 (1):\penalty0 32--74,
  1928.
\newblock ISSN 0025-5831.

\bibitem[Czarnota et~al.(2008)Czarnota, Jacques, Mercier, and
  Molinari]{CzarnotaSpallation}
C~Czarnota, N~Jacques, S~Mercier, and A~Molinari.
\newblock Modelling of dynamic ductile fracture and application to the
  simulation of plate impact tests on tantalum.
\newblock \emph{Journal of the mechanics and physics of solids}, 56\penalty0
  (4):\penalty0 1624--1650, 2008.
\newblock ISSN 0022-5096.

\bibitem[Deiterding et~al.(2006)Deiterding, Radovitzky, Mauch, Noels, Cummings,
  and Meiron]{deiterding2006}
Ralf Deiterding, Raul Radovitzky, Sean Mauch, Ludovic Noels, Julian~C.
  Cummings, and Daniel~I. Meiron.
\newblock A virtual test facility for the efficient simulation of solid
  material response under strong shock and detonation wave loading.
\newblock \emph{Engineering with Computers}, 22:\penalty0 325--347, 2006.

\bibitem[Deng et~al.(2018)Deng, Inaba, Xie, Shyue, and Xiao]{BVDTHINC}
Xi~Deng, Satoshi Inaba, Bin Xie, Keh-Ming Shyue, and Feng Xiao.
\newblock High fidelity discontinuity-resolving reconstruction for compressible
  multiphase flows with moving interfaces.
\newblock \emph{Journal of Computational Physics}, 371:\penalty0 945--966,
  2018.
\newblock ISSN 0021-9991.

\bibitem[Dorovskii et~al.(1983)Dorovskii, Iskol'dskii, and
  Romenskii]{RomenskiiEOS}
V.~Dorovskii, A.~Iskol'dskii, and E.~Romenskii.
\newblock Dynamics of impulsive metal heating by a current and electrical
  explosion of conductors.
\newblock \emph{Journal of Applied Mechanics and Technical Physics},
  24\penalty0 (4):\penalty0 454--467, 1983.
\newblock ISSN 0021-8944.

\bibitem[Favrie and Gavrilyuk(2012)]{FavriePlasticDiffuse}
N.~Favrie and S.L. Gavrilyuk.
\newblock Diffuse interface model for compressible fluid – compressible
  elastic–plastic solid interaction.
\newblock \emph{Journal of Computational Physics}, 231\penalty0 (7):\penalty0
  2695--2723, 2012.
\newblock ISSN 0021-9991.

\bibitem[Favrie et~al.(2009)Favrie, Gavrilyuk, and
  Saurel]{FavrieElasticDiffuse}
N.~Favrie, S.L. Gavrilyuk, and R.~Saurel.
\newblock Solid–fluid diffuse interface model in cases of extreme
  deformations.
\newblock \emph{Journal of Computational Physics}, 228\penalty0 (16):\penalty0
  6037--6077, 2009.
\newblock ISSN 0021-9991.

\bibitem[Fedkiw et~al.(1999)Fedkiw, Aslam, Merriman, and Osher]{FedkiwGFM}
Ronald~P Fedkiw, Tariq Aslam, Barry Merriman, and Stanley Osher.
\newblock A non-oscillatory {Eulerian} approach to interfaces in multimaterial
  flows (the ghost fluid method).
\newblock \emph{Journal of Computational Physics}, 152\penalty0 (2):\penalty0
  457--492, 1999.
\newblock ISSN 0021-9991.

\bibitem[Gokhale et~al.(2018)Gokhale, Nikiforakis, and Klein]{NandanCutCell}
Nandan Gokhale, Nikos Nikiforakis, and Rupert Klein.
\newblock A dimensionally split cartesian cut cell method for hyperbolic
  conservation laws.
\newblock \emph{Journal of Computational Physics}, 364:\penalty0 186--208,
  2018.
\newblock ISSN 0021-9991.

\bibitem[Hank et~al.(2017)Hank, Gavrilyuk, Favrie, and
  Massoni]{HankExperimental}
Sarah Hank, Sergey Gavrilyuk, Nicolas Favrie, and Jacques Massoni.
\newblock Impact simulation by an {Eulerian} model for interaction of multiple
  elastic-plastic solids and fluids.
\newblock \emph{International Journal of Impact Engineering}, 109\penalty0
  (C):\penalty0 104--111, 2017.
\newblock ISSN 0734-743X.

\bibitem[Johnsen and Colonius(2006)]{JohnsenColonius}
Eric Johnsen and Tim Colonius.
\newblock Implementation of {WENO} schemes in compressible multicomponent flow
  problems.
\newblock \emph{Journal of Computational Physics}, 219\penalty0 (2):\penalty0
  715--732, 2006.
\newblock ISSN 0021-9991.

\bibitem[Johnson and Cook(1985)]{JohnsonCook}
Gordon~R. Johnson and William~H. Cook.
\newblock Fracture characteristics of three metals subjected to various
  strains, strain rates, temperatures and pressures.
\newblock \emph{Engineering Fracture Mechanics}, 21\penalty0 (1):\penalty0
  31--48, 1985.
\newblock ISSN 0013-7944.

\bibitem[{Juanicotena}(2006)]{triboPairExperiment}
A.~{Juanicotena}.
\newblock {Experimental investigation of dynamic friction at high contact
  pressure applied to an aluminium/stainless steel tribo pair}.
\newblock \emph{Journal de Physique IV}, 134\penalty0 (1):\penalty0 559--564,
  August 2006.
\newblock \doi{10.1051/jp4:2006134086}.

\bibitem[Kachanov(1958)]{kachanov:1958}
L.~Kachanov.
\newblock On the creep fracture time.
\newblock \emph{Izv Akad, Nauk USSR Otd Tech.}, 8:\penalty0 26--31, 1958.

\bibitem[Kanel et~al.(1997)Kanel, Razorenov, Bogatch, Utkin, and
  Grady]{KanelSpallation}
G.I Kanel, S.V Razorenov, A~Bogatch, A.V Utkin, and Dennis~E Grady.
\newblock Simulation of spall fracture of aluminum and magnesium over a wide
  range of load duration and temperature.
\newblock \emph{International journal of impact engineering}, 20\penalty0
  (6):\penalty0 467--478, 1997.
\newblock ISSN 0734-743X.

\bibitem[Kapila et~al.(2001)Kapila, Menikoff, Bdzil, Son, and Stewart]{Kapila}
A.~K. Kapila, R.~Menikoff, J.~B. Bdzil, S.~F. Son, and D.~S. Stewart.
\newblock Two-phase modeling of deflagration-to-detonation transition in
  granular materials: Reduced equations.
\newblock \emph{Physics of Fluids}, 13\penalty0 (10):\penalty0 3002--3024,
  2001.
\newblock ISSN 1070-6631.

\bibitem[Klima et~al.(2020)Klima, Barlow, Kucharik, and
  Shashkov]{BarlowVoidOpening}
Matej Klima, Andrew Barlow, Milan Kucharik, and Mikhail Shashkov.
\newblock An interface-aware sub-scale dynamics multi-material cell model for
  solids with void closure and opening at all speeds.
\newblock \emph{Computers \& fluids}, 208\penalty0 (C):\penalty0 104578, 2020.
\newblock ISSN 0045-7930.

\bibitem[Lemaitre and Desmorat(2005)]{lemaitre:2005}
J.~Lemaitre and R.~Desmorat.
\newblock \emph{Engineering damage mechanics}.
\newblock Springer, 2005.

\bibitem[Massoni et~al.(2002)Massoni, Saurel, Nkonga, and Abgrall]{Massoni}
J~Massoni, R~Saurel, B~Nkonga, and R~Abgrall.
\newblock Proposition de méthodes et modèles eulériens pour les problèmes
  à interfaces entre fluides compressibles en présence de transfert de
  chaleur: Some models and eulerian methods for interface problems between
  compressible fluids with heat transfer.
\newblock \emph{International journal of heat and mass transfer}, 45\penalty0
  (6):\penalty0 1287--1307, 2002.
\newblock ISSN 0017-9310.

\bibitem[Michael and Nikiforakis(2016)]{MiNi16}
L.~Michael and N.~Nikiforakis.
\newblock A hybrid formulation for the numerical simulation of condensed phase
  explosives.
\newblock \emph{Journal of Computational Physics}, 316:\penalty0 193--217,
  2016.
\newblock ISSN 0021-9991.

\bibitem[Michael and Nikiforakis(2018)]{MiNi18}
L.~Michael and N.~Nikiforakis.
\newblock A multi-physics methodology for the simulation of reactive flow and
  elastoplastic structural response.
\newblock \emph{Journal of Computational Physics}, 367\penalty0 (C):\penalty0
  1--27, 2018.
\newblock ISSN 0021-9991.

\bibitem[Michael et~al.(2019)Michael, Millmore, and
  Nikiforakis]{4StatesOfMatter}
Louisa Michael, Stephen~T. Millmore, and Nikolaos Nikiforakis.
\newblock A multi-physics methodology for four-states of matter.
\newblock 2019.

\bibitem[Miller and Colella(2002)]{miller:2002}
G.H. Miller and P.~Colella.
\newblock A conservative three-dimensional {Eulerian} method for coupled
  solid–fluid shock capturing.
\newblock \emph{Journal of Computational Physics}, 183\penalty0 (1):\penalty0
  26 -- 82, 2002.
\newblock ISSN 0021-9991.
\newblock \doi{https://doi.org/10.1006/jcph.2002.7158}.
\newblock URL
  \url{http://www.sciencedirect.com/science/article/pii/S0021999102971585}.

\bibitem[Millett(2015)]{1DSpallationPaper}
Jeremy C.~F Millett.
\newblock Modifications of the response of materials to shock loading by age
  hardening.
\newblock \emph{Metallurgical and Materials Transactions A}, 46\penalty0
  (10):\penalty0 4506--4517, 2015.
\newblock ISSN 1073-5623.

\bibitem[Moler and Loan(2003)]{MatrixExponential}
Cleve Moler and Charles~Van Loan.
\newblock Nineteen dubious ways to compute the exponential of a matrix,
  twenty-five years later.
\newblock \emph{SIAM review}, 45\penalty0 (1):\penalty0 3--49, 2003.
\newblock ISSN 0036-1445.

\bibitem[Murakami(2012)]{MurakamiCDM}
S.~Murakami.
\newblock \emph{Continuum Damage Mechanics: A Continuum Mechanics Approach to
  the Analysis of Damage and Fracture}.
\newblock Springer, frist edition. edition, 2012.
\newblock ISBN 978-94-007-2665-9.

\bibitem[Ndanou et~al.(2015)Ndanou, Favrie, and
  Gavrilyuk]{NdanouDiffuseFracture}
S.~Ndanou, N.~Favrie, and S.~Gavrilyuk.
\newblock Multi-solid and multi-fluid diffuse interface model: Applications to
  dynamic fracture and fragmentation.
\newblock \emph{Journal of Computational Physics}, 295:\penalty0 523--555,
  2015.
\newblock ISSN 0021-9991.

\bibitem[Ottosen(2005)]{Ottosen2005}
Niels~Saabye Ottosen.
\newblock \emph{The mechanics of constitutive modeling / Niels Saabye Ottosen,
  Matti Ristinmaa.}
\newblock Elsevier, Amsterdam ; London, 1st ed. edition, 2005.
\newblock ISBN 008044606X.

\bibitem[Petitpas et~al.(2009)Petitpas, Massoni, Saurel, Lapebie, and
  Munier]{PetitpasCaviation}
Fabien Petitpas, Jacques Massoni, Richard Saurel, Emmanuel Lapebie, and Laurent
  Munier.
\newblock Diffuse interface model for high speed cavitating underwater systems.
\newblock \emph{International Journal of Multiphase Flow}, 35\penalty0
  (8):\penalty0 747--759, 2009.
\newblock ISSN 0301-9322.

\bibitem[Pirondi and Bonora(2003)]{PirondiAndBorona}
A~Pirondi and N~Bonora.
\newblock Modeling ductile damage under fully reversed cycling.
\newblock \emph{Computational Materials Science}, 26:\penalty0 129--141, 2003.
\newblock ISSN 0927-0256.

\bibitem[Poirier(2000)]{poirier:2000}
J.-P. Poirier.
\newblock \emph{Introduction to the physics of the Earth's interior}.
\newblock Cambridge University Press, second edition edition, 2000.

\bibitem[Rezaee-Hajidehi et~al.(2020)Rezaee-Hajidehi, Tuma, and
  Stupkiewicz]{PadeHenckyStrain}
M~Rezaee-Hajidehi, K~Tuma, and S~Stupkiewicz.
\newblock A note on pade approximants of tensor logarithm with application to
  hencky-type hyperelasticity.
\newblock \emph{Computational mechanics}, 2020.
\newblock ISSN 0178-7675.

\bibitem[Sambasivan and Udaykumar(2009)]{RiemannGFM}
Shiv~Kumar Sambasivan and H.~S Udaykumar.
\newblock Ghost fluid method for strong shock interactions part 1: Fluid-fluid
  interfaces.
\newblock \emph{AIAA Journal}, 47\penalty0 (12):\penalty0 2907--2922, 2009.
\newblock ISSN 0001-1452.

\bibitem[Saurel and Abgrall(1999)]{SaurelAbgrall}
R.~Saurel and R.~Abgrall.
\newblock A simple method for compressible multifluid flow.
\newblock \emph{SIAM Journal on Scientific Computing}, 21\penalty0 (3), 1999.
\newblock ISSN 1064-8275.

\bibitem[Schoch et~al.(2013)Schoch, Nordin-Bates, and
  Nikiforakis]{SchochCoupled}
Stefan Schoch, Kevin Nordin-Bates, and Nikolaos Nikiforakis.
\newblock An {Eulerian} algorithm for coupled simulations of
  elastoplastic-solids and condensed-phase explosives.
\newblock \emph{Journal of Computational Physics}, 252:\penalty0 163--194,
  2013.
\newblock ISSN 0021-9991.

\bibitem[Tavelli et~al.(2020)Tavelli, Chiocchetti, Romenski, Gabriel, and
  Dumbser]{DumbserDamage}
Maurizio Tavelli, Simone Chiocchetti, Evgeniy Romenski, Alice-Agnes Gabriel,
  and Michael Dumbser.
\newblock Space-time adaptive ader discontinuous galerkin schemes for nonlinear
  hyperelasticity with material failure.
\newblock \emph{Journal of computational physics}, 422:\penalty0 109758, 2020.
\newblock ISSN 0021-9991.

\bibitem[Thomas et~al.(2016)Thomas, Veeser, Turley, and
  Hixson]{ExperimentalPlateImpact}
S.~Thomas, L.~Veeser, W.~Turley, and R.~Hixson.
\newblock Comparisons of {CTH} simulations with measured wave profiles for
  simple flyer plate experiments.
\newblock \emph{Journal of Dynamic Behavior of Materials}, 2\penalty0
  (3):\penalty0 365--371, 2016.
\newblock ISSN 2199-7446.

\bibitem[Vitali et~al.(2012)Vitali, Lomov, Antoun, and Fujino]{VitaliXEM}
E.~Vitali, I.~N. Lomov, T.~H. Antoun, and D.~H. Fujino.
\newblock An extended {Eulerian} method for contacts in {Godunov} formulations.
\newblock \emph{International Journal for Numerical Methods in Engineering},
  92:\penalty0 1139--1156, 2012.

\bibitem[Vitali and Benson(2012)]{VitaliTransportDiffusion}
Efrem Vitali and David Benson.
\newblock Modeling localized failure with arbitrary {Lagrangian} {Eulerian}
  methods.
\newblock \emph{Computational Mechanics}, 49\penalty0 (2):\penalty0 197--212,
  2012.
\newblock ISSN 0178-7675.

\bibitem[Vogler et~al.(2003)Vogler, Thornhill, Reinhart, Chhabildas, Grady,
  Wilson, Hurricane, and Sunwoo]{oldOgiveTest}
Tracy~J Vogler, Tom~F Thornhill, William~D Reinhart, Lalit~C Chhabildas,
  Dennis~E Grady, Leonard~T Wilson, Omar~A Hurricane, and Anne Sunwoo.
\newblock Fragmentation of materials in expanding tube experiments.
\newblock \emph{International Journal of Impact Engineering}, 29\penalty0
  (1):\penalty0 735--746, 2003.
\newblock ISSN 0734-743X.

\bibitem[Wallis et~al.(2021)Wallis, Barton, and
  Nikiforakis]{WallisMultiPhysics}
Tim Wallis, Philip~T. Barton, and Nikolaos Nikiforakis.
\newblock A diffuse interface model of reactive-fluids and solid-dynamics.
\newblock \emph{Computers \& Structures}, 254:\penalty0 106578, 2021.
\newblock ISSN 0045-7949.
\newblock \doi{https://doi.org/10.1016/j.compstruc.2021.106578}.
\newblock URL
  \url{https://www.sciencedirect.com/science/article/pii/S0045794921001000}.

\bibitem[Wang et~al.(2006)Wang, Liu, and Khoo]{rGFM}
C.~W. Wang, T.~G. Liu, and B.~C. Khoo.
\newblock A real ghost fluid method for the simulation of multimedium
  compressible flow.
\newblock \emph{SIAM Journal on Scientific Computing}, 28\penalty0
  (1):\penalty0 278--302, 2006.
\newblock ISSN 1064-8275.

\bibitem[Xiao et~al.(2005)Xiao, Honma, and Kono]{XiaoTHINC}
F.~Xiao, Y.~Honma, and T.~Kono.
\newblock A simple algebraic interface capturing scheme using hyperbolic
  tangent function.
\newblock \emph{International Journal for Numerical Methods in Fluids},
  48\penalty0 (9):\penalty0 1023--1040, 7 2005.
\newblock ISSN 1097-0363.
\newblock \doi{10.1002/fld.975}.
\newblock URL \url{https://doi.org/10.1002/fld.975}.

\bibitem[Youngs(1984)]{YoungsNormal}
David Youngs.
\newblock An interface tracking method for a 3d {Eulerian} hydrodynamics code.
\newblock Technical report, AWRE, 1984.

\bibitem[Zhang et~al.(2019)Zhang, Almgren, Beckner, Bell, Blaschke, Chan, Day,
  Friesen, Gott, Graves, Katz, Myers, Nguyen, Nonaka, Rosso, Williams, and
  Zingale]{amrex}
Weiqun Zhang, Ann Almgren, Vince Beckner, John Bell, Johannes Blaschke,
  Cy~Chan, Marcus Day, Brian Friesen, Kevin Gott, Daniel Graves, Max Katz,
  Andrew Myers, Tan Nguyen, Andrew Nonaka, Michele Rosso, Samuel Williams, and
  Michael Zingale.
\newblock {AMReX}: a framework for block-structured adaptive mesh refinement.
\newblock \emph{Journal of Open Source Software}, 4\penalty0 (37):\penalty0
  1370, May 2019.
\newblock \doi{10.21105/joss.01370}.
\newblock URL \url{https://doi.org/10.21105/joss.01370}.

\end{thebibliography}

\end{document}